\documentclass[hyperpdf,bindnopdf]{hepthesis}
\input{preamble}

\title{Deeply Inelastic Scattering Revisited}
\author{Felix Hekhorn}

\makeatletter
\@ifpackageloaded{hyperref}{%
\hypersetup{%
  pdftitle = {Deeply Inelastic Scattering Revisited},
  pdfsubject = {JSS26 lecture},
  pdfkeywords = {physics, DIS, JSS},
  pdfauthor = {\textcopyright\ Felix Hekhorn}
}}{}
\makeatother

\begin{document}

\begin{frontmatter}
\titlepage[University of Jyväskylä]{Lecture notes for the 35th Jyväskylä Summer School 3 - 14 August 2026}

\begin{abstract}%
  Due to its conceptual simplicity, Deeply Inelastic Scattering (DIS) often serves as a textbook example for Quantum Chromo Dynamics (QCD) and partonic structure of nucleons.
  However, it is this supposed simplicity, which makes DIS an ideal case for discussing many additional effects and corrections.
  In these lectures we discuss DIS within the domain of perturbative QCD covering a range of topics ranging from mathematical problems to kinematical corrections, and to practical challenges when attempting to interpret experimental measurements.
  Although here we focus on the exemplary case of DIS, the discussed topcis are relevant to a larger class of processes.
  We also give an overview of existing and future DIS measurements, and discuss their implementation into modern extractions of parton distribution functions (PDFs) which quantify the partonic structure of nucleons.

  The discussed features are all implemented in the open source code \texttt{Yadism}~\cite{Candido:2024rkr,barontini_2026_18758473}, which has been used for PDF extractions from a wide range of the available world data on fully inclusive DIS.
\end{abstract}

\begin{acknowledgements}
  First, I want to thank Werner Vogelsang, who not only introduced me into the world of pQCD, DIS, and PDFs, but who also gave me the opportunity to explore all the glorious details, which we happily neglect in the rest of this work.
  It turns out, recently, there is a renewed interest in my PhD thesis~\cite{Hekhorn:2019nlf} which discusses a fairly complicated DIS process.

  Second, I want to thank the members of the NNPDF collaboration and in particular Alessandro Candido, Giacomo Magni, and Stefano Forte, with whom I sorted out in long discussions even more details of DIS.
  Together, we have eventually brought all our experience and expertise into \texttt{Yadism}~\cite{Candido:2024rkr}, which, hopefully, serves the community well in the coming years.

  Third, I want to thank Ilkka Helenius and Hannu Paukkunen, who have supported me in various ways, gave me a lot freedom to work on my projects, and have shown me how to apply my knowledge to other, new fields.
  Eventually, they have also provided me with this opportunity to talk about my research and teach in front of students.

  This work has been supported by the Academy of Finland project 358090 and was funded as a part of the Center of Excellence in Quark Matter of the Academy of Finland, project 346326.
  The \LaTeX{} template is based on \texttt{hepthesis} by A.~Buckley.
  The Feynman diagrams are drawn with the \href{https://pypi.org/project/feynman/}{\texttt{feynman}} Python package.
\end{acknowledgements}

\begin{preface}
  Here is the usual list of disclaimers:
  \begin{itemize}
    \item This is a highly personal selection of topics, which we consider relevant enough for presenting in this occasion and by no means a review of DIS, not even of fully inclusive DIS in the collinear framework using pQCD, which is the only topic discussed here.
    \item Since this is not a review also the list of reference is not exhaustive, but rather a starting point and the reader should consult the mentioned works for further references.
    \item This work may contain errors, such as typographical or grammatical, but hopefully no physical, mistakes. Their distribution and the overall normalization is currently unknown, but we hope the former is highly localized and the latter small.
  \end{itemize}

  The rest of the work is organized as follows: in \cref{chap:intro} we recall the basic principles of DIS and pQCD upon which we build in \cref{chap:rl}, where we discuss a range of advanced DIS features.
  As an outlook in \cref{chap:outlook} we highlight some more advanced topics, which are directly connected to DIS but beyond the scope of this course.
\end{preface}

\tableofcontents

\frontquote{%
  Heut ham' 'mer de Dampfmaschin'.\\
  Wat is 'n Dampfmaschin'?\\
  Da stelle' mir uns mal janz dumm und sag'n, 'n Dampfmaschin', dat is' 'n jroße', runde', schwarze' Raum.}%
  {Thomas Newcomen}
\thispagestyle{empty}

\end{frontmatter}

\begin{mainmatter}

\chapter{The basics}
\label{chap:intro}

\chapterquote{Und jedem Anfang wohnt ein Zauberer inne}%
{Rincewind}

\pagenumbering{arabic}

This chapter serves to bring everybody to the same starting point, by quickly recalling the most important features of the underlying theory and the required practical tools.
We refrain here from giving a detailed introduction, which instead can be found in any standard textbook~\cite{Leader:2011gpt,Leader:1996hm,Peskin:1995ev,Halzen:1984mc,Kronfeld:2010bx}.

\section{Introduction}

We aim in this course to make a meaningful comparison between the theoretical predictions and the experimental measurement of a fully inclusive Deeply Inelastic Scattering (DIS) cross section.
Instead of an abstract, theoretical discussion about DIS, which can be found in a standard lecture on particle physics, we're looking here to make a connection with real-life physics.

\subsection{Motivation and non-goals}
The main focus is on Parton Distribution Functions (PDFs), for which DIS plays a major role.
In fact, DIS was, is, and will be a very important constrain in determining PDFs - only recently, e.g.\ starting from NNPDF4.0~\cite{NNPDF:2021njg}, LHC data starts to dominate in PDF fits.
Stressing the importance and relevance of PDFs for any kind of perturbative QCD (pQCD) application and to understand the inner structure of hadrons is left to the literature~\cite{Gao:2017yyd,Heinrich:2020ybq,Amoroso:2022eow}.
Moreover, DIS can be used to answer unresolved questions about PDFs, e.g.\ whether the proton has an intrinsic charm component~\cite{Ball:2022qks,NNPDF:2023tyk}.

In order to obtain a more realistic DIS treatment, we review general pQCD features and exemplify them using DIS.
In fact, fully inclusive DIS is very handy for such problems, since here the effects mostly appear in their simplest form and when considering hadronic collisions, e.g.\ at the LHC, the strategies are mostly the same just more involved.
We keep a phenomenological point of view with a specific focus on the numerical implementation as, eventually, we want to compare to the experimental measurements.
We refrain from giving detailed proofs in this course and point to the standard textbooks~\cite{Leader:2011gpt,Leader:1996hm,Peskin:1995ev,Halzen:1984mc,Kronfeld:2010bx} for that.
Instead, we consider different features as given and focus on their practical implications.
However, a major complication in this course is the interconnection between almost all features and while we discuss them here in turn, we highlight also how they are intertwined with each other.

In recent years (and much more so hopefully in the coming years), DIS has become a major research topic again thanks to the advent of the Electron Ion Collider (EIC)~\cite{Accardi:2012qut}.
While the planned physics program of the EIC goes far beyond standard fully inclusive DIS, the simplest case discussed in this work can serve as an introduction into more complicated topics.

The structure of this course follows in large parts the implementation of \texttt{Yadism} (Yet Another DIS module)~\cite{Candido:2024rkr,barontini_2026_18758473}, which is developed and used by the NNPDF collaboration in their PDF determinations.
The online documentation, which is available at
\begin{center}
    \url{https://yadism.readthedocs.io/} \,,
\end{center}
may serve as additional, more technical introduction into the topic.
Note that \texttt{Yadism} is an open-source project to which everybody can contribute\footnote{e.g.\ if some documentation can be improved}.

Since DIS is a very wide field, it is worth spelling out explicitly some non-goals:
\begin{itemize}
    \item We only consider pQCD in the collinear framework here.
        However, because DIS is such a powerful experiment, which offers a clean access to the inner structure of the protons, it can be studied using other frameworks as well, such as Color Glass Condensates~\cite{Bertilsson:2026vtu,Casuga:2026xxt} \textrightarrow{} ask H.~Mäntysaari or T.~Lappi
    \item We only consider fully inclusive cross sections here. %
        In particular, we do not consider any more complicated final states, such as, e.g., single jet or di-jet production~\cite{H1:2014cbm,Helenius:2026odp}.
    \item We only consider pQCD in the collinear framework here.
        While, Monte Carlo Event generators, such as, e.g., Pythia~\cite{Bierlich:2022pfr}, are rooted in collinear pQCD they eventually go much beyond and we do not consider their details here~\cite{Helenius:2026odp} \textrightarrow{} ask O.~Fedkevych or I.~Helenius.
    \item We only consider DIS here.
        In particular, we do not consider photo-production or elastic or diffractive cross sections.
        We define the necessary kinematics for the distinction below in \cref{sec:basics}.
    \item We compute the theoretical predictions here.
        So while we try to keep the connection to the experimental side close, we do not discuss any specific details about how to measure cross sections.
\end{itemize}

\subsection{Requirements}

\begin{figure}[ht]
    \includegraphics[width=.8\textwidth]{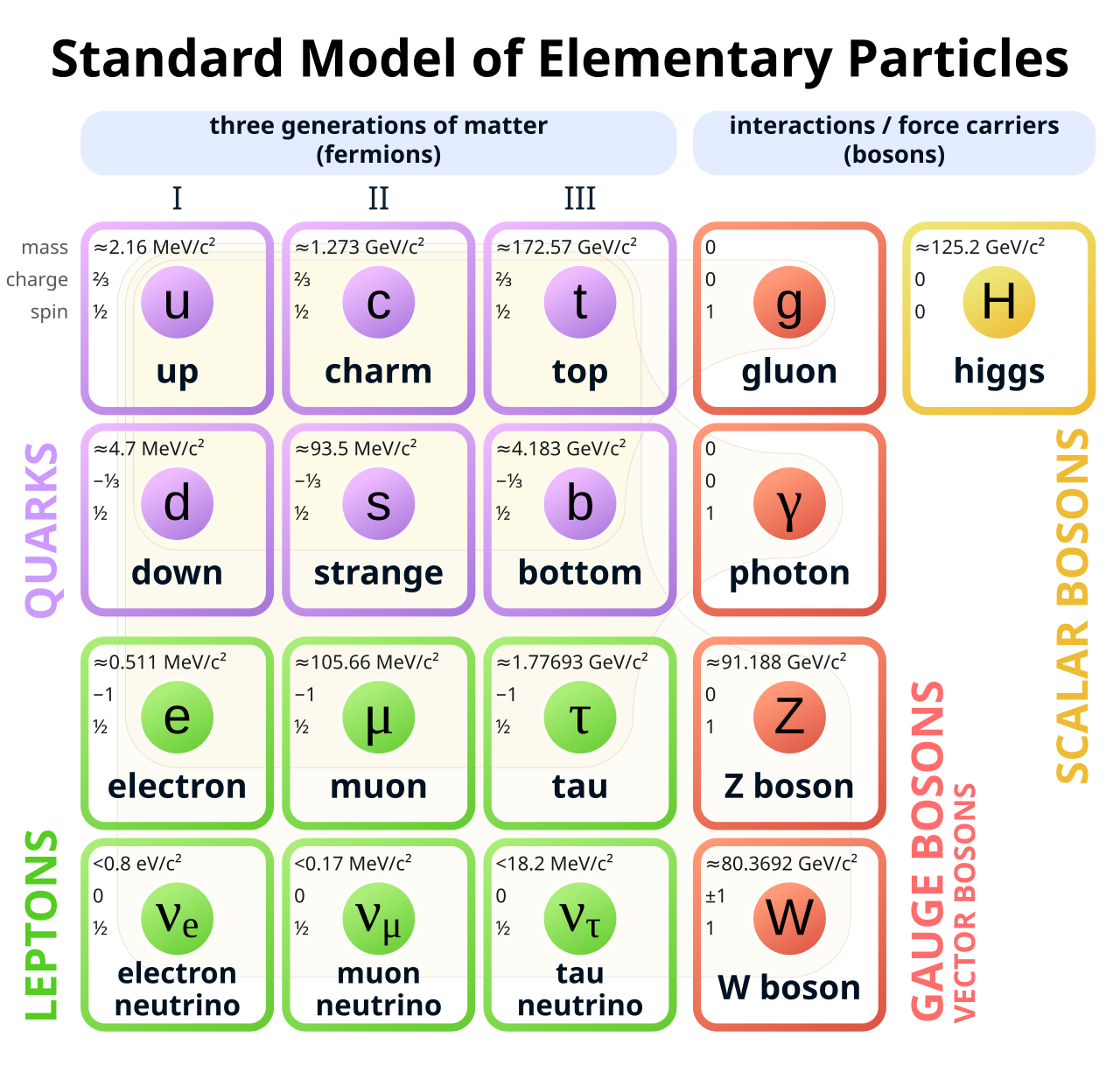}
    \caption{The Standard Model of Elementary Particles - taken from \cite{wikiSM}}
    \label{fig:SM}
\end{figure}

We need some basic knowledge from particle physics and, in particular, you need to be familiar with the Standard Model overview shown in \cref{fig:SM}.
We discuss in this course each and everyone of these particles, with the exception of the three heaviest in each category:
the heavies quark, the top, the heaviest lepton, the tau, and the heaviest boson, the Higgs.
They work, in principle, mostly like their lighter brothers and sisters, but in practice are basically always neglected.
We give a more thorough justification to discard them when we get to the actual numbers.
All other particles explicitly appear in this course and you need to know their properties.

\begin{conclusionbox}[label={box:particles}]{{Summary forces, particles, and charges}}
    \begin{itemize}
        \item The carrier of the strong force is the gluon
        \item Quarks have a strong charge, leptons don't
        \item The carriers of the electro-weak force are the photon, the Z boson, and the W boson
        \item Quarks and charged leptons have electro-magnetic charges
        \item Quarks and leptons have weak charges
    \end{itemize}
\end{conclusionbox}

We also need some basic knowledge from Quantum Field Theory (QFT).
To start the program we require a Lagrangian and while we focus mostly on $\mathcal L_\text{QCD}$ we need also the interaction with the electro-weak bosons, so, strictly speaking, we need the full standard model $\mathcal L_\text{SM}$.
Based on that we generate a set of Feynman rules and from that point onwards we are only talking in the language of diagrams.
As a first step we perform the renormalization of all couplings and fields to remove any ultra-violet (UV) divergencies.
\begin{conclusionbox}[label={box:pQCD}]{pQCD}
    \begin{itemize}
        \item To compute any perturbative object, we sum over all relevant diagrams and then multiply them with their complex conjugated counterpart.
        \item For measurable cross sections we need to integrate over all internal and final state particles and average over initial particles.
    \end{itemize}
\end{conclusionbox}

\subsection{The very basics of DIS}
\label{sec:basics}
\begin{conclusionbox}{DIS in words}
    fully inclusive DIS = fully inclusive deeply inelastic scattering
\end{conclusionbox}
What does that actually mean? We begin from the back:
\begin{itemize}
    \item \enquote{scattering} means, of course, we are scattering two particles with each other.
        In this particular case, we are scattering a lepton off a hadron and to make things simple in the beginning we consider an electron is scattered off a proton.
    \item \enquote{inelastic} means we are breaking the target, the proton, apart; it shatters into many pieces.
    \item \enquote{deeply} is an adverb\footnote{thus it is \enquote{deeply inelastic} or \enquote{deep-inelastic}, but not a list of adjectives \enquote{deep inelastic}} to inelastic and it means we are looking \enquote{deep} into the inside of the proton.
        In order to resolve small things, as usual, we need an highly energetic probe (i.e.\ with small wave length).
    \item \enquote{fully inclusive} means we are observing a single particle in the final state, which is the electron, but nothing else.
\end{itemize}
Thus, we can write the same information again with a formula
\begin{conclusionbox}[label={box:DISformula}]{DIS in a formula}
    fully inclusive DIS = $e^-(k) + p(P) \to e^-(k') + X $ with $-(k - k')^2 \gg \LQCD^2$
\end{conclusionbox}
\noindent{}where we have assigned the four-momenta $k,P,k'$ to the incoming electron, the incoming proton and the outgoing electron respectively.
As usual $X$ marks the inclusiveness, i.e.\ we don't observe anything else apart from the electron.
By requiring a large momentum-transfer between the incoming and the outgoing electron, i.e.\ their difference four-momentum vector has a large magnitude, we probe the proton with a highly energetic parton (which we can think of as photon for now).

Next, we need a way to actually compute theory predictions for the cross sections, which you can measure in an experiment.
We use the factorization theorem~\cite{Collins:1989gx},
\begin{conclusionbox}{Factorization Theorem}
    \begin{align}
        \sigma(x,Q^2) &= \sum_{j} \int\limits_x^1 \frac{\dd z}{z} C_j(z,Q^2) f_j(\xi = x/z,Q^2) + \order{\frac{\LQCD^2}{Q^2}} \label{eq:fact1} \\
         &= \sum_{j} \qty(C_j(Q^2)\otimes f_j(Q^2))(x) + \order{\frac{\LQCD^2}{Q^2}} \label{eq:fact2} \\
         &= \qty(\vb C^T(Q^2) \otimes \vb f(Q^2))(x) + \order{\frac{\LQCD^2}{Q^2}}\,, \label{eq:fact3}
    \end{align}
\end{conclusionbox}
\noindent{}which states how to compute a cross section $\sigma(x,Q^2)$ from coefficient function $\vb C(z,Q^2)$, computable in pQCD, and a universal PDF $\vb f(\xi,Q^2)$.
In the remaining part of the course we discuss, enhance and correct this formula in many different ways and basically challenge every aspect.
We define each and every variable and function in \cref{sec:xs}, so we can focus for the moment on three other important aspects: universality, measurability, and notation.

First, the crucial, physical aspect about \cref{eq:fact1} is universality: for any cross section $\sigma$ measured at any $x$ and $Q^2$ there is exactly one PDF $\vb f(\xi,Q^2)$.
Instead, the coefficient functions $\vb C(z,Q^2)$ may depend on the definition of the cross section, but they are computable from first principles.
This implies the predictive power of our framework: if $\vb C$ is fixed and $\vb f$ is universal, we can measure $\sigma$ in one set of experiments and then predict the outcome of another set of measurements $\sigma'$.

Second, it is important to keep in mind that \cref{eq:fact1} contains only a single measurable quantity: the cross section.
Instead, coefficient functions $\vb C$, PDFs $\vb f$ and also the strong coupling $\alpha_s$ are not directly measurable in a real experiment but have to be constructed consistently within the given framework of pQCD.
For example, while cross sections have to be positive by definition, this is less trivial for PDFs~\cite{Candido:2020yat,Collins:2021vke,Candido:2023ujx} (see \cref{sec:schemes}).

Third, a more practical aspect of \cref{eq:fact1,eq:fact2,eq:fact3} is the short-hand notation we introduce there.
While \cref{eq:fact1} is the most canonical form of the factorization theorem, which can be found like this in most textbooks, we introduce in \cref{eq:fact2} the multiplicative convolution symbol $\otimes$.
Note that we have suppressed the variable over which the convolution is performed and recall that convolution is associative, i.e.\ $\qty(f\otimes g)\otimes h = f\otimes\qty(g\otimes h)$.
Finally, in \cref{eq:fact3} we introduce the vector notation, where we denote objects, which have a non-trivial dependence on the parton flavor $j$, with bold font.
We use vector notation and the index notation interchangeably depending on the context.
\Cref{eq:fact3} looks nice and compact, but it worth recalling that it already hides a lot of details: specifically, $C_j(z,Q^2)$ is an object which depends on three different variables.

To Leading Order (LO) accuracy, the coefficient functions $\vb C$ greatly simplify and we find
\begin{equation}
    \sigma(x,Q^2) = \sum_j e_{q_j}^2 f_j(x,Q^2)\,, \label{eq:LOxs}
\end{equation}
i.e.\ the cross section is given by the sum of the quark PDFs weighted by their electrical charges squared.
Note that here the variables, which are used for the experimentally measured cross section on the left-hand-side, are also used as arguments of the PDFs on the right-hand-side.
This is a pure coincidence only valid in the simplest case at LO and two historical comments can be made here.
First, when studying (older) literature this equivalence can lead to confusion as, e.g., PDFs have been also called \enquote{structure functions} at the time, or what exactly we mean when talking about a general variable \enquote{$x$}.
Second, in the so-called \enquote{parton model}, which was popular before QCD was invented~\cite{Soper:2026kvo}, the index $j$ just referred to a \enquote{parton} with some properties.
Instead, within QCD we now know exactly, what the object in question is: a quark, which has a strict definition in QFT with well-defined properties and quantum numbers.
For simplicity, we still call an object \enquote{parton} if it can be either a quark or a gluon.

\section{Setting the stage}
\label{sec:xs}
\begin{figure}[ht]
    \includegraphics[width=.4\textwidth]{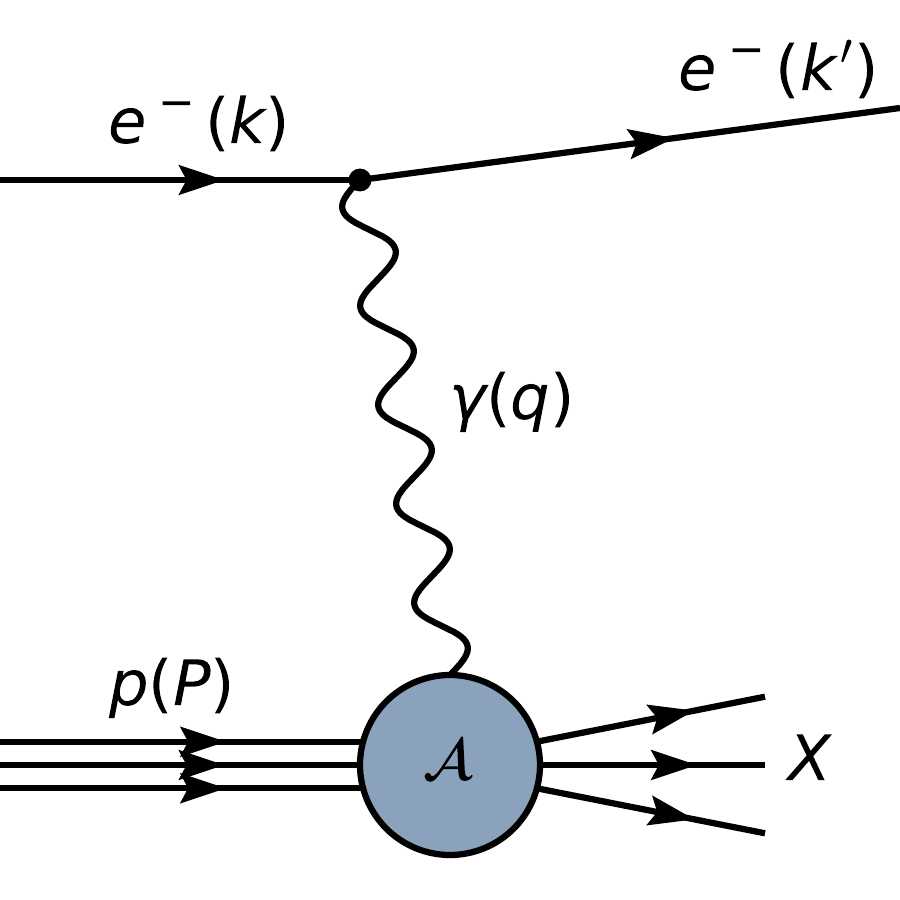}
    \caption{Fully inclusive DIS represented as Feynman diagram}
    \label{fig:DIS}
\end{figure}

In the following section we collect and define all ingredients needed for \cref{eq:fact1}.
As already said, DIS is given by
\begin{equation}
    e^-(k) + p(P) \to e^-(k') + X \label{eq:DISreaction}
\end{equation}
where the four-momenta $k,P,k'$ belong to the incoming electron, the incoming proton and the outgoing electron respectively as shown in \cref{fig:DIS}.
This allows us, first, to define four momentum of the exchanged photon
\begin{equation}
    q = k - k' \label{eq:defq}
\end{equation}
and from there the set of four common variables:
\begin{align}
    s_l &= (k + P)^2 = \frac{Q^2}{xy}\,, \label{eq:defslep} \\
    Q^2 &= - q^2 \label{eq:defQ2}\,, \\
    x &= \frac{Q^2}{2 q\cdot P}\,, \label{eq:defx} \\
    y &= \frac{q\cdot P}{k \cdot P}\,. \label{eq:defy}
\end{align}
We refer to $Q^2$ as virtuality (of the exchanged boson), to $x$ as Bjorken-$x$ (named after James Bjorken and often named $x_\text{Bj}$), and to $y$ as inelasticity.
Note that \cref{eq:defslep} gives an explicit relation among these, thus we have only three independent and one dependent variable.
In practice, the invariant mass of the electron-proton system $s_l$ is fixed by the experimental setup, i.e.\ by the energy of the colliding particles.
Also the measurable range of $y$ is in practice limited by the experimental geometry.
DIS cross sections are typically provided as double-differential cross sections in $x$ and $y$.
Finally, it is convenient to define the mass squared of the system X recoiling against the scatter electron,
\begin{equation}
    W^2 = (P + q)^2 = Q^2\left(\frac 1 x - 1\right) = s_h\,, \label{eq:defw2}
\end{equation}
which also corresponds to the invariant mass of the photon-proton system $s_h$.
\begin{conclusionbox}[label={box:DISreq}]{DIS requirements}
    We require $Q^2 \gg \LQCD^2$ (deep) and $W^2 \gg \LQCD^2$ (inelastic) to call a lepton-hadron scattering DIS.
\end{conclusionbox}

We can decompose the (double-differential) cross section into a leptonic interaction and a hadronic interaction, yielding each a tensor in Minkowski space, $L_{\mu\nu}$ and $W_{\mu\nu}$ respectively.
Thus, we can write
\begin{equation}
    \frac{\dd[2] \sigma}{\dd x \dd y} = \frac{2\pi y \aem^2}{Q^4} L_{\mu\nu}W^{\mu\nu}
\end{equation}
where we have introduced the electro-magnetic coupling $\aem$, which appears once on the leptonic side and once on the hadronic side.
We can express the leptonic tensor using plain Quantum Electrodynamics (QED) and provide a decomposition of the hadronic tensor in terms of scalar functions using $q$, $P$, and the metric tensor (see \cref{app:LO}).
Together, we obtain
\begin{equation}
    \frac{\dd[2] \sigma}{\dd x \dd y} = \frac{2\pi \aem^2}{xy Q^2} \left(\left(2 - 2y + y^2\right)F_2(x,Q^2) - y^2 F_L(x,Q^2)\right) \label{eq:dsigmadxdy}
\end{equation}
where we have introduced the two common scalar functions, now called structure functions $F_2$ and $F_L$, which are often measured in experiments.
It is convenient to also define a third, linearly dependent structure function via
\begin{equation}
    2xF_1 (x,Q^2) = F_2(x,Q^2) - F_L(x,Q^2)\,.
\end{equation}
Keep in mind that the discussion so far is a major simplification of DIS and in \cref{sec:diverse} we reintroduce the main complexity.

Let us continue to collect the remaining ingredient for \cref{eq:fact1}:
\begin{itemize}
    \item The sum over partons $j$ includes quarks and gluons and we refine in \cref{sec:hq} how large the sum actually is.
    \item The parton momentum fraction $\xi = x/z$ is the fraction of the four momentum the parton carries with respect to its parent proton.
        Recall that the convolution in \cref{eq:fact1} can also be written as an integral over $\xi$.
    \item The probability to find a parton $j$ with momentum $\xi P$ inside a proton with momentum $P$ in an experiment performed at the scale $Q^2$ is given by the PDF $f_j(\xi,Q^2)\dd \xi$.
    \item The integration over $z$ (or $\xi$) sums all configuration to generate the requested final state configuration.
\end{itemize}
Finally, the coefficient functions $\vb C(z,Q^2)$ are computable order-by-order in pQCD and thus can be decomposed as
\begin{equation}
    \vb C(z,Q^2) = \sum_{k=0} \alpha_s^k(Q^2) \vb C^{(k)}(z)\,. \label{eq:CpQCD}
\end{equation}
Note that the $Q^2$ dependency of $\vb C(z,Q^2)$ is fully captured by the strong coupling $\alpha_s(Q^2)$.

\section{Leading order and next-to-leading order coefficient functions}
\label{sec:nlo}
We keep the discussion in the rest of the work mostly general and do not focus on any particular observable.
However, here, for illustrative purposes we give the coefficient functions for the $F_2$ structure function to highlight some generic features.

\begin{figure}[ht]
    \centering
    \begin{subcaptionblock}{.2\textwidth}
        \includegraphics[width=\textwidth]{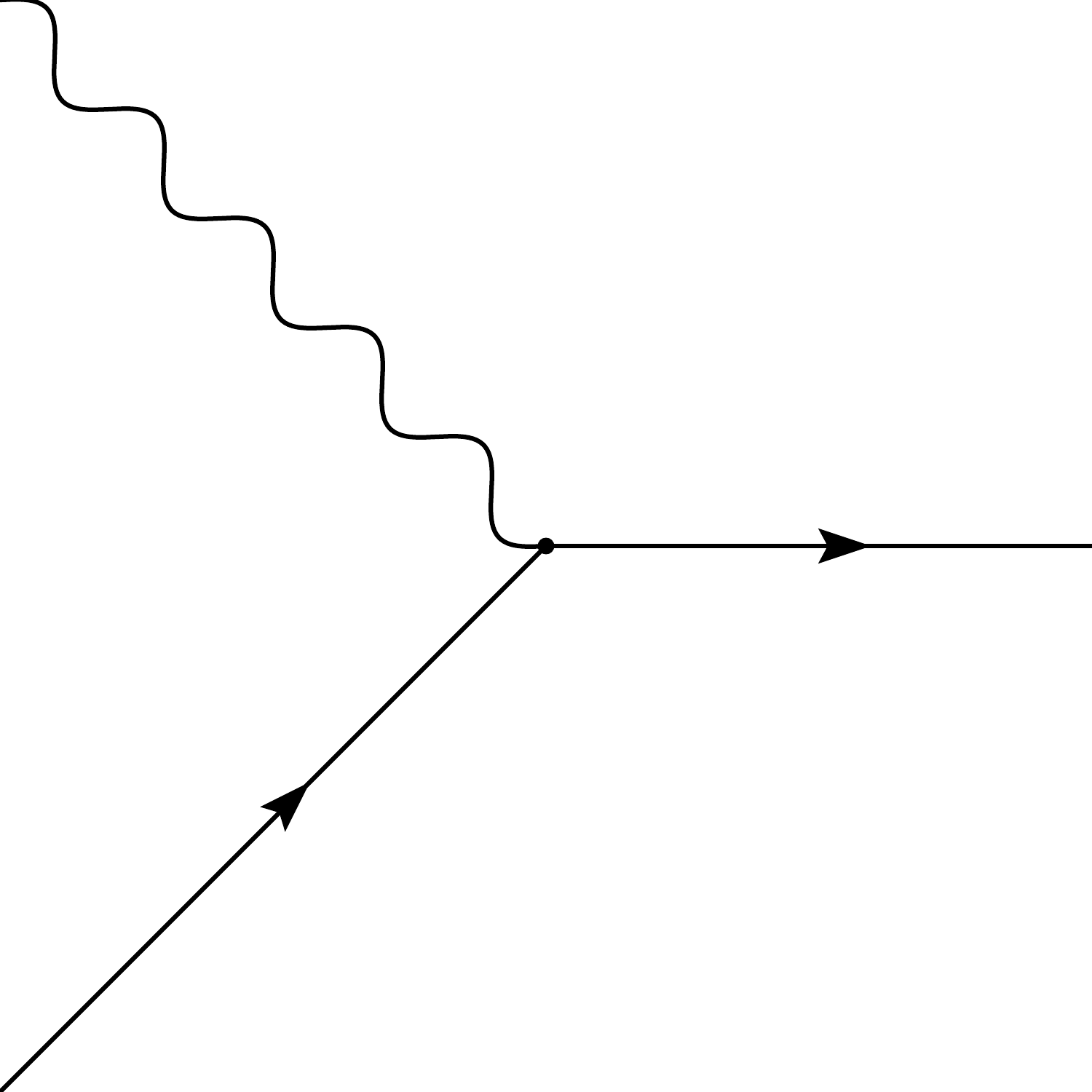}
        \caption{}
        \label{fig:LO}
    \end{subcaptionblock}%
    \begin{subcaptionblock}{.2\textwidth}
        \includegraphics[width=\textwidth]{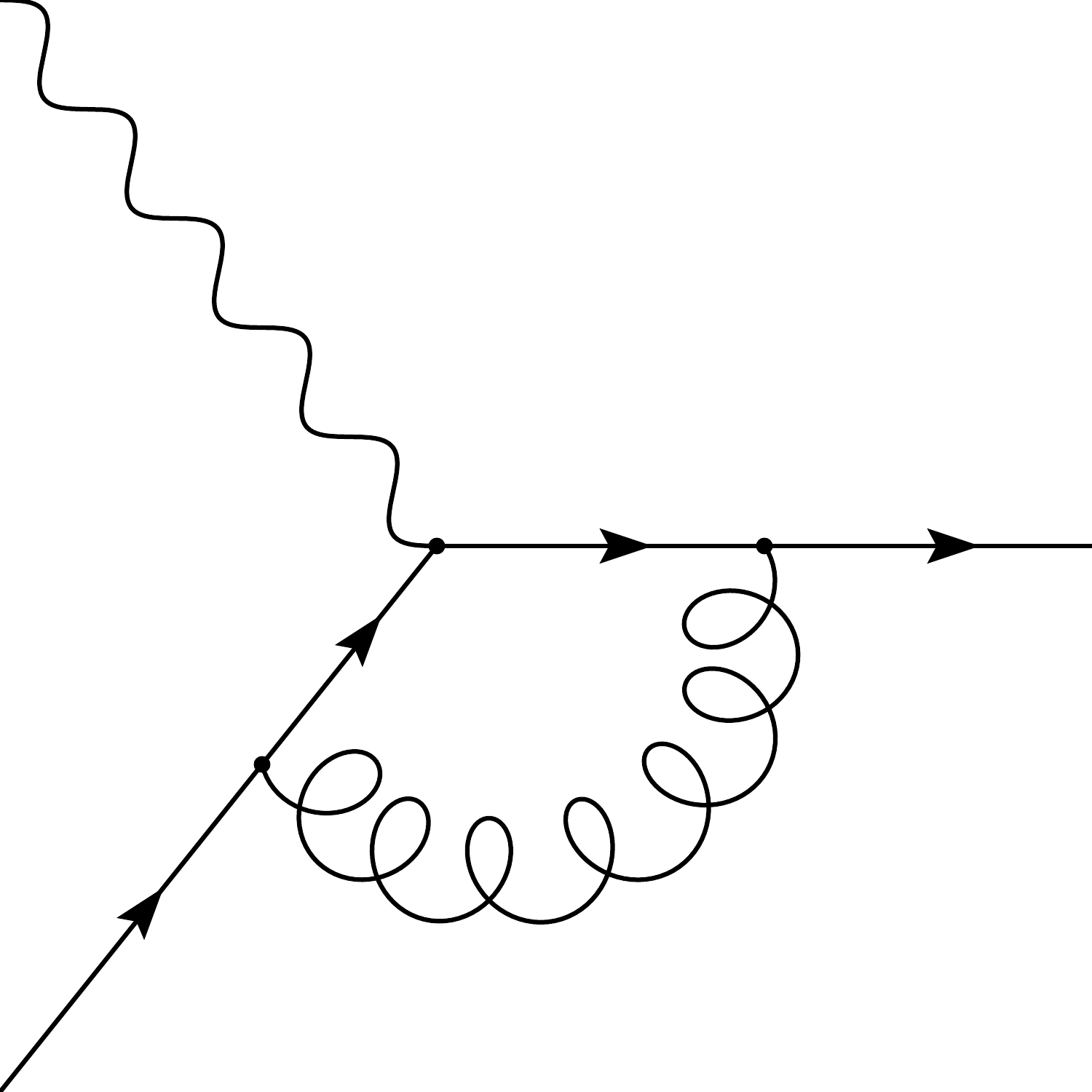}
        \caption{}
        \label{fig:NLOqv}
    \end{subcaptionblock}%
    \begin{subcaptionblock}{.2\textwidth}
        \includegraphics[width=\textwidth]{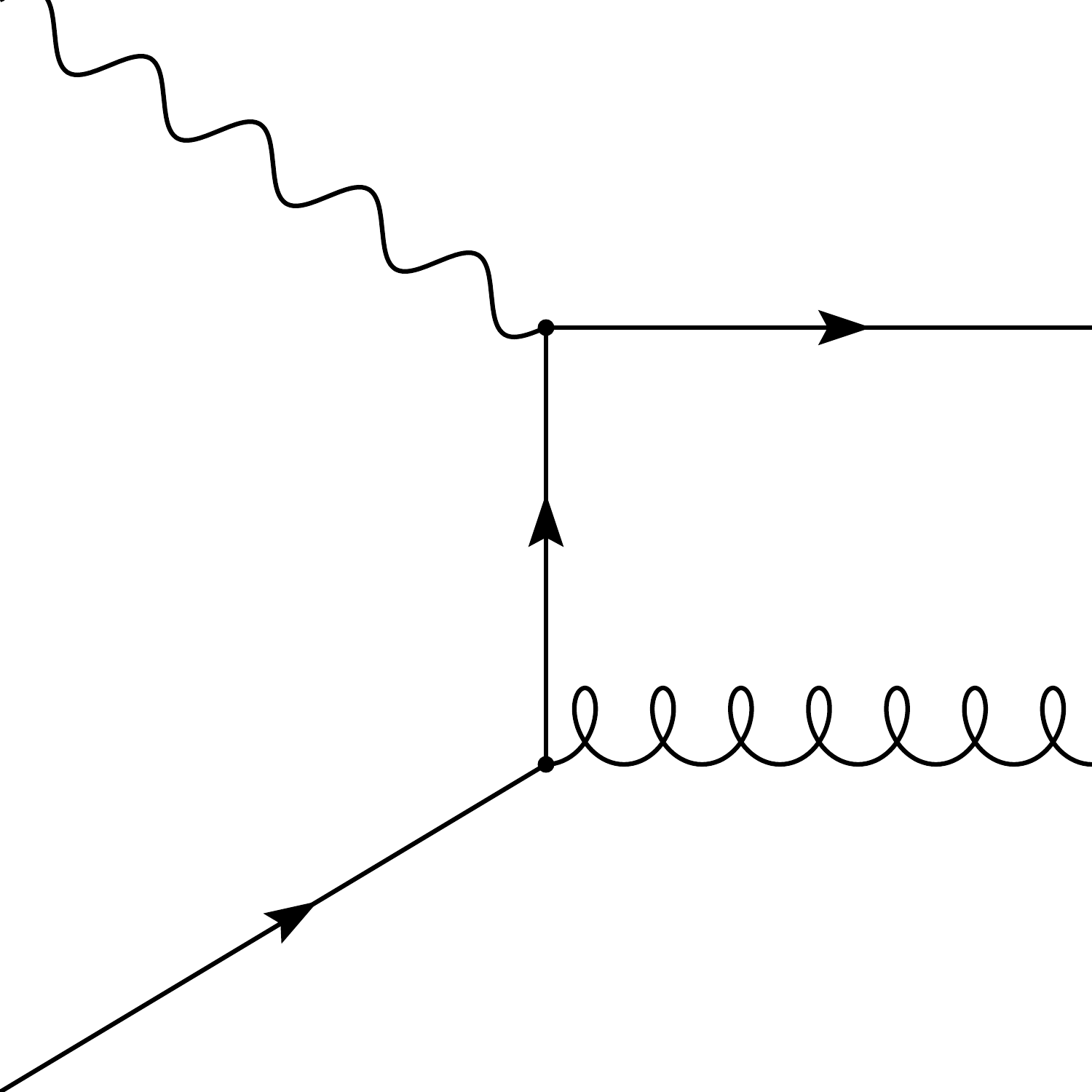}
        \caption{}
        \label{fig:NLOqrt}
    \end{subcaptionblock}%
    \begin{subcaptionblock}{.2\textwidth}
        \includegraphics[width=\textwidth]{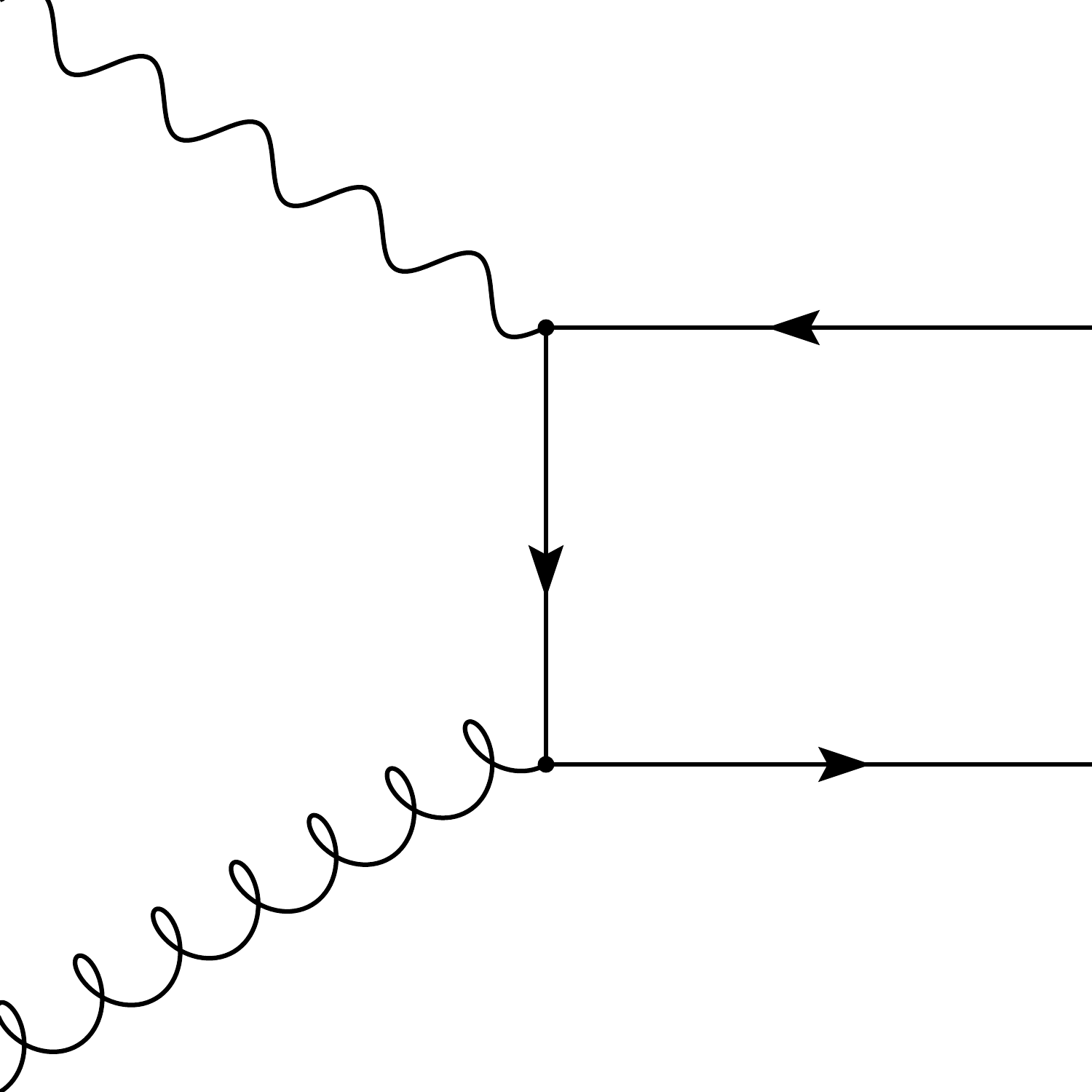}
        \caption{}
        \label{fig:NLOg}
    \end{subcaptionblock}
    \captionsetup{subrefformat=parens}
    \caption{Representative LO and NLO DIS Feynman diagrams.
        While at LO there is a single diagram \subref{fig:LO}, starting from NLO there are several type of diagrams: virtual corrections \subref{fig:NLOqv}, radiative corrections \subref{fig:NLOqrt}, and the gluon as initial state parton \subref{fig:NLOg}.
        Note that the list is not exhaustive.
    }
    \label{fig:Feynman}
\end{figure}

At LO accuracy, DIS is trivial, \cref{fig:LO}, and we find
\begin{equation}
    C_q^{(0)}(z) = C_{\bar q}^{(0)}(z) = e_q^2 \delta(1 - z) ~\forall q\,, \quad C_g^{(0)}(z) = 0 \label{eq:LOC}
\end{equation}
where $q$ represents any quark and $g$ represent the gluon.
Inserting \cref{eq:LOC} into \cref{eq:fact1} yields \cref{eq:LOxs}.
In \cref{app:LO} we cover more details on LO DIS.

Starting from next-to-leading order (NLO), the diagrams become more complicated and we encounter new features:
\begin{itemize}
    \item We have to perform renormalization at the requested order to remove all UV divergencies.
    \item We encounter loop corrections, which introduce infra-red (IR) divergencies, \cref{fig:NLOqv}.
    \item We can radiate additional particles, which introduce soft and/or collinear divergencies, e.g.\ \cref{fig:NLOqrt}.
    \item The initial parton can also be a gluon now, e.g.\ \cref{fig:NLOg}.
\end{itemize}
As requested in \cref{box:pQCD} we have to sum over all possible diagrams and since we only consider the fully inclusive case here, we can apply the Kinoshita-Lee-Nauenberg theorem~\cite{Bloch:1937pw,Kinoshita:1962ur,Lee:1964is} straightforwardly and cancel all soft and IR divergencies.
Finally, we apply collinear factorization and subtract the remaining collinear divergencies from the coefficient functions and reabsorb them into the PDF definition, \cref{eq:fact1}.
Any divergency which we remove introduces a distribution, such as the Dirac delta distribution or the usual +-distribution, into the coefficient functions, i.e.\ there is a direct mapping.

Eventually, we get~\cite{Vermaseren:2005qc} for all quarks $q$
\begin{align}
    C_q^{(1)}(z) &= \frac{e_q^2 C_F}{4\pi} \left[ 4 \qty(\frac{\ln(1-z)}{1-z})_+ - 3 \qty(\frac{1}{1-z})_+ - \qty(9+4\zeta_2)\delta(1-z)\right. \nonumber\\
    & \hspace{40pt} \left. - 2(1+z)\ln\left(\frac{1-z}{z}\right) - 4 \frac{\ln(z)}{1-z} + 6 + 4z\right] \label{eq:NLOCq}\\
     &= C_{\bar q}^{(1)}(z)
\end{align}
and for the gluon
\begin{align}
    C_g^{(1)}(z) &=\frac 1 {4\pi} \left(\sum_q e_q^2\right) \left[ \qty(2 - 4 z (1 - z)) \ln\left(\frac{1 - z} z\right)  - 2 + 16 z (1 - z)  \right] \label{eq:NLOCg}
\end{align}
where $C_F$ is the second Casimir operator of the fundamental representation, $C_F = (N_C^2 - 1)/(2 N_C)$, with $N_C=3$ the number of colors.

At a given perturbative order we explore a certain space of mathematical (special) functions and, e.g., at NLO we encounter for the first time the Riemann Zeta function $\zeta(n)$, which is defined at integer values $n$ by
\begin{equation}
    \zeta_n = \zeta(n) = \sum\limits_{j=1}^\infty \frac 1 {j^n}\,.
\end{equation}
The definition of the usual +-distribution is given in \cref{eq:PlusDistr} and discussed further in \cref{sec:rsl}.

  \chapter{DIS in real life}
\label{chap:rl}

\chapterquote{Drü Phüsükür müt düm Vürlüsüngsscrüpt,\\
süßen vür dür Tüfül ünd ürzühltün süch wüs.}%
{Forschungsalltag}

In the following we expose the complexity that DIS requires, when we want to confront actual experimental measurements with our theoretical predictions, or the other way round, if we want to use them to extract realistic PDFs.
We start by reviewing a small part of the available world data in \cref{sec:diverse}, which introduces new particles and thus new possibilities into the game.
In \cref{sec:interpolation} we examine how to implement the coefficient function in an efficient way, such that PDF fitting groups can use them in a simple way.
We review the pQCD property of resummation in \cref{sec:resum}, which we then use to discuss two other major features: first, scale variations in \cref{sec:sv} and then the treatment of heavy quarks in \cref{sec:hq}.
Kinematic corrections are discussed in \cref{sec:TMC} and in \cref{sec:schemes} we turn instead to more theoretical considerations.
Finally, in \cref{sec:summary} we collect all considered corrections and spell the hidden variables in the textbook version of \cref{eq:fact1}.

All topics are either explicitly or at the very least always implicitly correlated with each other, making a given feature actually more complicated then it may seem on first glance.
Thus the sorting of the sections is challenging, but hopefully follows now an order of importance.
We try to spell the interplay explicitly where it is appropriate.

\section{DIS is diverse}
\label{sec:diverse}
In this section we give an overview of some of the different DIS datasets commonly used in PDF fits.
The various experiments probe different physics and are thus necessary for a realistic PDF extraction.

\subsection{HERA}
\begin{figure}[ht]
    \includegraphics[width=.7\textwidth]{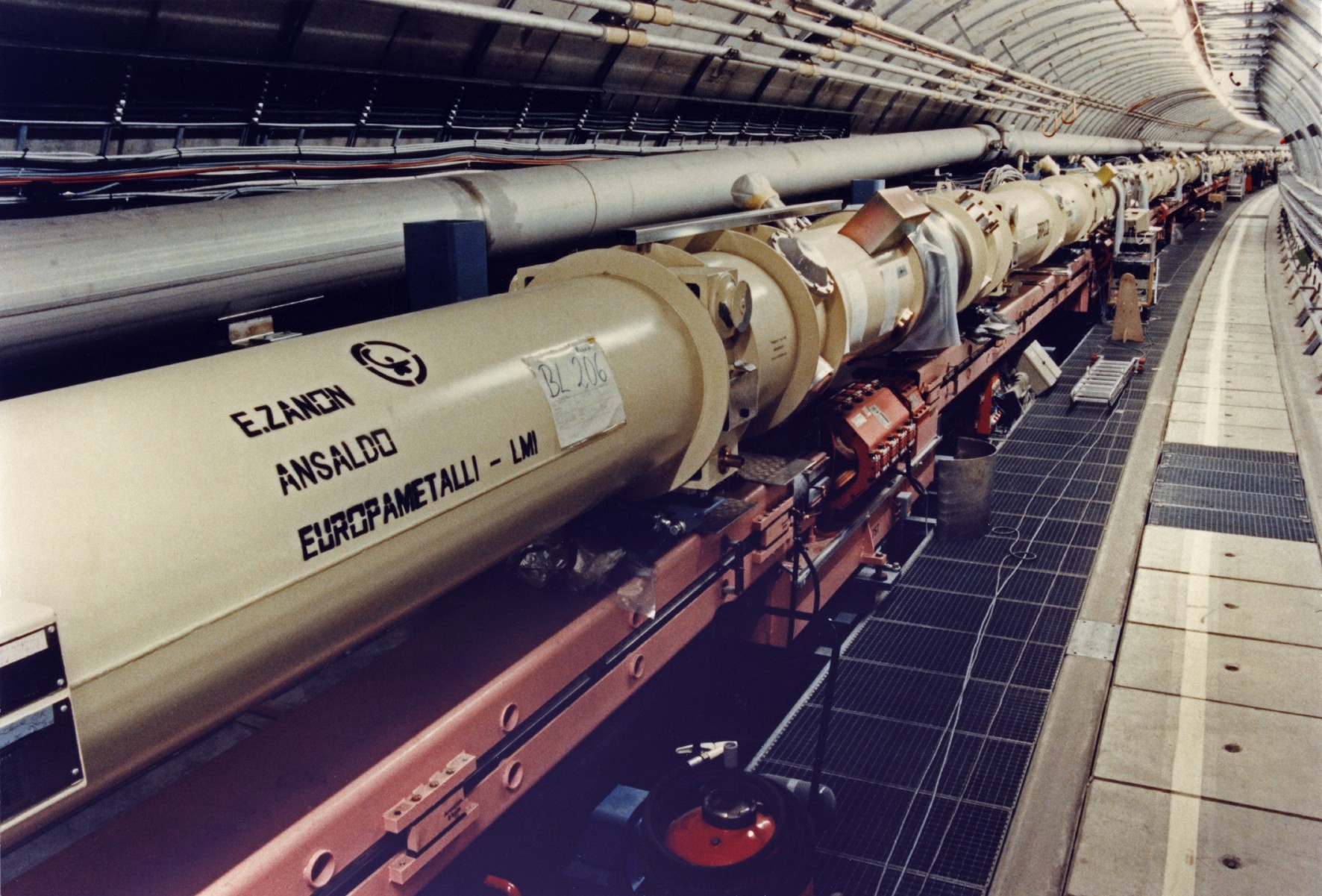}
    \caption{\enquote{The view into the 6.3-kilometre-long HERA tunnel shows the superconducting magnets for guiding the protons on top and the electron ring beneath them.} - taken from~\cite{wikiHERA}}
    \label{fig:HERA}
\end{figure}

We start with the Hadron-Elektron-Ringanlage (HERA), which was in operation between 1992 and 2007.
HERA was hosted by DESY in Hamburg, Germany and \cref{fig:HERA} shows an image from the HERA tunnel.
While the promoted physics discovery, leptoquarks, eventually did not happen, HERA has become one of the most influential DIS machines so far.
In fact, it is until this day the only lepton-proton collider in the world.
It recorded a huge amount of data, which is still analyzed nowadays, 20 years after the shutdown.
HERA had four experiments (H1, ZEUS, HERA-B and HERMES), but we focus here on the combined analysis of H1 and ZEUS~\cite{H1:2015ubc}.
In \cref{tab:HERA} we given an overview of the most important fully-inclusive DIS measurements from HERA.

\begin{table}
    \caption{Fully inclusive lepton-proton scattering measurements at HERA~\cite{H1:2015ubc}.
    We quote the scattered lepton, the measured kinematic ranges, the total number of data points $N_\text{dat}$, and the measured observable.
    $N_\text{dat}^\text{fit}$ refers to the number of data points used in NNPDF4.0~\cite{NNPDF:2021njg}.}
    \label{tab:HERA}
    \begin{tabular}{ll ll ll}
        projectile & $\sqrt{s_l}$ [\si{\GeV}] & x & $Q^2$ [\si{\GeV^2}] & $N_\text{dat}^\text{fit}/N_\text{dat}$ & obs.\\
        \hline
        $e^-$ & 318 & \numrange[range-phrase = --]{0.0008}{0.65} & \numrange[range-phrase = --]{60}{5e4} & 159/159 & $\sigma_r^\text{NC}$ \\
        $e^+$ & 318 & \numrange[range-phrase = --]{5.02e-06}{0.65} & \numrange[range-phrase = --]{0.15}{3e4} & 377/485 & $\sigma_r^\text{NC}$ \\
        $e^+$ & 300 & \numrange[range-phrase = --]{6.21e-07}{0.4} & \numrange[range-phrase = --]{0.045}{3e4} & 70/112 & $\sigma_r^\text{NC}$ \\
        $e^+$ & 251 & \numrange[range-phrase = --]{2.79e-5}{0.65} & \numrange[range-phrase = --]{1.5}{800} & 254/260 & $\sigma_r^\text{NC}$ \\
        $e^+$ & 225 & \numrange[range-phrase = --]{4.64e-5}{0.65} & \numrange[range-phrase = --]{2}{800} & 204/209 & $\sigma_r^\text{NC}$ \\
        $e^-$ & 318 & \numrange[range-phrase = --]{0.008}{0.65} & \numrange[range-phrase = --]{3e2}{3e4} & 42/42 & $\sigma_r^\text{CC}$ \\
        $e^+$ & 318 & \numrange[range-phrase = --]{0.008}{0.4} & \numrange[range-phrase = --]{3e2}{3e4} & 39/39 & $\sigma_r^\text{CC}$ 
    \end{tabular}
\end{table}

We start by looking at the projectile column and immediately we realize that HERA was most of the time not accelerating electrons, but positrons.
This is to remind us that DIS is lepton-hadron scattering and, indeed, we have to consider all leptons.
We expand our selection further below in \cref{sec:nuclearneutrino}, but remain for the moment with the electron and the positron, which are exact copies of each other, but with opposite charge.

Next, we turn to the energy column $\sqrt{s_l}$, which spreads across four different energies.
In practice the lepton beam was operated with a fixed beam energy ($E_e = \SI{27.5}{\GeV}$) and varying energies for the protons ($E_p=\SIlist{920;820;575;460}{\GeV}$).
If we then look to the $x$ and $Q^2$ columns and recall \cref{eq:defslep}, we see that the table only indicates the maximal available range of those variables.
Specifically, HERA was able to measure activity in the detector for $0.005 < y < 0.95$.

However, not only the accessible kinematical phase space limits our kinematic coverage, but also our requirement to study DIS, \cref{box:DISreq}.
In practice, e.g.\ NNPDF4.0~\cite{NNPDF:2021njg} applies a cut of $Q^2 \geq \SI{8}{\GeV^2}$ and $W^2 \geq \SI{12.5}{\GeV^2}$.
Mathematically, the $Q^2$ cut ensures we are in a perturbative regime, \cref{eq:CpQCD}, while the $W^2$ cut has multiple motivations.
First, it ensures we are in an inelastic regime, i.e.\ we are not finding resonances or excitations of the proton.
Second and related to the former, we know it suppresses the correction term $\order{\frac{\LQCD^2}{Q^2}}$ in \cref{eq:fact1}.
Studying and trying to estimate the correction term is a topic in current research~\cite{Ball:2025xtj,Harland-Lang:2025wvm,Cerutti:2025yji} and we return to it in \cref{sec:TMC}.
Looking to the column with the number of data points, we note that even after the kinematic cuts we are left with a total of 1145 measured fully inclusive DIS cross sections.

Finally, we turn to the observables column for which we see two different entries exist:
the reduced neutral current (NC) cross section $\sigma_r^\text{NC}$ and the reduced charge current (CC) cross section $\sigma_r^\text{CC}$.
The \enquote{current} here refers to the current between the leptonic part and the hadronic part, in particular the exchanged boson.
The \enquote{reduced} cross section refers to a slight rescaling of the double differential cross section, \cref{eq:dsigmadxdy},
\begin{equation}
    \sigma_r(x,Q^2) = \frac{xy Q^2}{2\pi \aem^2} \frac 1 {\left(2 - 2y + y^2\right)}\frac{\dd[2] \sigma}{\dd x \dd y} \approx F_2(x,Q^2)
\end{equation}
such that it refers to a specific PDF combination at LO accuracy, i.e.\ \cref{eq:LOxs} is actually the reduced cross section.

Next, we examine the CC case by recalling two important facts from \cref{box:particles}:
first, charged leptons, and thus also electrons and positrons, have weak charges and, second, $W$ bosons have an electric charge.
Thus, by CC we consider the lepton interacts with the hadron via the exchange of a $W$ boson, which behaves quite different then a photon.
First and foremost, due to the electric charge conservation the $W$ boson can only interact with specific leptons and quarks.
On the leptonic side this implies that, first, e.g., the positron always emits a $W^+$ boson, and, second, the recoiling lepton is a neutrino, which, in practice, is impossible to track.
Specifically, it is impossible to measure it's momentum four-vector $k'$, \cref{eq:DISreaction}, and thus measure the $W$ boson momentum $q$, \cref{eq:defq}, or any other derived variable ($x,Q^2,y$), from that.
Instead, the full reconstruction of the hadronic final state also allows to extract the relevant kinematic variables, which is used instead~\cite{H1:2015ubc}.
Although the production of CC events is suppressed by the propagator of the $W$ boson, $\sim (Q^2 + m_W^2)^{-1}$, it is a LO effect.
Moreover, since the $W$ are maximally parity violating this allows a third, independent structure function (typically referred to as $F_3$) to appear in the decomposition of the cross section, \cref{eq:dsigmadxdy}.
We review the problems introduced by $\gamma_5$, which causes the parity violation, further in \cref{sec:g5}.

We now consider the NC case, which is a generalization of the photon exchange discussed in \cref{chap:intro}.
As we recall from \cref{box:particles} the $Z$ boson is practically a massive copy of the photon $\gamma$ and they share all other quantum numbers.
This implies whenever a photon is exchanged, also a $Z$ boson can be exchanged, with the main difference that its contributions are suppressed by the respective propagator, $\sim (Q^2 + m_Z^2)^{-1}$.
More specifically, we get three contributions to our final cross section: a pure photon contribution $\sigma^\gamma$ (where the superscript $\gamma$ is often suppressed), a pure $Z$ contribution $\sigma^Z$, and an interference contribution\footnote{the interference term typically contains the required factor of 2 from the binomial formula inside} $\sigma^{\gamma Z}$, which have to be summed together:
\begin{equation}
    \sigma^\text{NC} = \sigma^\gamma + \sigma^{\gamma Z} + \sigma^{Z}\,.
\end{equation}
While at low virtualities $Q^2 \ll m_Z^2$ the $Z$ contributions can be neglected, this does not apply to the HERA measurements with largest $Q^2$.
Similar to $W$ boson, also $Z$ bosons have an axial-vectorial coupling to fermions, $\sim \gamma^\mu\gamma^5$, which introduces a contribution from $F_3$ to \cref{eq:dsigmadxdy}.
The interference contribution, $F_{1,2}^{\gamma Z}$, yields an electric charge (from the photon) and a vectorial coupling (from the $Z$ boson) both on the leptonic and the hadronic side, whereas the pure $Z$ contribution, $F_{1,2}^Z$, yields squares of both vectorial and axial-vectorial couplings.
On the other side, as $F_3$ is per-se a parity-violating observable, there the interference contribution $F_3^{\gamma Z}$ yields an electric charge (from the photon) and an axial-vectorial coupling (from the $Z$) and the pure $Z$ contribution, $F_3^Z$ yields a product of the vectorial and axial-vectorial couplings.

The fact that NC and CC couple completely different to fermions, i.e.\ quarks and leptons, and the definition of three independent structure functions in both cases allows one to fully determine PDFs from DIS~\cite{Lappi:2024dvv}.
We discuss more technical details in \cref{app:LO}.

Let us return to the key point of this section here.
We need to consider all leptons and all gauge bosons for DIS, i.e.\ all particles in \cref{fig:SM}, but based on the discussion above, we can now better justify why we actually exclude some.
According on our definition of DIS, \cref{box:DISformula}, in theory also the Higgs boson would contribute to NC cross section, but since its coupling is proportional to the mass of the particle and we always\footnote{almost always~\cite{Spezzano:2025rfy} - but even so Higgs would only contribute a very large $Q^2$} consider massless leptons it does in practice not contribute.
As the tau has a very short lifetime, they are in practice impossible to accelerate in a collider and thus not suitable for DIS collider experiments.
Instead, while muons are also difficult to accelerate, concrete experiments for muon-proton scattering have been proposed~\cite{Akturk:2025ubm}.

\subsection{Nuclear and Neutrino DIS}
\label{sec:nuclearneutrino}

After having considered all available bosons and a few more leptons, we now need to consider more hadrons.
While we focus here on the proton PDF point-of-view, DIS is equally important for nuclear PDFs~\cite{Eskola:2021nhw,AbdulKhalek:2022fyi,Klasen:2025ekj,Helenius:2021tof}, which describe the internal structure of nuclei.
Moreover, the two are typically directly intertwined: proton PDF extractions often contain measurements on nuclear targets~\cite{NNPDF:2021njg}, while nuclear PDF extraction often require a proton baseline.
While the main focus on the EIC~\cite{AbdulKhalek:2022fyi} is far beyond the fully inclusive case, the first years of the early physics program, scheduled currently in less then ten years from now, contain some dedicated runs of nuclear DIS~\cite{Abbott:2026eqp}.

At the same time, using nuclear targets also allows for a new set of leptons as projectile: neutrinos.
Since neutrinos only interact weakly (literally and metaphorically), but nuclear targets have a large cross section the two effects yield measurable cross sections.
Neutrinos interact in practice always via CC DIS, which makes them a valuable resource in PDF fits.
For example CHORUS~\cite{CHORUS:2005cpn} and NuTeV~\cite{NuTeV:2001dfo,Mason:2006qa} provide valuable information on the flavor decomposition of PDFs~\cite{Helenius:2026uuz}.

However, neutrino CC DIS is typically not done at neutrino colliders, since in practice it is impossible to accelerate neutrinos directly.
Instead, experiments use other sources of neutrinos, for example hadron colliders such as the recently started FASER experiment~\cite{FASER:2022hcn} at CERN, which studies the neutrinos generated as a side effect in the ATLAS experiment.
Actually, with the successful start of FASER~\cite{Hayakawa:2025xra} the Forward Physics Facility (FPF) as an extension of the HL-LHC program has been proposed~\cite{FPF:2025bor}.

Another source of highly accelerated neutrinos are astronomical objects, which are studied, e.g., at IceCube~\cite{IceCube:2013low} at the south pole or at KM3NET~\cite{KM3Net:2016zxf} in the Mediterranean Sea.
Possible studies include the determination of neutrino structure functions~\cite{Candido:2023utz}, or the constraining power on an intrinsic charm component in the proton~\cite{Das:2025snq}.
This proves the usefulness of DIS also in a beyond-collider context.

\subsection{Other experiments}
PDF fitting groups often also use a set of older measurements, e.g., from NMC~\cite{NewMuon:1996uwk,NewMuon:1996fwh}, SLAC~\cite{Whitlow:1991uw}, or BCDMS~\cite{BCDMS:1989qop}.
SLAC deserves a special mention, since this is where the original idea of the parton model was born~\cite{Bloom:1969kc,Feynman:1969wa,Bjorken:1968dy,Bjorken:1969ja}, which eventually led to pQCD as we know it today.
Note that these experiments provide directly the structure function $F_2$ instead of the actually measured cross section $\sigma$, which are related via \cref{eq:dsigmadxdy}.
Thus in order to provide a structure function at fixed $x$ and $Q^2$ this requires a cross section measurement at different collision energies $s_l$ due to \cref{eq:defslep}; this technique is referred to as Rosenbluth separation~\cite{Rosenbluth:1950yq}.
However, even if experiments provide cross sections, the overall normalization can be experiment-specific.

Finally, we mention the ongoing experiments at the Thomas Jefferson National Accelerator Facility (JLab) in Virginia, USA, which provide new DIS measurements~\cite{Accardi:2026hdv} focussing on the low energy region.

\section{Code implementation}
\label{sec:interpolation}
If we want to compare any theoretical prediction to an experimental measurement we eventually need to implement the coefficient functions into a computer program.
Specifically, we need to find a representation of our results, which is suitable for computers and which can be evaluated numerically, preferably in a fast and efficient way.

\subsection{Distributions}
\label{sec:rsl}

Coefficient functions may contain distributions, for example the quark $F_2$ coefficient function at LO is just a Dirac delta distribution:
\begin{equation}
    C_q^{(0)}(z) = e_q^2 \delta(1-z) \,.
\end{equation}
To better compare with our next step, let us repeat the definition of the Dirac delta distribution:
\begin{equation}
    \int\limits_0^1 \! \dd z\, \delta(1-z) f(z) = f(1). \label{eq:DiracDelta}
\end{equation}
However, this is only the simplest case, which appears at LO, and at higher order we get more complicated distributions.
For example the quark $F_2$ coefficient function at NLO, \cref{eq:NLOCq}, contains many more distributions
\begin{align}
    C_q^{(1)}(z) &\supset 4 \qty(\frac{\ln(1-z)}{1-z})_+ - 3 \qty(\frac{1}{1-z})_+ - \qty(9+4\zeta_2)\delta(1-z)
\end{align}
where the usual +-distributions are defined via
\begin{equation}
    \int\limits_0^1 \! \dd z\, a(z)_+ f(z) = \int\limits_0^1 \! \dd z\, a(z) \left(f(z) - f(1)\right) \,. \label{eq:PlusDistr}
\end{equation}
We review in \cref{sec:nlo} briefly how these distributions appear, but here we turn to more practical matters.
The problem is: distributions are (complicated) math objects, defined only via integral relations, \cref{eq:DiracDelta,eq:PlusDistr}, which are unsuitable for numeric evaluation.
Moreover, \cref{eq:DiracDelta,eq:PlusDistr} are defined via an integral spanning the whole domain of $z$, i.e.\ in particular, the lower integration bound is 0, but \cref{eq:fact1} requires a convolution, where instead the lower integration bound depends on the measured kinematics $x$.
Note that the distributional content of $\vb C(z)$ is the main motivation to spell \cref{eq:fact1} as an integral in $z$ rather than $\xi$.
In the following paragraphs we drop the dependence on the parton index $j$ and the virtuality $Q^2$ as the strategy simply applies to everything.

It is convenient to introduce a different representation for coefficient functions: the Regular-Singular-Local (RSL) representation\footnote{This is not an official name and other names and conventions exist in literature.}.
In the RSL representation any coefficient function $C(z)$ is mapped onto three functions,
\begin{equation}
    C(z) \leftrightarrow (C^R(z), C^S(z), C^L(x))
\end{equation}
where $C(z)$ may contain distributions, but neither $C^R(z)$, $C^S(z)$, or $C^L(x)$.
The three functions are defined via their action under the convolution integral,
\begin{equation}
    \qty(C\otimes f)(x) =  \int\limits_x^1 \! \frac{dz}{z} \, f(x/z) C^R(z) + \int\limits_x^1 \! dz \, \left(\frac{f(x/z)}{z} - f(x)\right) C^S(z) + f(x) C^L(x) \label{eq:RSL}
\end{equation}
for any sufficiently smooth test function $f$ (such as PDFs in their usual place).
Note that \cref{eq:RSL} only contains integrals with the lower integration bound $x$.
Also note that while the argument of $C^S$ and $C^L$ are intentionally called $z$, referring to the partonic momentum fraction over which we integrate, the argument of $C^L$ is denoted by $x$, which refers to the argument of the convolution on the right-hand-side.

Let us consider some simple examples.
\begin{itemize}
    \item If the coefficient function is a regular function $C(z) = r(z)$ we find
        \begin{equation}
            C^R(z) = r(z), \qquad C^S(z) = 0, \qquad C^L(x) = 0
        \end{equation}
    \item If the coefficient function is a Dirac delta distribution $C(z) = \delta(1-z)$ we find
        \begin{equation}
            C^R(z) = 0, \qquad C^S(z) = 0, \qquad C^L(x) = 1
        \end{equation}
    \item If the coefficient function is a \enquote{raw} +-distribution $C(z) = (p(z))_+$ we find
        \begin{equation}
            C^R(z) = 0, \qquad C^S(z) = p(z), \qquad C^L(x) = -\int_0^x \! \dd z\, p(z) \label{eq:RSLD}
        \end{equation}
\end{itemize}
More examples can be found in \cref{app:RSL}.
Eventually, we also face the problem that +-distributions can absorb regular functions.
For example, \cref{eq:NLOCq} can also be written as
\begin{align}
    C_q^{(1)}(z) &= \frac{e_q^2 C_F}{4\pi} \left[ 4 \frac{\ln(1-z)}{1-z} - 3 \frac{1}{1-z} - 2(1+z)\ln\left(\frac{1-z}{z}\right) - 4 \frac{\ln(z)}{1-z} + 6 + 4z\right]_+ \label{eq:NLOCqp}
\end{align}
where we have absorbed all functional dependence under the distribution.
Thus before changing the representation we also need to fix the actual form and one possible choice is to only allow the minimal functional dependence under the distribution sign with a numeric prefactor, e.g.\ as in \cref{eq:NLOCq}.

\subsection{Interpolation}
The RSL representation together with a suitable implementation of \cref{eq:RSL} provides already a practical way to implement arbitrary convolutions.
However, we can make an additional step by applying a further discretization, which is desirable for computers.

We assume there is a minimum Bjorken-$x$ $x_\text{grid,min}$ for which we need to compute observables, which implies that we also only need coefficient functions $\vb C(z,Q^2)$ and PDFs $\vb f(\xi,Q^2)$ above this limit, \cref{eq:fact1}.
Next, we choose a set of $N_\text{grid}$ grid points,
\begin{equation}
    \mathbbm G = \{ x_k : x_\text{grid,min} < x_k <= 1, \quad k = 0,\ldots, N_\text{grid} - 1 \} \label{eq:grid}
\end{equation}
which we then use to interpolate the PDF
\begin{equation}
    f_j(\xi,Q^2) \sim \sum\limits_{k=0}^{N_{grid} - 1 } f_j(x_k,Q^2) p_k(\xi) \label{eq:interpolation}
\end{equation}
with a suitable set of interpolation polynomials $p_k(\xi)$.
In practice, piecewise Lagrange interpolation polynomials are a good choice~\cite{Candido:2022tld,Carrazza:2020gss}.

Inserting \cref{eq:interpolation} into \cref{eq:fact1} yields
\begin{align}
    \sigma(x,Q^2) &= \sum_{j} \int\limits_x^1 \frac{\dd z}{z} C_j(z,Q^2) \left( \sum\limits_{k=0}^{N_{grid} - 1 } f_j(x_k,Q^2) p_k(x/z) \right)\\
     &=\sum_{j} \sum\limits_{k=0}^{N_{grid} - 1 } f_j(x_k,Q^2) \hat \sigma_{jk}(x,Q^2) \label{eq:sigmainterpolation}
\end{align}
where we defined
\begin{equation}
    \hat \sigma_{jk}(x,Q^2) = \qty(C_j(Q^2) \otimes p_k)(x)\,.
\end{equation}
Choosing our grid $\mathbbm G$ fixes immediately the interpolation polynomials, $p_k(\xi)$, and thus we can tabulate $\hat \sigma_{jk}(x,Q^2)$ without having knowledge about the actual PDF $f_j(\xi,Q^2)$.
Note that the interpolation in momentum fraction space is independent from the flavor index $j$ and for simplicity we can choose the same grid for all flavors.
This makes both $f_j(x_k,Q^2)$ and $\hat \sigma_{jk}(x,Q^2)$ two two-dimensional tensors in \cref{eq:sigmainterpolation}, which get multiplied in such a way that they yield a plain number.
Note that interpolation is also independent from the RSL representation and in practice \cref{eq:RSL} is executed on $p_k(\xi)$.

\subsection{Grids}
\label{sec:grids}
Being able to compute $\hat \sigma_{jk}(x,Q^2)$ without knowledge about the actual PDF is advantageous since, we recall, PDFs are non-perturbative objects, encoding low-scale physics, which are not computable from first principles inside pQCD.
Instead, PDFs are obtained from a fitting procedure, which can be summarized as follows:
\begin{conclusionbox}{PDF fitting}
    \begin{enumerate}
        \item Guess a candidate PDF
        \item Compute the theory predictions with this candidate PDF
        \item Compare the theory predictions to the experimental measurement
    \end{enumerate}
    Repeat the above steps until the procedure converges, i.e.\ the comparison is \enquote{good enough}.
\end{conclusionbox}
The specific details on how to guess a candidate PDF, how to exactly compute the theory predictions, which experimental data to use, and what \enquote{good enough} exactly means is a choice.
This is the reason why competing PDF fitting groups exist~\cite{NNPDF:2021njg,Hou:2019efy,Bailey:2020ooq,PDF4LHCWorkingGroup:2022cjn,Accardi:2026hdv} and why they do not necessarily obtain the same PDF, although we all assume the latter to be universal.

The main point for us here is that we need to compute theory predictions very many times for slightly different candidate PDFs.
This is a serious problem for complicated hadronic observables, such as, e.g., jet cross sections~\cite{NNLOJET:2025rno}, where the computation can take many hours.
The solution is to compute fast interpolation grids, using, e.g., PineAPPL~\cite{Carrazza:2020gss,Jezo:2026adf,christopher_schwan_2025_15635174}, APPLgrid~\cite{Carli:2010rw}, or fastNLO~\cite{Wobisch:2011ij}, which store the partonic matrix elements in an efficient way, such that any subsequent convolution with an arbitrary PDF is fast, as in \cref{eq:sigmainterpolation}.
Having all observables in a PDF fit sharing the same technology simplifies their consistent treatment and inside the NNPDF collaboration this is realized in the pineline framework~\cite{Barontini:2023vmr}.
In practice we organize the discretization of the coefficient functions in a slightly different way then above.

Interpolation grids are higher dimensional tensors which we can recombine in a suitable way a posteriori by performing fast linear algebra operations.
The first three major dimensions are:
\begin{description}
    \item[bin] This is the list of cross section that we need to compute simultaneously.
        In DIS this typically corresponds to all $N_\text{dat}$ data point measured in a given configuration, e.g.\ in \cref{tab:HERA}.
        Thus, in practice the actual list is given by the experimental measurement.
    \item[order] This is the list of perturbative orders, i.e.\ we store each $\vb C^{(k)}(z)$, \cref{eq:CpQCD}, separately.
        Having the full list available allows to use a given grid also at lower orders and thus study, e.g., perturbative stability.
        In practice, the actual list is given by the requested accuracy of the user (provided we can actually fulfill that request).
    \item[channel] This is the list of PDF flavor combinations, i.e.\ the sum over $j$ in \cref{eq:fact1}.
        Having the full list available allows to study the impact of different PDF combination and, e.g., determine which PDF the observable can potentially constrain most.
        In DIS it is given by the coupling structure of the exchanged boson, i.e.\ $\gamma/Z$ vs.\ $W^\pm$.
        In practice, the actual list is the easiest given directly in terms of individual parton flavors as the available combinations turn out to be too complex.
\end{description}

These three dimensions are present in any interpolation grid and in DIS there are two additional dimensions:
\begin{description}
    \item[scale] This is the list of factorization and renormalization scales, discussed in \cref{sec:sv}.
        In DIS this dimension turns out to be trivial, i.e.\ it has support by a single value, as in fully inclusive DIS there is only a single scale, $Q^2$ (or a multiplicative there of if we do not set $\mu_F^2 = \mu_R^2 = Q^2$).
        In practice, the actual list is given by the $Q^2$ specified in the bin configuration.
    \item[momentum fraction] This is the list of momentum fractions, \cref{eq:grid}, over which the coefficient functions are convolved with the PDF.
        Recall that LO, \cref{eq:LOC}, and NLO, e.g.\ \cref{eq:NLOCq}, are quite different and thus they use different ranges of momentum fractions.
        In practice, the actual list is given by a suitable choice of $\mathbbm G$, \cref{eq:grid}.
\end{description}

For fully inclusive DIS we end up with a five dimensional tensor, where each inner dimension may depend on the outer.
A typical example is the fact that the gluon only contributes starting from NLO, \cref{eq:LOC,eq:NLOCg}.

\section{Resummation}
\label{sec:resum}
In preparation to \cref{sec:sv,sec:hq} we need to review two main ingredients for pQCD: the strong coupling $\alpha_s(Q^2)$, \cref{eq:CpQCD}, and the PDFs $\vb f(\xi,Q^2)$, \cref{eq:fact1}.
Specifically, we need to remind ourselves on their scale dependency also referred to as evolution.
While the two objects obey very similar equations, the strong coupling $\alpha_s(Q^2)$ is conceptually simpler and, thus, we start with that.

\subsection{The strong coupling}
\label{sec:as}

The strong coupling $\alpha_s$ is directly linked to the coupling of the gluons to quarks,
\begin{equation}
    \mathcal L_\text{QCD} \supset g T_a \bar{\psi}_q \gamma^\mu A^a_\mu \psi_q, \qquad \alpha_s = \frac{g^2}{4\pi}
\end{equation}
and is the central object in pQCD around which we organize our perturbative series.
If one considers higher order corrections to the gluon-quark couplings, one encounters quickly divergencies of ultra-violet origin, i.e.\ for large loop momenta.
This forces the introduction of a regularization procedure, typically dimensional regularization, which eventually leads to a scale dependent definition of the renormalized strong coupling $\alpha_s(\mu_R^2)$.
The introduced scale, called renormalization scale $\mu_R$ henceforth, is an arbitrary scale, which is an artifact of our regularization procedure and which effectively represents a separation between perturbative and non-perturbative regimes.
However, the scale dependence of $\alpha_s(\mu_R^2)$ is rigorously predicted in pQCD, organizing the terms in such a way that within the requested accuracy any dependence of $\mu_R$ is cancelled.
Recall that the scale dependence can be computed once the Lagrangian $\mathcal L$ is defined and does not require any actual observable.

Eventually, we obtain the renormalization group equation (RGE) of the strong coupling,
\begin{equation}
    \mu_R^2 \dv{\alpha_s(\mu_R^2)}{\mu_R^2} = \beta(\alpha_s) = - \sum_{j=0}\beta_j\qty(\alpha_s(\mu_R^2))^{2+j} \label{eq:beta}
\end{equation}
which is also the definition of the (perturbative) beta function $\beta(\alpha_s)$.
The upper limit of the sum in \cref{eq:beta} defines the perturbative accuracy, to which all other objects in pQCD refer to, which is left implicit here and in the rest of this work.

At LO accuracy we find
\begin{equation}
    \mu_R^2 \dv{\alpha_s(\mu_R^2)}{\mu_R^2} = - \beta_0 \qty(\alpha_s(\mu_R^2))^2, \qquad \beta_0 = \frac{11 C_A - 2 n_f}{12\pi} \label{eq:LObeta}
\end{equation}
where we used the second Casimir constant of the adjoint representation $C_A = N_c = 3$ and the number of active flavors $n_f$, discussed in \cref{sec:hq}.
In this specific case we can solve \cref{eq:LObeta} and obtain
\begin{equation}
    \alpha_s^\text{LO}(\mu_R^2) = \frac{\alpha_s(\mu_{R,0}^2)}{1 + \alpha_s(\mu_{R,0}^2)\beta_0\ln(\mu_R^2/\mu_{R,0}^2)} \label{eq:LOalphas}
\end{equation}
for a given boundary condition $\alpha_s(\mu_{R,0}^2)$ at the scale $\mu_{R,0}^2$.
Recall, that \cref{eq:LOalphas} has not two free parameters, $\mu_{R,0}$ and $\alpha_s(\mu_{R,0}^2)$ as one would naively think, but only one $\LQCD$ (see \cref{box:LOalphas}).
Also remember that $\LQCD$ corresponds to the pole in \cref{eq:LOalphas} and pQCD only makes sense above that scale, $\mu_R > \LQCD$.
Moreover, in practice it is often convenient to rewrite \cref{eq:beta} in terms of $t_R = \ln(\mu_R^2/\mu_{R,0}^2)$.

Now, coming back to the title of this section, note that \cref{eq:LOalphas} is a resummation:
\begin{equation}
    \alpha_s^\text{LO}(\mu_R^2) = \alpha_s(\mu_{R,0}^2) \sum_{k=0}^\infty \qty(-\alpha_s(\mu_{R,0}^2)\beta_0\ln(\mu_R^2/\mu_{R,0}^2))^k \label{eq:LOalphasresum}
\end{equation}
i.e.\ it collects an infinite tower of logarithms, here $\ln(\mu_{R,0}^2/\mu_R^2)$, into a compact expression, \cref{eq:LOalphas}.
Since $\mu_{R,0}$ is an arbitrary scale, often chosen to coincide with the mass of the $Z$ boson $\mu_{R,0}=m_Z$, it is basically always required to perform the resummation and use \cref{eq:LOalphas}.
However, in some situations, to be discussed in \cref{sec:sv}, we need \cref{eq:LOalphas} truncated to a finite order, i.e.\ unraveling \cref{eq:LOalphasresum}, and we get
\begin{equation}
    \alpha_s^\text{LO}(\mu_R^2) = \alpha_s(\mu_{R,0}^2) - \qty(\alpha_s(\mu_{R,0}^2))^2\beta_0\ln(\mu_R^2/\mu_{R,0}^2) + \ldots \label{eq:LOalphasexp}
\end{equation}
While for the special case $\mu_{R,0} = \mu_R$ we need to be consistent of course and we thus find that $\alpha_s^\text{LO}(\mu_R^2)$ is $\alpha_s(\mu_{R,0}^2)$ at the LO truncation, we note that they differ by higher order terms.
\Cref{eq:LOalphasexp} is the first, simplest case of a perturbative expansion of a resummed expression, \cref{eq:LOalphas}, and can also be obtained by using the Taylor expansion of $\alpha_s(\mu_R^2)$ around $\mu_{R,0}^2$ together with \cref{eq:LObeta}.
In other words, \cref{eq:LOalphasexp} is a perturbative expansion of \cref{eq:LOalphas} in powers of $\alpha_s(\mu_{R,0}^2)$.

Starting from NLO accuracy the RGE, e.g.\ in this specific case given by
\begin{equation}
    \mu_R^2 \dv{\alpha_s(\mu_R^2)}{\mu_R^2} = - \beta_0 \qty(\alpha_s(\mu_R^2))^2 - \beta_1 \qty(\alpha_s(\mu_R^2))^3\,, \label{eq:NLObeta}
\end{equation}
can no longer be solved exactly with elementary functions.
However, it is still possible to find an analytic solution which solves \cref{eq:NLObeta} at the requested accuracy, i.e.\ in this case at NLO accuracy (see \cref{box:NLOalphas}).
This solution still performs a resummation, as we have to, but it only collects the so-called next-to-leading logarithms (NLL) (see \cref{box:NLOalphasLL}).
Note that the RGE has a given perturbative accuracy, e.g.\ NLO, but the solution has a given logarithmic accuracy, e.g.\ NLL, however, in the common language one often just uses only the former.

A completely different way to solve \cref{eq:NLObeta} is using numerical tools instead.
We refer to the literature on how to implement a specific solution in practice and remain here just with the fact that it is a complicated problem on its own, which has no unique solution and for which approximations are mandatory.
\begin{conclusionbox}[label={box:RGE}]{Solution of RGE}
    \begin{itemize}
        \item The LO RGE of the strong coupling, \cref{eq:LObeta}, can be solved exactly, \cref{eq:LOalphas}
        \item The RGE of the strong coupling at higher orders, e.g.\ \cref{eq:NLObeta}, can not be solved exactly and one needs to choose a solution method
    \end{itemize}
\end{conclusionbox}
Any proper solution of the RGE must solve the RGE itself at the requested accuracy else it is not a solution, but they can and do differ by terms beyond - we refer to the uncertainty which is associated with this freedom as Missing Higher Order Uncertainty (MHOU).
Truncating the perturbative series is a fundamental part of pQCD and thus we are always bound to the accuracy which we request at the beginning.
While in many simple application MHOU are often neglected they are an important ingredient in modern PDF fits~\cite{NNPDF:2024dpb} and in \cref{sec:sv} we demonstrate one possible way to estimate them.

\subsection{DGLAP}
\label{sec:DGLAP}

For PDFs $\vb f(\xi, \mu_F^2)$ the situation is essentially the same as for the strong coupling $\alpha_s(\mu_R^2)$, except the latter is just a number and the former are a set of functions.
The scale dependence is a again predicted by pQCD, but now we need an arbitrary physical observable to do so.
The specific observable does not matter since PDFs are universal, but due to the linear dependence on PDF and the \enquote{simplicity} of fully inclusive DIS it often serves as reference case.
The arbitrary scale that is introduced in the regularization procedure, typically called collinear factorization, is referred to as factorization scale $\mu_F$ henceforth.

Eventually, we obtain the RGE of PDFs, often referred to as DGLAP equation named after the researchers who established it~\cite{Altarelli:1977zs,Gribov:1972ri,Dokshitzer:1977sg},
\begin{equation}
    \mu_F^2 \dv{\vb f(\xi,\mu_F^2)}{\mu_F^2} = \qty(\vb P (\mu_F^2) \otimes \vb f(\mu_F^2))(\xi) = \sum_{j=0} \qty(\alpha_s(\mu_F^2))^{1+j} \qty(\vb P^{(j)} \otimes \vb f(\mu_F^2))(\xi) \label{eq:DGLAP}
\end{equation}
where $\vb P (y, \mu_F^2)$ is the matrix of Altarelli-Parisi splitting functions.
Note that the upper limit of the sum in \cref{eq:DGLAP} must be set to the requested perturbative accuracy, i.e.\ consistent with \cref{eq:LObeta}.

When attempting to solve \cref{eq:DGLAP} we must worry about the two complications in PDFs with respect to the strong coupling.
First, since PDFs are a vector and, thus, the Altarelli-Parisi splitting functions are a matrix $\vb P$, we need to worry about non-commutativity.
Indeed, while at LO accuracy we only have a single matrix $\vb P^{(0)}$, which trivially commutates with itself, starting from NLO we encounter a second matrix $\vb P^{(1)}$, which does not commutate with the former.
Second, at any fixed scale PDFs are a function of the momentum fraction $\xi$ and, moreover, the right-hand-side of \cref{eq:DGLAP} features an integral (with the definition of the convolution $\otimes$ in \cref{eq:fact1,eq:fact2}).

To overcome the second problem, one possibility is to apply an integral transformation, the so-called Mellin transformation, defined by
\begin{align}
    \vb{\tilde f}(N,\mu_F^2) &= \int\limits_0^1 \dd \xi\,\xi^{N-1} \vb f(\xi,\mu_F^2) \label{eq:Mellinf} \\
    \gamma(N,\mu_F^2) &= \int\limits_0^1 \dd y\,y^{N-1} \vb P(y,\mu_F^2) \label{eq:Mellinad}
\end{align}
where the Mellin-transformed splitting functions $\gamma$ are often referred to as anomalous dimensions.
Inserting \cref{eq:Mellinf,eq:Mellinad} into \cref{eq:DGLAP} yields a \enquote{simple} first order differential equation as convolutions turn into ordinary products and which can be solved by a \enquote{simple} exponentiation.
However, in practice PDFs are mostly needed in momentum fraction space, $\vb f(\xi,\mu_F^2)$, and not in Mellin space, $\vb{\tilde f}(N,\mu_F^2)$, thus if we use the Mellin transform, we must also use its inverse transformation.

We can write the general solution to \cref{eq:DGLAP} in terms of an Evolution Kernel Operator (EKO)~\cite{Candido:2022tld} $\vb E$,
\begin{equation}
    \vb f (\xi,\mu_F^2) = \qty(\vb E(\mu_F^2 \leftarrow \mu_{F,0}^2) \otimes \vb f(\mu_{F,0}^2))(\xi)\,, \label{eq:EKO}
\end{equation}
but we skip spelling out an explicit solution here and refer to the literature instead~\cite{Candido:2022tld}.
The important part for us is that the operator (which is almost an exponential) again captures a resummation, in this case of $\ln(\mu_F^2/\mu_{F,0}^2)$.
Again, we can perform a perturbative expansion of the LO resummed solution to see the logarithm explicitly,
\begin{equation}
    \vb f^\text{LO}(\xi,\mu_F^2) = \vb f(\xi,\mu_{F,0}^2) + \alpha_s(\mu_{F,0}^2) \ln(\mu_F^2/\mu_{F,0}^2) \qty(\vb P^{(0)} \otimes \vb f(\mu_{F,0}^2))(\xi) + \ldots \label{eq:LOPDFexp}
\end{equation}
and starting from NLO accuracy, we face the same issues as for the strong coupling and \cref{box:RGE} applies in the same way.

\section{Scale variations}
\label{sec:sv}
We now attempt to address a problem we first see in \cref{sec:resum}: Missing Higher Order Uncertainty (MHOU).
We repeat: the truncation of the perturbative series is a fundamental part of our framework and we are bound to it.
The naive solution of just computing the next higher correction is not a solution, as we would face the problem just again there.
Thus the task is to get an estimate of the missing terms based on current knowledge, which by definition is an ill-posed problem.
Several possible strategies have been suggested in the literature~\cite{NNPDF:2024dpb,Lim:2024nsk,Tackmann:2024kci,Kassabov:2022orn,Bonvini:2020xeo,Cacciari:2011ze,Duhr:2021mfd,Ghosh:2022lrf}, which are based on resummation of specific variables and/or statistical arguments.

From the practical point of view as a PDF fitter, however, there is one big problem with many of the proposed algorithms: they only work for some specific set of observables and require a detailed analysis of the kinematics.
Thus, we review here the most common approach to estimate MHOU, which can be rigorously derived in pQCD and applied to any observable: scale variations.
On the other side, especially in recent years, scale variations have been challenged as suitable tool, due to the inherent arbitrariness and the lack of statistical interpretation (see \cref{sec:svdiscussion}).
For now we focus on their practical implementation in DIS.

\subsection{Renormalization scale}
\label{sec:rensv}

We start again from the simpler case of the renormalization scale $\mu_R$ associated with the strong coupling $\alpha_s(\mu_R^2)$ via the RGE, \cref{eq:LObeta}.
As we say above, the renormalization scale captures the ultra-violet divergencies, which arise, e.g., from self-energy diagrams and which are thus associated with the partonic diagram itself.
Moreover, as we see from \cref{eq:CpQCD}, we expand the partonic coefficient functions in powers of the strong coupling.
\begin{conclusionbox}{Renormalization scale dependence}
    The renormalization scale dependence is captured by the partonic coefficient function $\vb C(z,Q^2,\mu_R^2)$.
\end{conclusionbox}
Since we have to choose something for the arbitrary scale $\mu_R$ and we only have a single scale available, the virtuality $Q^2$, we start by setting them equal to each other $\mu_R^2 = Q^2$, which is also the implicit choice in \cref{eq:fact1}.
We then need to find a way to estimate the missing terms in $\vb C(z,Q^2,\mu_R^2=Q^2)$, i.e.\ something that is identical at the requested perturbative accuracy, but differs from it beyond.
This can be achieved by using \cref{eq:LOalphasexp} together with a suitable redefinition of the perturbative coefficients of the partonic coefficient function $\vb C^{(j)}(z)$.
Specifically, we require
\begin{equation}
    \vb C(z,Q^2,\mu_R^2) = \vb C(z,Q^2,\mu_R^2=Q^2)\qty(1 + \order{\alpha_s(Q^2)})
\end{equation}
with
\begin{equation}
    \vb C(z,Q^2,\mu_R^2) = \sum_{j=0} \qty(\alpha_s(\mu_R^2))^j \vb C^{(j)}(z,\ln(\mu_R^2/Q^2)) \,. \label{eq:CbarRpQCD}
\end{equation}
Before we continue with deriving explicit expressions for $\vb C^{(j)}(z,\ln(\mu_R^2/Q^2))$ some comments are in order:
\begin{itemize}
    \item $\vb C(z,Q^2)$ in \cref{eq:fact1} is an abbreviation of $\vb C(z,Q^2,\mu_R^2=Q^2)$ and $\vb C^{(j)}(z)$ in \cref{eq:CpQCD} of $\vb C^{(j)}(z,\ln(\mu_R^2/Q^2)=0)$
    \item in \cref{eq:CbarRpQCD} $\mu_R$ has its intended meaning: it is the scale of the strong coupling
    \item from \cref{sec:as} we know $\alpha_s(\mu_R^2)$ is a pure resummation of logarithms, thus $\vb C^{(j)}$ may only depend on $\ln(\mu_R^2/Q^2)$
    \item $\vb C^{(j)}(z,\ln(\mu_R^2/Q^2)=0)$ can be computed from diagrams by assuming explicitly $\mu_R = Q$ there
\end{itemize}

We now derive the explicit next-to-next-to-leading order (NNLO) expression for fully inclusive DIS.
In particular, we recall that LO is given by $\vb C^{(0)}(z)$, i.e.\ proportional to $\qty(\alpha_s(Q^2))^0$ in \cref{eq:CpQCD}, NLO is given by $\vb C^{(1)}(z)$ proportional to $\qty(\alpha_s(Q^2))^1$, and NNLO by $\vb C^{(2)}(z)$ proportional to $\qty(\alpha_s(Q^2))^2$.
Moreover, we assume that the coefficient functions at the central scale $\mu_R=Q$ are known and we can write \cref{eq:CpQCD} up to NNLO
\begin{equation}
    \vb C(z,Q^2) = \vb C^{(0)}(z) + \alpha_s(Q^2) \vb C^{(1)}(z) + \qty(\alpha_s(Q^2))^2 \vb C^{(2)}(z) \label{eq:CNNLO}
\end{equation}
Expanding \cref{eq:CbarRpQCD} up to the same accuracy we get
\begin{align}
    \vb C(z,Q^2,\mu_R^2) &= \vb C^{(0)}(z,\ln(\mu_R^2/Q^2)) + \alpha_s(\mu_R^2) \vb C^{(1)}(z,\ln(\mu_R^2/Q^2)) \nonumber\\
    &\hspace{20pt} + \qty(\alpha_s(\mu_R^2))^2 \vb C^{(2)}(z,\ln(\mu_R^2/Q^2)) \label{eq:CbarRNNLO}
\end{align}
We are now left with the task to make \cref{eq:CNNLO} and \cref{eq:CbarRNNLO} perturbatively equivalent, and although we are considering NNLO the equivalence must hold at any lower order.
Thus, we can first consider LO $\vb C^{(0)}$: since there is no dependency on the strong coupling in the first place, the scale-varied coefficient function may not depend on the renormalization scale and we find
\begin{equation}
    \vb C^{(0)}(z,\ln(\mu_R^2/Q^2)) = \vb C^{(0)}(z)\,.
\end{equation}
Starting from NLO an explicit dependency on the strong coupling appears, however, from \cref{eq:LOalphasexp} we know that the two choices, $\alpha_s(Q^2)$ and $\alpha_s(\mu_R^2)$ differ by higher order terms.
Specifically, we have
\begin{equation}
    \alpha_s(Q^2) = \alpha_s(\mu_R^2) + \order{\qty(\alpha_s(\mu_R^2))^2} \qq{and} \alpha_s(\mu_R^2) = \alpha_s(Q^2) + \order{\qty(\alpha_s(Q^2))^2}
\end{equation}
Thus, if \cref{eq:CNNLO} and \cref{eq:CbarRNNLO} have to be the same up to $\order{\qty(\alpha_s(Q^2))^2}$, we find
\begin{equation}
    \vb C^{(1)}(z,\ln(\mu_R^2/Q^2)) = \vb C^{(1)}(z)\,.
\end{equation}
Since we are here interested in the perturbative equivalence, we must use the perturbative expansion of the evolution.

Note that with this identification \cref{eq:CNNLO} and \cref{eq:CbarRNNLO} (truncated to NLO) are not the same in practice, they are only perturbatively equivalent.
In any actual calculation with $\mu_R\neq Q$ also $\alpha_s(\mu_R^2)$ and $\alpha_s(Q^2)$ are different, since we must resum for both of them.
They differ by an amount which is beyond our fixed-order framework and if that difference is \enquote{small} depends on the actual values of the strong coupling of the respective scales.
However, with that we achieve the main purpose of this game: we generate some higher order terms, which we can then use to estimate MHOUs.

Starting from NNLO we see that the matching becomes non-trivial.
In the case at hand we can insert \cref{eq:LOalphasexp} into \cref{eq:CNNLO} (with the choice $(\mu_R,\mu_{R,0})\to (Q,\mu_R)$) and re-expand in powers of $\alpha_s(\mu_R^2)$ as requested.
We find
\begin{equation}
    \vb C^{(2)}(z,\ln(\mu_R^2/Q^2)) = \vb C^{(2)}(z) + \beta_0 \ln(\mu_R^2/Q^2) \vb C^{(1)}(z)\,, \label{eq:CbarRimpl}
\end{equation}
so the scale-varied coefficient function receives a correction from a lower-order coefficient function.

Let us take a step back and review the general features.
\begin{conclusionbox}[label={box:svren}]{Renormalization scale variation}
    Renormalization scale-varied coefficient functions
    \begin{itemize}
        \item can always be computed a posteriori from their non-scale-varied counterparts
        \item are a linear combination of their lower order non-scale-varied counterparts
        \item are derived by re-expanding the strong coupling in the perturbative sum up to the required perturbative accuracy
    \end{itemize}
\end{conclusionbox}
We also know that the dependency on the scale variation $\ln(\mu_R^2/Q^2)$ must be a polynomial dependence since the running of strong coupling is a pure resummation of this logarithm.
By re-expanding we mean a Taylor expansion of the strong coupling at the default scale, here $\alpha_s(Q)$, in terms of the value at the renormalization scale, $\alpha_s(\mu_R^2)$.
For computing the strong coupling at any scale $\alpha_s(\mu_R^2)$ we must use the beta function at the required perturbative accuracy, but for deriving the scale-varied coefficient functions we only need one order less.
In the special case of fully inclusive DIS, which has no dependence on the strong coupling at LO, we even need only two orders lower then the coefficient functions.

For the sake of correcting \cref{eq:fact3}, we can improve it by writing
\begin{equation}
    \sigma(x,Q^2,\mu_R^2) = \qty(\vb C^T(Q^2,\mu_R^2) \otimes \vb f(Q^2))(x) \,.
\end{equation}

\subsection{Factorization scale}

As in \cref{sec:DGLAP} the general algorithm repeats for the factorization scale $\mu_F$, which is the intrinsic scale of PDFs, except everything is more complicated.
However, it is not only more complicated because PDFs are a set of function, but also because the PDF explicitly appears in \cref{eq:fact1}.
This implies, that the factorization scale dependence is not only be captured by the coefficient function, in contrast to the renormalization scale.
In particular, the factorization scale captures collinear divergencies, which arise in diagrams, and re-attributes them to the PDF, thus there is a direct connection between the two.
Recall our main goal here: we want to estimate MHOU by generating higher order terms to the measurable observable and we simply have more freedom here to do so.
In order to simplify the discussion we explicitly neglect renormalization scale variations for the moment (and delay the combination to \cref{sec:svcomb}).

The most common way to deal with factorization scale variation is to directly modify the coefficient functions, as in the case for \cref{eq:CbarRpQCD}, and we refer to this choice as \enquote{factorization scale variation scheme C}~\cite{NNPDF:2024dpb}.
Specifically, we make a similar ansatz as for the renormalization scale variation and write
\begin{equation}
    \sigma(x,Q^2,\mu_F^2) = \qty(\vb C^T(Q^2,\mu_F^2) \otimes \vb f(\mu_F^2))(x) \label{eq:sigmaFC}
\end{equation}
where now $\mu_F$ has its intended place and the usual perturbative expansion of the coefficient function still holds, i.e.\
\begin{equation}
    \vb C(z,Q^2,\mu_F^2) = \sum_{j=0} \qty(\alpha_s(Q^2))^j \vb C^{(j)}(z,\ln(\mu_F^2/Q^2)) \,. \label{eq:CbarFpQCD}
\end{equation}

However, in order to derive an explicit expression for $\vb C(z,Q^2,\mu_F^2)$ it is instructive to start from \enquote{factorization scale variation scheme B}~\cite{NNPDF:2024dpb}.
In this scheme, we make a different choice and attribute the variation instead to the PDF:
\begin{equation}
    \sigma(x,Q^2,\mu_F^2) = \qty(\vb C^T(Q^2) \otimes \vb f^{SV}(\mu_F^2,\ln(\mu_F^2/Q^2)))(x) \label{eq:sigmaFB}
\end{equation}
Note that \cref{eq:sigmaFB} only uses the central coefficient functions $\vb C(Q^2)$, but requires the use of a special PDF $\vb f^{SV}$.
In a PDF fit, where full control over the PDF is required any way and partonic matrix elements (coefficient functions) come from a variety of places, this scheme can be actually preferable.

In order to define $\vb f^{SV}$ we start by observing, that we can write \cref{eq:LOPDFexp} also as a convolution with a factorization scale operator $\vb K$
\begin{equation}
    \vb f(\xi,\mu_F^2) = \qty(\vb K(\mu_F^2 \leftarrow \mu_{F,0}^2) \otimes \vb f(\mu_{F,0}))(\xi)
\end{equation}
where $\vb K$ admits a perturbative expansion
\begin{equation}
    \vb K(y,\mu_F^2 \leftarrow \mu_{F,0}^2) = \sum_{j=0} \qty(\alpha_s(\mu_{F,0}^2))^j \vb K^{(j)}(y,\ln(\mu_F^2/\mu_{F,0}^2))
\end{equation}
and is given up to NLO accuracy by
\begin{equation}
    \vb K(y,\mu_F^2 \leftarrow \mu_{F,0}^2) = \mathbbm{1}\delta(1-y) + \alpha_s(\mu_{F,0}^2) \ln(\mu_F^2/\mu_{F,0}^2) \vb P^{(0)}(y) + \ldots \label{eq:NLOK}
\end{equation}
Here, $\mathbbm 1$ refers to an identity operation in flavour space and $\delta(1-y)$ is the identity in the function convolution space as $\vb K$ is acting on both.

Note that $\vb K$ is a fixed order equivalent of the EKO $\vb E$, but the two are fundamentally different: the latter is a true solution to DGLAP, which must resum all necessary terms, but the former is only an approximation for a small amount of evolution.
Specifically, $\vb K$ only is meaningful when we consider scale variations and, in particular, we use the interplay between $\vb K$ and $\vb E$ to generate the missing higher order terms we are looking for, because we get
\begin{equation}
    \vb f(\xi,\mu_1^2) = \qty(\vb K(\mu_1^2 \leftarrow \mu_2^2) \otimes \vb E(\mu_2^2 \leftarrow \mu_1^2) \otimes \vb f(\mu_1^2) )(\xi) + \ldots
\end{equation}
where $\ldots$ denotes the missing higher order terms.
Inserting the equation back into \cref{eq:fact3} with $(\mu_1,\mu_2)\to(Q,\mu_F)$ we obtain the definition of our sought-after special PDF
\begin{equation}
    \vb f^{SV}(\xi,\mu_F^2,\ln(\mu_F^2/Q^2)) = \qty(\vb K(Q^2 \leftarrow \mu_F^2)) \otimes \vb f(\mu_F^2))(\xi) \label{eq:fSV}
\end{equation}
where we apply \cref{eq:EKO} for the PDF.
We find $\vb f^{SV}(\xi,\mu_F^2,\ln(\mu_F^2/Q^2)) = \vb f(\xi,Q^2) + \ldots$ as we want to, because then \cref{eq:fact3} and \cref{eq:sigmaFB} differ by higher order terms.

We can now return to the definition of scheme C: we insert \cref{eq:fSV} into \cref{eq:sigmaFB} and obtain
\begin{align}
    \sigma(x,Q^2,\mu_F^2) &= \qty[\vb C^T(Q^2) \otimes \qty(\vb K(Q^2 \leftarrow \mu_F^2) \otimes \vb f(\mu_F^2))](x)\\
     &= \qty[\qty(\vb C^T(Q^2) \otimes \vb K(Q^2 \leftarrow \mu_F^2)) \otimes \vb f(\mu_F^2)](x) \label{eq:sigmaFCdef}
\end{align}
where we use the associative property of the convolution.
However, we can now re-expand terms in a different order.
First, it is convenient to use the perturbative expansion of the strong coupling to rewrite $\vb K$ in powers of $\alpha_s(Q^2)$ (as opposed to $\alpha_s(\mu_F^2)$, \cref{eq:NLOK}):
\begin{equation}
    \vb K'(y,Q^2 \leftarrow \mu_F^2) = \sum_{j=0} \qty(\alpha_s(Q^2))^j \vb {K'}^{(j)}(y,\ln(\mu_F^2/Q^2))
\end{equation}
with
\begin{equation}
    \vb K'(y,Q^2 \leftarrow \mu_F^2) = \left(\vb K(Q^2 \leftarrow \mu_F^2)\otimes\left[\mathbbm{1} + \order{\alpha_s(Q^2)}\right]\right)(y)
\end{equation}
i.e.\ $\vb K'$ and $\vb K$ differ by higher order terms.
Effectively, this is performing the algorithm of \cref{sec:rensv} for $\vb K$.
Up to NLO accuracy we find from \cref{eq:NLOK}
\begin{equation}
    \vb K'(y,Q^2 \leftarrow \mu_F^2) = \mathbbm{1}\delta(1-y) - \alpha_s(Q^2) \ln(\mu_F^2/Q^2) \vb P^{(0)}(y) + \ldots \label{eq:NLOKp}
\end{equation}

Second, we identify $\vb C(z,Q^2,\mu_F^2)$ from \cref{eq:sigmaFC} as
\begin{equation}
    \vb C^T(z,Q^2,\mu_F^2) = \qty(\vb C^T(Q^2) \otimes \vb K'(Q^2 \leftarrow \mu_F^2))(z)
\end{equation}
but which is yet another reorganization of the perturbative orders since it turns the product of sums in \cref{eq:sigmaFCdef} into a single sum, \cref{eq:CbarFpQCD}.
Let us see this in practice.

As before, we start from LO accuracy and from \cref{eq:NLOKp} we find immediately
\begin{equation}
    \vb C^{(0)}(z,\ln(\mu_F^2/Q^2)) = \vb C^{(0)}(z), \label{eq:CbarFLO}
\end{equation}
i.e., also here scale variations do not alter the LO coefficient function.
However, unlike in the case for renormalization scale variations, factorization scale variations have always an impact starting from LO accuracy, because PDFs are present per-se in the factorization formula, \cref{eq:fact1}.
Specifically, in any actual calculation with $\mu_F\neq Q$ also $\vb f(\mu_F^2)$ and $\vb f(Q^2)$ are different, since we must resum for both of them.
They differ by an amount which is beyond our fixed-order framework and if that difference is \enquote{small} depends on the actual values of the PDF of the respective scales.
However, with that we achieve the main purpose of this game: we generate some higher order terms, which we can then use to estimate MHOUs.
This is an intentional copy from \cref{sec:rensv}.

Starting from NLO the matching becomes non-trivial and, e.g., at this order we find
\begin{equation}
    \vb C^{(1),T}(z,\ln(\mu_F^2/Q^2)) = \vb C^{(1),T}(z) -  \ln(\mu_F^2/Q^2) \qty(\vb C^{(0),T} \otimes \vb P^{(0)})(z)\,. \label{eq:CbarFNLO}
\end{equation}
Always remember that for PDF operations, such as acting with the matrix of splitting functions $\vb P(y,\mu_F^2)$, non-commutativity is crucial, i.e., the flavor index dictated by the left-hand-side of \cref{eq:CbarFNLO} is matched by the second term on the right-hand-side by a vector-matrix multiplication (from the left).
Note that starting from NNLO also explicit coefficients of the beta function start to appear, since the logarithm $\ln(\mu_F^2/Q^2)$ is coming directly from the running of the strong coupling, \cref{eq:LOalphasresum}.
In other words: PDF evolution resums logarithms through the running of the strong coupling as the scale dependence of $\vb P(y,\mu_F^2)$ is entirely captured by the strong coupling, \cref{eq:DGLAP}.

In the following we assume scheme C as our default choice, thus \cref{box:svren} holds also for factorization scale variations and we can improve \cref{eq:fact3} with \cref{eq:sigmaFC}.
The case of scheme B\footnote{scheme A is doing yet something different - see Ref.~\cite{NNPDF:2024dpb}} is slightly more complicated, as explained above, but straightforward to obtain.

\subsection{Combined variation}
\label{sec:svcomb}
We recall, the renormalization and factorization scales are two independent scales, related to two different kind of physics.
The former captures ultra-violet and the latter collinear divergencies.
While we discuss them one at a time above, in practice we have to consider them simultaneously.
In order to do so, we must first consider factorization scale variation, which yields \cref{eq:sigmaFC},
\begin{equation}
    \sigma(x,Q^2,\mu_F^2,\mu_R^2=Q^2) = \qty(\vb C^T(Q^2,\mu_F^2,\mu_R^2=Q^2) \otimes \vb f(\mu_F^2))(x)
\end{equation}
but now we make the renormalization scale dependence explicit and $\vb C(Q^2,\mu_F^2,\mu_R^2=Q^2)$ corresponds to \cref{eq:CbarFpQCD}.
Then in the second step, we apply the renormalization scale variation onto $\vb C(Q^2,\mu_F^2,\mu_R^2=Q^2)$, which still depend on $\alpha_s(Q^2)$, and get
\begin{equation}
    \sigma(x,Q^2,\mu_F^2,\mu_R^2) = \qty(\vb C^T(Q^2,\mu_F^2,\mu_R^2) \otimes \vb f(\mu_F^2))(x)\,. \label{eq:sigmaFR}
\end{equation}
We can make the dependence on the various scales in \cref{eq:sigmaFR} more explicit by rewriting it as
\begin{equation}
    \sigma(x,Q^2,\mu_F^2,\mu_R^2) = \qty(\vb C^T(\alpha_s(\mu_R^2),\ln(\mu_F^2/Q^2),\ln(\mu_R^2/Q^2)) \otimes \vb f(\mu_F^2))(x) \label{eq:sigmaFRln}
\end{equation}
where, again, finally $\mu_F$ and $\mu_R$ fulfill their dedicated role.
Note that in \cref{eq:sigmaFRln} two additional scales, $\mu_{F,0}$ and $\mu_{R,0}$ are left implicit and those eventually guarantee that it is a dimensionless equation.

Let us see this in practice and remember that in order to see renormalization scale logarithms we need to play at NNLO.
We find
\begin{align}
    &\vb C(z,\alpha_s(\mu_R^2),\ln(\mu_F^2/Q^2),\ln(\mu_R^2/Q^2)) \nonumber\\
    &= \vb C^{(0)}(z) \nonumber\\
     &\hspace{10pt} + \alpha_s(\mu_R^2) \left[\vb C^{(1)}(z) + \vb C^{(1),F}(z)\ln(\mu_F^2/Q^2) \right] \nonumber\\
     &\hspace{10pt} + \left(\alpha_s(\mu_R^2)\right)^2 \left[\vb C^{(2)}(z) + \vb C^{(2),F}(z)\ln(\mu_F^2/Q^2) + \vb C^{(2),R}(z)\ln(\mu_R^2/Q^2) \right.\nonumber\\
     &\hspace{70pt} \left. \vb C^{(2),FF}(z)\ln^2(\mu_F^2/Q^2) + \vb C^{(2),FR}(z)\ln(\mu_F^2/Q^2)\ln(\mu_R^2/Q^2) \right] + \ldots
\end{align}
An explicit expression for $\vb C^{(1),F}$ is given in \cref{eq:CbarFNLO} and for $\vb C^{(2),R}$ in \cref{eq:CbarRimpl}.
We see that the number of terms per perturbative order quickly grows as the respective algorithms collect more of the full behavior.
The appearance of $\vb C^{(2),FR}$ also demonstrate how the two variations become intertwined.

\subsection{Discussion}
\label{sec:svdiscussion}
The renormalization and factorization scale are remnants of the necessary regularization of UV and collinear divergencies.
Although, scale variations are a genuine pQCD property they have some features, which questions their usefulness for estimating MHOUs.

When introducing the scales, we are required to set them to \enquote{a typical scale of the process}, which makes them inherently arbitrary from the start before doing any variation.
In fully inclusive DIS there is a single scale available, the virtuality $Q^2$, and thus the common choice is $\mu_F^2 = Q^2 = \mu_R^2$, but any multiplicative there of could work just as well\footnote{For example CT18X~\cite{Hou:2019efy} makes a different choice.}.
If we want to use scale variations to estimate MHOU, we need to make a further arbitrary choice about which range we want to use for the respective scales and if we want to correlate them.
As we see in \cref{box:svren} the scale-varied coefficient functions are always a linear combination of the existing ones, i.e.\ they can not introduce new channels (new PDF combinations).
However, it is often straightforward to show that a new channel should appear although its size is unknown and scale variations are unable to predict this.
An example for such a scenario in fully inclusive DIS can be seen at NLO: starting from NLO the gluon can participate in the scattering, but the LO only contains quark PDFs.
In other cases, we know that higher orders must contain certain kinematic logs, e.g. from particles radiating at small and/or large transverse momentum or in the soft limit and in fully inclusive DIS the latter is the case\footnote{More mathematically speaking: soft gluon resummation is a double logarithmic resummation $\tilde c_q^{(j)} \sim \alpha_s^j \ln^{2j}(N)$ for $N\to\infty$ in Mellin space, \cref{eq:Mellinf}. Double logarithmic structures can not be predicted by scale variations.}.
Also those corrections can (in many cases) not be predicted by scale variations.

Scale variations also provide no statistical interpretation, which is in stark contrast to experimental uncertainties, where we usually assume Gaussian uncertainties and thus the provided error has a clear interpretation.
However, strictly speaking even there this only applies to statistical errors but not to systematic errors and the question which of the two dominates depends very much on the specific measurement.

Despite all these (potential) problems, scale variations have a solid foundation in QFT and can be applied (straightforwardly) to any pQCD process, which makes them an easy tool to estiate MHOU.
In order to make a faithful statement, we can then simply say: when we compute this process with this specific choice of scale, we obtain this number; if we vary that scale in this range, we obtain this range of numbers.

\subsection{Implementation and flavor space}
\label{sec:svimpl}
Let us return to more practical consequences.
As we see above, renormalization scale variations are just a scalar operation, but factorization scale variations are more complicated and it is worth spelling out their explicit construction to see what is actually happening.

The factorization scale variation is directly related to the PDFs themselves and, in particular, eventually, they derive from \cref{eq:DGLAP}.
However, the splitting functions $\vb P$ are encoding how partons couple among each other, which is through the gluon\footnote{for quark-photon interaction see Ref.~\cite{NNPDF:2024djq}} of course.
This is in contrast to DIS, where instead the coefficient functions are encoding how partons couple to the electro-weak vector boson, which is completely different, see \cref{box:particles}.

At LO accuracy the problem does not appear yet, because there are no scale-varied coefficient functions, see \cref{eq:CbarFLO}, but starting from NLO accuracy we need to worry.
In particular, we know that the gluon couples the same to all quarks, thus, if we assume four flavors for the moment (see \cref{sec:hq}) and combine them in the following way~\cite{NNPDF:2024djq}\footnote{for the more general case see Ref.~\cite{NNPDF:2024djq}},
\begin{align}
    f_{\Sigma_\Delta}(x,Q^2) &= f_u(x,Q^2) + f_{\bar u}(x,Q^2) + f_c(x,Q^2) + f_{\bar c}(x,Q^2)\nonumber\\
    & \hspace{10pt} - f_d(x,Q^2) - f_{\bar d}(x,Q^2) - f_s(x,Q^2) - f_{\bar s}(x,Q^2)  \label{eq:SigmaDelta}
\end{align}
we know that this combination does not interact with the gluon.
In order to see more clearly what happens we need to go back to the vector notation from \cref{eq:fact3}.
We repeat: we denote objects, which have a non-trivial dependence on the parton flavors, with bold font.
This applies to vectors in flavor space, e.g.\ coefficient functions $C_j \to \vb C$ or PDFs $f_j \to \vb f$, but also to matrices, e.g.\ splitting functions $P_{jk} \to \vb P$ or the factorization scale operator $K_{jl}\to \vb K$.
By introducing the vector notation we hide any reference to a given basis in the flavor space, which now becomes relevant (see also \cref{sec:hq}).

Let us be more formal.
We define the flavor space $\mathcal F$ as the vector space spanned by the gluon and all quark flavors.
We consider the naive flavor basis as our canonical basis and write
\begin{equation}
    \mathcal F = \mathrm{span}(g, u, \bar u, d, \bar d, s, \bar s, c, \bar c, b, \bar b, t, \bar t)\,. \label{eq:flspace}
\end{equation}
We have a 13 dimensional vector space, $\dim{\mathcal F} = 13$\footnote{one can add the photon and make it 14 dimensional~\cite{NNPDF:2024djq}}, consisting of the gluon, six quarks, and six anti-quarks.
Our bold vectors are an element in that space, e.g.\ $\vb C \in \mathcal F$, or more plainly,
\begin{equation}
    \vb C^T \sim (C_g, C_u, C_{\bar u}, C_d, C_{\bar d}, C_s, C_{\bar s}, C_c, C_{\bar c}, C_b, C_{\bar b}, C_t, C_{\bar t})
\end{equation}
and, of course, any other basis works just as well.
The same applies also to the PDF $\vb f$, which is the other vector in \cref{eq:fact3}, and any operator, e.g.\ splitting functions $\vb P$, just lives in the product space, $\vb P \in \mathcal F \otimes \mathcal F$.
The flavor basis is handy for computing LO and NLO DIS coefficient functions, i.e.\ we consider each quark individually and we can find easily \cref{eq:LOC,eq:LOxs}.

For the discussion in \cref{sec:DGLAP} and in particular for dealing with splitting functions, $\vb P$, a different basis is more convenient: the evolution basis.
The two choices are both orthogonal bases, i.e.\ they are simply related by a rotation\footnote{actually, it is a rotation and a scaling, i.e.\ the evolution basis is not orthonormal, which may cause some trouble} in flavor space.
Specifically, $\vb P$ has a much simpler representation if one of the basis vectors is the flavor singlet
\begin{align}
    f_\Sigma(x,Q^2) &= f_u(x,Q^2) + f_{\bar u}(x,Q^2) + f_c(x,Q^2) + f_{\bar c}(x,Q^2)\nonumber\\
    & \hspace{10pt} + f_d(x,Q^2) + f_{\bar d}(x,Q^2) + f_s(x,Q^2) + f_{\bar s}(x,Q^2) \label{eq:Singlet}
\end{align}
and all other basis vectors can be chosen such that they do not couple to the gluon, e.g. \cref{eq:SigmaDelta}.

Let us see this in practice for LO photon DIS: we start from \cref{eq:LOC}, which is written in flavor basis, i.e.\ for each given quark flavor and the gluon.
Inserting \cref{eq:LOC} into \cref{eq:fact1} yields \cref{eq:LOxs} and we find explicitly
\begin{align}
    \sigma(x,Q^2) &= e_u^2\qty(f_u(x,Q^2) + f_{\bar u}(x,Q^2) + f_c(x,Q^2) + f_{\bar c}(x,Q^2)) \nonumber\\
    & \hspace{10pt} + e_d^2\qty(f_d(x,Q^2) + f_{\bar d}(x,Q^2) + f_s(x,Q^2) + f_{\bar s}(x,Q^2))\,.
\end{align}
Using \cref{eq:Singlet,eq:SigmaDelta} we can rewrite this in evolution basis
\begin{equation}
    \sigma(x,Q^2) &= \frac{e_u^2 + e_d^2} 2 f_\Sigma(x,Q^2) + \frac{e_u^2 - e_d^2} 2 f_{\Sigma_\Delta}(x,Q^2)
\end{equation}
or equivalently
\begin{equation}
    C_\Sigma^{(0)}(z) = \frac{e_u^2 + e_d^2} 2 \delta(1-z),\quad C_{\Sigma_\Delta}^{(0)}(z) = \frac{e_u^2 - e_d^2} 2 \delta(1-z), \quad C_j(z) = 0 \,\forall j\neq\{\Sigma,\Sigma_\Delta\} \label{eq:LOCev}
\end{equation}
where the parton index $j$ in \cref{eq:fact1} now also runs over the evolution basis.
We stress that \cref{eq:LOC,eq:LOCev} are two equivalent representation of the coefficient function $\vb C^{(0)}$ using two different bases.

Moving to NLO, we have the same structure in the quark sector, \cref{eq:NLOCq}, and we encounter for the first time the gluon channel, \cref{eq:NLOCg}.
Thus, when we want to execute \cref{eq:CbarFNLO}, we need to consider three directions in flavor space: the gluon, the singlet $f_\Sigma$ and $f_{\Sigma_\Delta}$, \cref{eq:SigmaDelta}.
The preferred basis for $\vb P$ is manifest by the following statements,
\begin{equation}
    \mathrm P_{\Sigma \Sigma_\Delta}(y) = \mathrm P_{g \Sigma_\Delta}(y) = \mathrm P_{\Sigma_\Delta \Sigma}(y) = \mathrm P_{\Sigma_\Delta g}(y) = 0
\end{equation}
which do not occur in flavor basis.
Eventually, \cref{eq:CbarFNLO} becomes explicitly
\begin{align}
    C^{(1)}_{\Sigma_\Delta}(z,\ln(\mu_F^2/Q^2)) &= C^{(1)}_{\Sigma_\Delta}(z) - \ln(\mu_F^2/Q^2) \qty(C^{(0)}_{\Sigma_\Delta} \otimes \mathrm P_{\Sigma_\Delta \Sigma_\Delta}^{(0)})(z) \label{eq:NLOCbarFNS}\\
    C^{(1)}_{\Sigma}(z,\ln(\mu_F^2/Q^2)) &= C^{(1)}_{\Sigma}(z) - \ln(\mu_F^2/Q^2) \qty(C^{(0)}_{\Sigma} \otimes \mathrm P_{\Sigma \Sigma}^{(0)})(z)\\
    C^{(1)}_{g}(z,\ln(\mu_F^2/Q^2)) &= C^{(1)}_{g}(z) - \ln(\mu_F^2/Q^2) \qty(C^{(0)}_{\Sigma} \otimes \mathrm P_{\Sigma g}^{(0)})(z) \label{eq:NLOCbarFg}
\end{align}
where the required splitting functions can be found in the literature~\cite{Leader:2011gpt,Leader:1996hm,Peskin:1995ev}.

Actually, also LO splitting functions $\vb P^{(0)}$ obey symmetry relations, in particular all quarks couple identically to the gluon and it is not possible (yet) to change the quark flavor.
This implies $\mathrm P_{\Sigma_\Delta \Sigma_\Delta}^{(0)}(y) = \mathrm P_{\Sigma \Sigma}^{(0)}(y) = \mathrm P_{qq}^{(0)}(y)$ and thus \cref{eq:CbarFNLO} can also be executed straightforwardly in flavor space.
However, this symmetry breaks down starting from NLO splitting functions $\vb P^{(1)}$, which would be required for NNLO DIS cross sections.

\subsection{Interplay with previous content}

\begin{itemize}
    \item Renormalization scale variation is a reshuffling of lower order coefficient functions and factorization scale variation adds convolutions between splitting functions and lower order coefficient functions.
        This implies that the mathematical complexity remains the same or in other words: we never encounter new mathematical objects in scale variations; see \cref{sec:nlo}.
        However, note that this holds in a distributional sense, i.e.\ e.g.\ in \cref{eq:NLOCbarFNS} both $C^{(0)}_{\Sigma_\Delta}(z)$ and $\mathrm P_{\Sigma_\Delta \Sigma_\Delta}^{(0)}(y)$ contain distributions and thus does their convolution.
    \item Although scale-varied coefficient functions can always be computed a posteriori, see \cref{box:svren}, in practice it usually makes sense to tabulate them into interpolation grids as well, see \cref{sec:grids}.
        These additional coefficient functions can be easily accommodated when we generalize our concept of \enquote{order}: it can not only represent the perturbative order per se, but also additional powers of renormalization or factorization logarithms: $\alpha_s^k(\mu_R^2)\ln^m(\mu_R^2/Q^2)\ln^n(\mu_R^2/Q^2)$.
    \item The necessary projection onto a singlet component is particularly tricky for CC DIS, see \cref{sec:diverse}.
        Recall that the $W^\pm$ boson only couple to specific PDFs and thus if we want to project the coefficient functions into evolution basis we need the full basis (and not just \cref{eq:SigmaDelta,eq:Singlet}).
    \item Note that the coupling structure is always retained, i.e.\ e.g.\ in \cref{eq:NLOCbarFg}
        \begin{equation}
            C^{(1)}_{g}(z) \sim e_u^2 + e_d^2 \sim C^{(1)}_{g}(z,\ln(\mu_F^2/Q^2))\,.
        \end{equation}
        This implies that a given \enquote{channel} in an interpolation grid (\cref{sec:grids}) has a unique coupling, though recall that some channels only appear at high enough perturbative order.
\end{itemize}

\section{Heavy quarks}
\label{sec:hq}
When we compute LO DIS from \cref{fig:LO} we say it is easy to identify the respective coefficient function, \cref{eq:LOC}.
However, what exactly do we mean with the \enquote{$\forall q$}?
Which quarks are we exactly talking about?
This is directly linked to the flavor space $\mathcal F$.
However, unlike the problem we discuss in \cref{sec:svimpl} about different bases (flavor basis vs.\ evolution basis), here we are more directly concerned what which elements are active in the first place, or if there are trivial directions in $\mathcal F$.

The question we need to address here is: which quarks exactly can participate in \cref{eq:LOxs,eq:LOC}?
We refer to this number as the number of active quarks $n_f$.
This dependency is completely hidden in \cref{eq:fact1}, but, as we demonstrate in the following, is a major concern for both coefficient functions and PDF.
Note that the gluon is always considered active and contributes to all diagrams always.

The number of flavors $n_f$ is per se arbitrary and there is no absolutely correct answer, since there is no theory argument, which would fix this choice.
However, in practice for a given observable $\sigma(x,Q^2)$ we can consider, which options are phenomenologically reasonable and if some option would yield unstable results.

For the sake of the argument, let us consider a fully inclusive DIS cross section $\sigma(x,Q^2=\SI{25}{\GeV^2})$, so we probe the probe the proton at $\mu_F=Q=\SI{5}{\GeV}$.
We can now consider the different quark masses as a guideline: on the one side the masses of up, down, and strange are significantly below the non-perturbative regime of QCD $m_u,m_d,m_s \ll \LQCD \sim \SI{1}{\GeV}$ and we must assume we can find them in the proton.
On the other side the mass of the top is significantly above the probed scale, $m_t \gg Q$, thus it is very unlikely to find it in the proton and we can safely neglect it.
We can shut down such contributions by setting $f_t(\xi, Q^2) = 0 = f_{\bar t}(\xi, Q^2)$ and this way the top quark does not contribute to \cref{eq:fact1}.
In addition, we can also define $C_t(z, Q^2) = 0 = C_{\bar t}(z, Q^2)$ to explicitly say that we don't let the top contribute, e.g.\ if we consider the PDF an external ingredient.

Finally, we consider the two remaining quarks, charm and bottom, for which maybe there is no clear answer.
First, for the charm quark we have $m_c \sim \SI{1.5}{\GeV}$ and we can assume reasonably that it can contribute to this cross section.
On the other side, for the bottom quark we have $m_b \sim \SI{4.5}{\GeV}$ and maybe we can both allow or deny it.
The specific choice on which quarks we allow to participate and in which way they do so is referred to as flavor number scheme (FNS).

\subsection{ZM-VFNS}

One very popular choice is the Zero-Mass Variable Flavor Number Scheme (ZM-VFNS), which is a telling name and which we have silently adopted in the previous parts.
Formally, we can define it in the following way:
first, whenever the probed scale $Q$ crosses a quark mass that quark becomes active;
second; in coefficient functions all active quarks are considered massless and all non-active quarks are considered infinitely massive.
The infinite mass ensures $f_H(\xi, Q^2) = 0 = f_{\bar H}(\xi, Q^2)$ and $C_H(z, Q^2) = 0 = C_{\bar H}(z, Q^2)$ for all non-active quarks $H$ and the zero mass ensures that all formulas in \cref{sec:nlo} hold.
If we denote the new dependency on the number of active flavors $n_f$ by a superscript $(n_f)$ we can define the ZM-VFNS in the following way
\begin{equation}
    \sigma^{\text{ZM-VFNS}}(x,Q^2) = \left\{ \begin{array}{ll}
        \sigma^\mathrm{(3)}(x,Q^2) &\text{if}~ Q^2 < m_c^2\\
        \sigma^\mathrm{(4)}(x,Q^2) &\text{if}~ m_c^2 \leq Q^2 < m_b^2\\
        \sigma^\mathrm{(5)}(x,Q^2) &\text{if}~ m_b^2 \leq Q^2 < m_t^2\\
        \sigma^\mathrm{(6)}(x,Q^2) &\text{if}~ Q^2 \geq m_t^2
    \end{array} \right. \label{eq:zmvfns}
\end{equation}
where
\begin{equation}
    \sigma^{(n_f)}(x,Q^2) = \left(\vb C^{T,(n_f)}(Q^2) \otimes \vb f^{(n_f)}(Q^2)\right)(x) \label{eq:factnf}
\end{equation}
i.e.\ each contribution has its own factorization formula and we have, e.g.,
\begin{equation}
    \begin{array}{l}
        C_H^{(3)}(z, Q^2) = 0 = C_{\bar H}^{(3)}(z, Q^2)\\
        f_H^{(3)}(\xi, Q^2) = 0 = f_{\bar H}^{(3)}(\xi, Q^2)
    \end{array}
    \quad H\in\{c,\bar c, b, \bar b, t, \bar t\}
\end{equation}
and similar for the others.
Specifically, we can improve \cref{eq:LOC} by writing
\begin{align}
    C_q^{(0),(3)}(z) &= e_q^2 \delta(1-z) & q\in\{d,\bar d, u, \bar u, s, \bar s\}\\
    C_H^{(0),(3)}(z) &= 0 & H\in\{c,\bar c, b, \bar b, t, \bar t\}\\
    C_g^{(0),(3)}(z) &= 0
\end{align}
where we still use the full flavor space $\mathcal F$, \cref{eq:flspace}, but finally spell all elements explicitly.

Note that it does not make sense to consider less then three quarks, since $m_u,m_d,m_s \ll \LQCD \sim \SI{1}{\GeV}$ and pQCD only works for $Q > \LQCD$.
All quarks are considered with their actual mass for the decision if they are in or out, \cref{eq:zmvfns}, but then after that their mass is discarded when computing $\vb C^{(n_f)}(z,Q^2)$.
Thus, the VFNS indicates that the number of active quarks $n_f$ depends on the probed scale $Q$, i.e.\ it is variable, and the ZM explains how to compute diagrams.
However, once we have an actual scale at hand, $n_f$ is also fixed.

\begin{conclusionbox}{ZM-VFNS}
    In the ZM-VFNS
    \begin{itemize}
        \item the number of active quarks $n_f$ depends on the probed scale $Q^2$
        \item contain only massless quarks in the diagrams for the coefficient functions
    \end{itemize}
\end{conclusionbox}

\subsection{FFNS}
\label{sec:ffns}

For the number of active quarks we can make another simple choice: fix $n_f$ once and forever - the Fixed Flavor Number Scheme (FFNS).
Here, the probed scale $Q$ does not matter, but we use the same coefficient functions and PDFs for any scale.
Actually, saying FFNS is not sufficient, we still have some choices and the most obvious is $n_f$.

However, we can also improve with respect to other parts: while it is significantly easier to compute diagrams with only massless quarks, we can compute them also with massive quarks.
For simplicity we assume in the following we have $n_f$ massless quarks plus exactly one\footnote{the case of two massive quarks is e.g.\ discussed in Ref.~\cite{Barontini:2024xgu}} massive quark with mass $m^2$.
While we keep the notation general we can always consider as a concrete example $n_f=3$ (up, down, strange) and thus the massive quark as the charm with $m^2 = m_c^2$.
We recall two important features of the propagator of a massive quark,
\begin{equation}
    D_q(p) \sim \frac{\slashed{p} + m\mathbbm{1}}{p^2 - m^2}\,. \label{eq:quarkprop}
\end{equation}
First, the mass appears explicitly in the numerator, which makes computing Dirac traces significantly more lengthy, and, second, the mass appears in the denominator, which shields the propagator from a collinear pole.
When we consider a finite mass we eventually have three different scales in our diagrams: the total energy of the colliding partons $\hat s=(\xi P + q)^2$, the photon virtuality $Q^2$, and the quark mass $m^2$.
This means that unlike the massless case discussed so far, the coefficient functions now depend on two scales, for example the partonic momentum fraction $z = x / \xi = Q^2/(\hat s + Q^2)$ and $\xi_q = Q^2 / m^2$ or any other bijective combination thereof\footnote{For example, for NC DIS on can show~\cite{Hekhorn:2019nlf} that $\chi = \frac{1-\sqrt{1-4m^2/\hat s}}{1+\sqrt{1-4m^2/\hat s}}$ and $\chi_q = \frac{\sqrt{1+4m^2/Q^2}-1}{\sqrt{1+4m^2/Q^2}+1}$ are good combinations.}.
This in turn implies that the available mathematical function space is much bigger, which makes the computation of these coefficient functions $\vb C(z,Q^2,m^2)$ significantly more complex.

We focus here on the new dependency on $\xi_q$, which captures the new features.
First, we consider the case $Q^2 \gg m^2$ or $\xi_q \to \infty$ where the denominator of \cref{eq:quarkprop} plays the important role.
Indeed, since it is shielding the propagator from a collinear divergency, we encounter some signals thereof when we put $m^2 \to 0$.
Specifically, we can rewrite the coefficient functions in this limit,
\begin{equation}
    \lim_{Q^2/m^2\to\infty}\vb C^{(n_f),T}(z,Q^2,m^2) = \qty(\vb C^{(n_f+1),T}(Q^2) \otimes \vb A^{(n_f+1)}(Q^2,m^2))(z) \label{eq:Cmatching}
\end{equation}
i.e.\ we can connect the coefficient functions using two neighboring schemes and a new, universal operator $\vb A$, which admits a perturbative series
\begin{equation}
    \vb A^{(n_f+1)}(Q^2,m^2) = \sum_{j=0} \left(\alpha_s^{(n_f+1)}(Q^2)\right)^j \vb A^{(j),(n_f+1)}(\ln(Q^2/m^2))\,.
\end{equation}
We discuss more details of this new operator in a moment, but the important thing to stress here is the universality.
This universality derives from the universality of PDFs, which makes the operator closely connected to collinear splitting functions as we expect.
Moreover, \cref{eq:Cmatching} puts strong constraints on the computation of any coefficient function\footnote{or any partonic matrix elements as a similar equations hold, e.g., in hadron-hadron collisions} once the universal operator $\vb A$ is known at the requested accuracy.
Note that the coefficient functions in \cref{eq:Cmatching} depend on different sets of variables, since we consider the \enquote{+1} quark massless in the $(n_f+1)$ scheme.
Thus, as expected, the full mass dependence on the right-hand-side of \cref{eq:Cmatching} is carried by the universal operator $\vb A$ and it is a pure, actual polynomial, dependency on $\ln(Q^2/m^2)$.
So, the remnant of the collinear divergency for a massive quark is a logarithm $\ln(Q^2/m^2)$ as we may have guessed from the beginning.

Next, we consider the case $Q^2 \sim m^2$ where the numerator of \cref{eq:quarkprop} plays the important role.
While we conclude above that the FFNS coefficient functions contain logarithms $\ln(Q^2/m^2)$ they also naturally contain more terms.
Specifically, if we expand the FFNS coefficient function around $Q^2 \sim m^2$ we obtain
\begin{equation}
    \vb C^{(n_f)}(z,Q^2,m^2) = \vb C_{thr,0}^{(n_f)}(z) + \vb C_{thr,1}^{(n_f)}(z) \left(\frac{Q^2}{m^2}\right) + \vb C_{thr,2}^{(n_f)}(z) \left(\frac{Q^2}{m^2}\right)^2 + \ldots \label{eq:Cpower}
\end{equation}
with some coefficients $\vb C_{thr,k}^{(n_f)}(z)$.
Note that the equation is to be understood as a series expansion at the requested point as in practice the actual dependency on $\xi_q$ is very intricate\footnote{for example via $\chi_q$}.
It is clear that in the region $Q^2 \gg m^2$ these terms can be neglected, but in the region $Q^2 \sim m^2$ these terms may not be small and thus neglecting such contributions may yield wrong results.
That is the precisely the advantage of FFNS over ZM-VFNS: while in the latter we explicitly deny any massive quark in the diagrams, the former allows them to be included in the sum over diagrams.

\begin{conclusionbox}{FFNS}
    In the FFNS
    \begin{itemize}
        \item we need to fix the number of massless and massive quarks
        \item massive quarks are then allowed in the diagrams for the coefficient functions
        \item we encounter universal logarithms $\ln(Q^2/m^2)$ for $Q^2\to m^2$ due in the coefficient functions to shielded collinear divergencies
    \end{itemize}
\end{conclusionbox}

But if the FFNS calculation is actually better than ZM-VFNS, why don't we always use the former?
Before answering the question, we need to return to the new operator $\vb A$.

\subsection{Matching schemes}
\label{sec:matching}

The number of active quarks is not only relevant for coefficient functions, but, in fact, for the whole pQCD program and in particular for the strong coupling $\alpha_s$ and PDFs $\vb f$.
As we see from \cref{eq:factnf} also PDFs depend on the number of flavors, e.g.\ $\vb f^{(3)}(\xi,\mu_F^2)$, and we can improve \cref{eq:CpQCD} with
\begin{equation}
    \vb C^{(n_f)}(z,Q^2,m^2) = \sum_{k=0} \left(\alpha_s^{(n_f)}(Q^2)\right)^k \vb C^{(k),(n_f)}(z,\xi_q)
\end{equation}
where we now also explicitly denote the flavor dependence of the strong coupling.
Actually, from \cref{eq:LObeta} we already known that the beta coefficients depend explicitly on $n_f$ and thus $\alpha_s$ must depend on it as well.
The same applies to PDFs since also splitting functions carry an explicit $n_f$ dependence and all appearances above, indeed, correspond to $\vb P^{(n_f)}$.

The question is: how are those different schemes, e.g.\ $\alpha_s^{(3)}$ and $\alpha_s^{(4)}$, connected with each other?
For coefficient functions the difference originates from the different diagrams which are included into the calculation of the coefficient functions, $\vb C^{(3)}$ and $\vb C^{(4)}$, but what about the strong coupling $\alpha_s$ and the PDFs $\vb f$?
In order to answer this question it is important to return to the basic principles of \cref{sec:basics}: we know that only cross sections $\sigma$ are physical objects for which we can derive physical properties, e.g.\ that they must be positive.
In contrast, the strong coupling $\alpha_s$, PDFs $\vb f$, and coefficient functions $\vb C$ are not physical objects, which can not be measured directly in an experiment and which only are defined within our framework of pQCD.
The concept of having different number of active flavors $n_f$ is a theoretical idea, which we employ to make a consistent approximation in our calculation, but it is irrelevant to nature.
In pQCD we can still derive certain rules for our relevant objects in order to obtain a consistent theory, which we can then use to make predictions.

The important physical properties, which we need here is consistency and perturbative equivalence.
Specifically, we require that the cross section in two different scheme are perturbatively equivalent, i.e.\
\begin{equation}
    \sigma^{(n_f)}(x,Q^2) = \sigma^{(n_f+1)}(x,Q^2)\left[1 + \order{\alpha_s^{(n_f+1)}(Q^2)}\right]\,, \label{eq:sigmamatching}
\end{equation}
where the two cross section are consistently computed in their respective flavor number scheme.
In particular, they use their own coefficient functions, their own strong coupling, and their own PDFs.
However, since the coefficient functions, the beta coefficients, and the splitting functions are computed from the respective relevant diagrams this makes the condition quite restrictive.
Moreover, if we recall \cref{sec:resum} we know that both, $\alpha_s$ and $\vb f$, perform a resummation, so \cref{eq:sigmamatching} is to be understood in a logarithmic sense and thus, in practice, it is convenient to consider \cref{eq:sigmamatching} in the limit $Q^2 \to \infty$.

One can show~\cite{Barontini:2024xgu} that the connection between two different schemes can be done by a universal matching procedure for the strong coupling and the PDFs.
Specifically, for the strong coupling we find
\begin{equation}
    \alpha_s^{(n_f+1)}(\mu_R^2) = \zeta^{(n_f+1)}(\mu_R^2, m^2)\alpha_s^{(n_f)}(\mu_R^2) \label{eq:alphasmatching}
\end{equation}
with a universal matching operator $\zeta$, which admits a perturbative series and has a polynomial dependence on $\ln(\mu_R^2/m^2)$.
Analogously, for the PDFs we find
\begin{equation}
    \vb f^{(n_f+1)}(\xi, \mu_F^2) = \left(\vb A^{(n_f+1)}(\mu_F^2, m^2) \otimes \vb f^{(n_f)}(\mu_F^2)\right)(\xi) \label{eq:pdfmatching}
\end{equation}
with a universal matching operator $\vb A$, which is also used in \cref{eq:Cmatching}.
The scheme change, \cref{eq:alphasmatching} and \cref{eq:pdfmatching}, can be done at any scale, but, following \cref{eq:zmvfns}, is conventionally done at the mass of the heavy quark, $\mu_R^2 = m^2 = \mu_F^2$.

Let us see this in practice: consider a cross section $\sigma(x,Q^2)$ with $Q^2 > m^2$.
As we say above the number of active quarks is a free choice, we can compute the cross section either in the $n_f$ scheme, \cref{eq:factnf}, or in the $(n_f+1)$ scheme,
\begin{equation}
    \sigma^{(n_f+1)}(x,Q^2) = \left(\vb C^{T,(n_f+1)}(Q^2) \otimes \vb f^{(n_f+1)}(Q^2)\right)(x)\,. \label{eq:sigmaup}
\end{equation}
Inserting \cref{eq:pdfmatching} into \cref{eq:sigmaup} we obtain
\begin{equation}
    \sigma^{(n_f+1)}(x,Q^2) = \left(\vb C^{T,(n_f+1)}(Q^2) \otimes \vb A^{(n_f+1)}(Q^2,m^2) \otimes \vb f^{(n_f)}(Q^2)\right)(x) \label{eq:sigmaupmatched}
\end{equation}
and if \cref{eq:Cmatching} holds, the two expressions, \cref{eq:factnf} and \cref{eq:sigmaup}, are actually perturbatively equivalent.
Recall that \cref{eq:Cmatching} only holds in the $Q^2 \gg m^2$ limit and so in practice for finite $Q^2$ the two expression are only perturbatively equivalent.
Note that here we matched the two schemes at $Q^2$ in contrast to the usual procedure.

\Cref{eq:alphasmatching,eq:pdfmatching} imply that the two quantities are not continuous across the matching, but they differ by the application of the matching operator.
Actually, at low enough perturbative order the matching operator, $\zeta$ and $\vb A$, are trivial (e.g.\ $\zeta(\mu_R^2,m^2) = 1+\order{\alpha_s(\mu_R^2)}$) and the discontinuity is a higher-order effect.
The strong coupling, the PDFs, and the coefficient function, who are all discontinuous on their own, conspire together in such a way that the cross section is continuous at the requested accuracy, \cref{eq:sigmamatching}.
This maps directly to the statement from \cref{sec:basics} that the cross section is a measurable object, which obviously must be continuous, but the others not.

\subsection{FONLL as an example of a GM-VFNS}

Let us return to the question on why we not always consider the FFNS.
Looking to \cref{eq:sigmaupmatched}, we note that we have matched the PDF at $Q^2$, which is what we have needed there to make the argument of consistency work.
However, we also say that the usual thing instead is to match at the heavy quark mass $m^2$, so let us see how that plays out.
Starting from \cref{eq:sigmaup}, we find
\begin{align}
    &\sigma^{(n_f+1)}(x,Q^2) \nonumber\\
    &= \left(\vb C^{T,(n_f+1)}(Q^2) \otimes \vb f^{(n_f+1)}(Q^2)\right)(x)\\
    &= \left(\vb C^{T,(n_f+1)}(Q^2) \otimes \vb E^{(n_f+1)}(Q^2 \leftarrow m^2) \otimes \vb f^{(n_f+1)}(m^2)\right)(x)\\
    &= \left(\vb C^{T,(n_f+1)}(Q^2) \otimes \vb E^{(n_f+1)}(Q^2 \leftarrow m^2) \otimes \vb A^{(n_f+1)}(m^2,m^2) \otimes \vb f^{(n_f)}(m^2)\right)(x) \label{eq:sigmaup0}
\end{align}
where in the second equation we have inserted the EKO in the $(n_f+1)$ scheme, $\vb E^{(n_f+1)}$, from \cref{eq:EKO} and then in the third equation the matching, \cref{eq:pdfmatching}, which is now in the customary place.
We can massage \cref{eq:factnf} in a similar way and find
\begin{align}
    \sigma^{(n_f)}(x,Q^2,m^2) &= \left(\vb C^{T,(n_f)}(Q^2,m^2) \otimes \vb f^{(n_f)}(Q^2)\right)(x)\\
    &= \left(\vb C^{T,(n_f)}(Q^2,m^2) \otimes \vb E^{(n_f)}(Q^2 \leftarrow m^2) \otimes \vb f^{(n_f)}(m^2)\right)(x) \label{eq:sigma0}
\end{align}
with the EKO now in the $n_f$ scheme, $\vb E^{(n_f)}$.
Note that \cref{eq:sigmaup0,eq:sigma0} now contain the same PDF $\vb f^{(n_f)}(m^2)$.

As we discuss in \cref{sec:resum}, the EKO, $\vb E$, is performing a resummation and in this particular case a resummation of the logarithm $\log(Q^2/m^2)$ as can be seen from the arguments in \cref{eq:sigmaup0,eq:sigma0}.
This is true in either scheme, but the amount of logarithms, meaning the prefactor that is resummed along side, is different.
The resummation is directly proportional to the splitting functions, \cref{eq:LOPDFexp}, which is different between the two scheme.
In particular, in the $(n_f)$ scheme we are resumming the collinear logs of $n_f$ light quarks, but in the $(n_f+1)$ scheme we are resumming the collinear logs of $(n_f+1)$ light quarks.
Thus the latter contains effectively more logarithms.
Recall that only massless quarks can develop a collinear pole and can thus participate in (collinear) evolution.

But then how could we find an equivalence between the two schemes in \cref{sec:matching}?
Examining \cref{eq:factnf}, we see the equivalence is valid for cross sections, where coefficient functions explicitly enter.
Indeed, from \cref{sec:ffns} we know that $\vb C^{(n_f)}(z,Q^2,m^2)$ contains logarithms $\ln(Q^2/m^2)$, which are precisely the ones we are hunting here.
However, the big difference is that $\vb C^{(n_f)}(z,Q^2,m^2)$ only contains a finite amount of logarithms, where as $\vb E^{(n_f+1)}$ is resumming those logarithms.
This means that for $Q^2 \gg m^2$ FFNS is indeed a bad choice as the logarithms $\ln(Q^2/m^2)$ are big and should better be resummed.
Instead, any power-suppressed terms, \cref{eq:Cpower}, which are important in the region $Q^2 \sim m^2$ and made FFNS preferable over ZM-VFNS, are irrelevant in this region.

\begin{conclusionbox}[label=box:gmvfns]{GM-VFNS}
    A General Mass-Variable Flavor Number Scheme (GM-VFNS) must approximate a cross section $\sigma(x,Q^2)$
    \begin{itemize}
        \item in the region $Q^2 \sim m^2$ with the FFNS result $\sigma^{(n_f)}(x,Q^2,m^2)$
        \item in the region $Q^2 \gg m^2$ with the resummed ZM-VFNS result $\sigma^{(n_f+1)}(x,Q^2)$
    \end{itemize}
\end{conclusionbox}

\Cref{box:gmvfns} gives no exact prescription on how the approximation has to be done and, indeed, there is some freedom on how to do this.
Different PDF fitting groups make different choices~\cite{Aivazis:1993pi,Thorne:1997uu,Kramer:2000hn,Tung:2001mv,Nadolsky:2009ge,Forte:2010ta,Guzzi:2011ew,Bonvini:2016fgf,Barontini:2024xgu} and as an exemplary case we consider here FONLL~\cite{Forte:2010ta,Barontini:2024xgu}, which has a simple idea:
in order to match the two regions, we simply add the two expressions together and subtract any overlap.
Let us consider the concrete case of charm:
\begin{equation}
    \sigma^\text{FONLL}(x,Q^2,m_c^2) = \sigma^{(3)}(x,Q^2,m_c^2) + \sigma^{(4)}(x,Q^2) - \sigma^{(3\cap 4)}(x,Q^2,m_c^2)\,. \label{eq:sigmaFONLL}
\end{equation}
Here,
\begin{itemize}
    \item $\sigma^{(3)}(x,Q^2,m_c^2)$ represents the massive FFNS result, which contains power-like $Q^2/m_c^2$ terms and a finite amount of $\ln(Q^2/m_c^2)$ associated with the charm in the coefficient function $\vb C^{(3)}(z,Q^2,m_c^2)$
    \item $\sigma^{(4)}(x,Q^2)$ represents the massless ZM-VFNS result, which resums all logarithms $\ln(Q^2/m_c^2)$, including the ones associated with the charm
    \item $\sigma^{(3\cap 4)}(x,Q^2,m_c^2)$ collects all terms, which appear in both expressions simultaneously, and for which we have borrowed the notation from set theory
\end{itemize}
One can show~\cite{Barontini:2024xgu} that the overlap contribution can be written as
\begin{equation}
    \sigma^{(3\cap 4)}(x,Q^2,m_c^2) = \left(\vb C^{(3\cap 4)}(Q^2,m_c^2)\otimes \vb f^{(3)}(Q^2)\right)(x)
\end{equation}
where the corresponding coefficient function is given by
\begin{equation}
    \vb C^{(3\cap 4),T}(z,Q^2,m_c^2) = \left(\vb C^{(4),T}(Q^2)\otimes \vb A^{(4)}(Q^2,m_c^2)\right)(z)\,,
\end{equation}
i.e.\ the right-hand-side of \cref{eq:Cmatching} (without any limit).
It is immediately clear that in the limit $Q^2 \gg m_c^2$ the coefficient functions $\vb C^{(3)}$ and $\vb C^{(3\cap 4)}$ become identical and, thus, in \cref{eq:sigmaFONLL} only $\sigma^{(4)}$ contributes as requested by \cref{box:gmvfns}.
On the other side, since $\sigma^{(3\cap 4)}$ contains exactly the logarithmic terms of $\sigma^{(3)}$ and those are also contained in $\sigma^{(4)}$, which actually only contains logarithms, we also match the $Q^2 \sim m_c^2$ region.
The yet additional logarithms in $\sigma^{(4)}$, which are resummed there, are small in this region and a part of the approximation FONLL takes.

\subsection{Interplay with previous content}
\begin{itemize}
    \item The relevant parameter in \cref{box:gmvfns} is the scale $Q^2$, which, as we explain in \cref{sec:diverse}, is very different between various experiments.
        While for NC DIS at HERA heavy quark mass effects are very relevant and can be up to $\SI{20}{\percent}$~\cite{H1:2015ubc,H1:2018flt} in some kinematic regime, for CC DIS at HERA there are practically irrelevant since $Q^2$ is large, see \cref{tab:HERA}.
        However, when we consider CC DIS at small $Q^2$, as is present e.g.\ in neutrino experiments, they are again very relevant~\cite{Gao:2017kkx,Helenius:2026uuz,Caola:2026kvl}.
    \item At this point we can properly justify to neglect the top quark, since typical virtualities are below the top quark threshold, see \cref{eq:zmvfns}.
    \item The mathematical structure in massless and massive diagrams is very different, e.g.\ the former yields collinear divergencies, see \cref{sec:rsl,sec:nlo}, but the latter regulates them with explicit logarithms $\ln(Q^2/m^2)$, which makes them very different for interpolation grids, \cref{sec:grids}.
        Moreover, the massive diagrams contain far more special functions, which already makes the NLO diagrams no longer analytically solvable but only numerically tractable~\cite{Hekhorn:2019nlf}.
    \item Since the number of active flavors $n_f$ is a direct dependency of the strong coupling $\alpha_s^{(n_f)}$ and PDFs $\vb f^{(n_f)}$ it appears explicitly in scale variations, \cref{sec:sv}.
        While for the renormalization scale variation it reduces to choosing the correct $\beta_0^{(n_f)}$ in \cref{eq:CbarRimpl}, for the factorization scale variations it is more complicated as usual.
        First, it requires choosing the correct $\vb P^{(n_f)}$ in \cref{eq:CbarFNLO}, but second it also requires the correct definition of the evolution basis.
        For example, now we can identify \cref{eq:Singlet} as the four flavor singlet $f_\Sigma^{(4)}$, where as in three flavor we would get
        \begin{equation}
            f_\Sigma^{(3)}(x,Q^2) &= f_d(x,Q^2) + f_{\bar d}(x,Q^2) + f_u(x,Q^2) + f_{\bar u}(x,Q^2) + f_s(x,Q^2) + f_{\bar s}(x,Q^2)
        \end{equation}
        since, e.g., the charm is not an active quark and does not participate in \cref{eq:DGLAP}.
    \item For massive diagrams we now have two scales in hand, the virtuality $Q^2$ and the mass $m^2$, thus the choice of the \enquote{typical} scale of the process is no longer obvious, \cref{sec:sv}.
        While e.g.\ the NC diagrams~\cite{Hekhorn:2019nlf} prefer on their own $\mu_F^2 = Q^2 + 4m^2 = \mu_R^2$, we often choose still $\mu_F^2 = Q^2 = \mu_R^2$ since they are combined with the massless diagrams and we prefer a common scale.
\end{itemize}

\section{Target mass corrections}
\label{sec:TMC}
In all our discussion so far we have implicitly assumed that the target, which often is the proton, see \cref{sec:diverse}, is a massless particle.
However, we can relax that condition and consider a target with finite mass $M$, which yields \enquote{Target Mass Corrections} (TMCs).
We refer to Refs.~\cite{Schienbein:2007gr,Ruiz:2023ozv} for a detailed review on the topic and instead focus here on the practical consequences.

Our master formula, \cref{eq:fact1}, gives the so-called leading twist expression, which is dominant with respect to higher twist contributions, which are summarized as $\order{\LQCD^2/Q^2}$.
While these higher twist terms are in principle not accessible in pQCD as PDFs are by definition leading twist objects, they have recently gained renewed interest by PDF fitting groups~\cite{Cerutti:2025yji,Ball:2025xtj,Harland-Lang:2025wvm}.
TMCs are formally a part of these higher twist contributions, but since they can be computed systematically and consistently they are typically included in modern PDF extractions.

In practice, TMCs amount to a shift in kinematics and a reweighting.
We define
\begin{equation}
    r = \sqrt{1 + \frac{4x^2 M^2}{Q^2}} \label{eq:TMCr}
\end{equation}
and the so-called Nachtmann variable~\cite{Nachtmann:1973mr}
\begin{equation}
    \xi_N = \frac{2x}{1 + r}\,. \label{eq:Nachtmann}
\end{equation}
Eventually, we find for the structure function $F_2$
\begin{equation}
    F_2^\text{TMC}(x,Q^2) = \frac{x^2}{\xi_N^2 r^3} F_2^\text{noTMC}(\xi_N, Q^2) + \frac{6M^2 x^3}{Q^2 r^4}h_2(\xi_N,Q^2) + \frac{12M^4 x^4}{Q^4 r^5}g_2(\xi_N,Q^2) \label{eq:TMC}
\end{equation}
where $F_2^\text{noTMC}$ refers to the setup we have discussed so far and
\begin{align}
    h_2(\xi_N,Q^2) &= \left(\tilde h_2 \otimes F_2^\text{noTMC}(Q^2)\right)(\xi_N) &\text{with}\quad \tilde h_2(z) &= \frac z {\xi_N}\,, \label{eq:TMCh2}\\
    g_2(\xi_N,Q^2) &= \left(\tilde g_2 \otimes F_2^\text{noTMC}(Q^2)\right)(\xi_N) &\text{with}\quad \tilde g_2(z) &= 1 - z\,. \label{eq:TMCg2}
\end{align}
For the other structure functions $F_1$, $F_L$, and $F_3$ one can find similar equations~\cite{Schienbein:2007gr}.
TMCs are relevant for large $x$ and/or small $Q^2$ as follows from \cref{eq:TMCr,eq:Nachtmann}.
Examining \cref{eq:TMC} we note that the structure function at the experimentally measured Bjorken-$x$ corresponds to first approximation to a structure function computed at the Nachtmann variable $\xi_N$.
This is a simple consequence from the reduced phase space, which is available for massive particles.

Note that \cref{eq:TMCh2,eq:TMCg2} are convolutions, i.e.\ they require the evaluation of $F_2^\text{noTMC}$ at intermediate momentum fraction, which can be challenging in practice.
Thus one often applies the following approximation~\cite{Schienbein:2007gr}
\begin{equation}
    F_2^\text{TMC}(x,Q^2) \approx \frac{x^2}{\xi_N^2 r^3} F_2^\text{noTMC}(\xi_N, Q^2) \left(1 + \frac{6 M^2 x \xi_N}{Q^2 r}(1-\xi_N)^2\right)
\end{equation}
which only requires a single evaluation.

TMCs introduce an explicit dependence on the target mass $M$ into \cref{eq:fact1} and capture a consistent part of the higher twist contributions $\order{\LQCD^2/Q^2}$.
TMCs are independent from the requested perturbative accuracy or the exchanged boson since they refer to a kinematic correction and thus apply to every calculation.
While they don't interfere directly with our previous content they must be applied separately on top of everything as a final operation after all other corrections have been taken into account.

Note that \cref{box:DISreq} should read more precisely
\begin{conclusionbox}[label={box:DISreqM}]{Improved DIS requirements}
    We require $Q^2 \gg M^2$ (deep) and $W^2 \gg M^2$ (inelastic) to call a lepton-hadron scattering DIS.
\end{conclusionbox}
\noindent{}We recall again the definition of $W^2$, \cref{eq:defw2}, and notice that a cut in $W^2$ removes precisely the relevant region for TMC of small $Q^2$ and large $x$.
Thus, we are eventually double cautions about this region: first, we exclude data per se to ensure we are in an inelastic scattering and, second, in addition, we apply TMC to our remaining predictions.

\section{PDF schemes}
\label{sec:schemes}
Finally, we consider a (mostly) theoretical issue.
If we look to our master formula, \cref{eq:fact1}, we recall it is a convolution between two unphysical objects: the coefficient function, $\vb C$, and the PDF, $\vb f$.
Since only the convolution, the cross section, $\sigma$, is a physical object, there is some freedom in the definition of $\vb C$ and $\vb f$.
We discuss this freedom in \cref{sec:sv} for factorization scale variation schemes, but here we take a more fundamental point of view: how do we define PDFs in the first place?

This freedom is referred to as PDF schemes and DIS is playing a key role here.
We find \cref{eq:LOC} at LO accuracy and \cref{eq:NLOCq,eq:NLOCg} at NLO in the $\msbar$ scheme, which is the most common PDF scheme and implicitly assumed so far.
In the DIS scheme~\cite{Diemoz:1987xu} we define for the coefficient functions of the structure function $F_2$
\begin{equation}
    C_q^{\mathrm{DIS}}(z,Q^2) = C_{\bar q}^{\mathrm{DIS}}(z,Q^2) = e_q^2 \delta(1-z)\,\forall q \qq{and} C_g^{\mathrm{DIS}}(z,Q^2) = 0 \label{eq:CDIS}
\end{equation}
which holds to all perturbative orders.
Since the structure function has to remain invariant of course, i.e.\
\begin{equation}
    F_2(x,Q^2) = \left(\vb C^{T,\msbar}(Q^2) \otimes \vb f^{\msbar}(Q^2) \right)(x)
        = \left(\vb C^{T,\mathrm{DIS}} \otimes \vb f^{\mathrm{DIS}}(Q^2) \right)(x)\,, \label{eq:F2DIS}
\end{equation}
this implies, first, that also the PDFs are different and, second, as this must be true at any $Q^2$, also the PDF evolutions are different.
In other words: the PDF scheme defines what we call a quark or a gluon.
In practice all modern PDFs are determined in the $\msbar$ scheme and thus the DIS scheme is not of any practical relevance, but it can provide some insights based on theoretical considerations.

Before making explicit use of the DIS scheme, we review the interplay with the previous content.
\begin{itemize}
    \item We need all possible DIS combinations (NC vs.\ CC) to define all quark combinations, \cref{sec:diverse}.
    \item The prescription only defines quarks, but not the gluon, which is explicitly defined to be not contributing to fully inclusive DIS.
        Instead, one needs an additional definition of the gluon distribution, for which various options exist~\cite{Diemoz:1987xu,Soar:2009yh,Altarelli:1998gn,Candido:2023ujx}.
    \item \Cref{eq:F2DIS} explicitly defines $F_2$ and thus any observable other then $F_2$, e.g.\ $F_L$ from \cref{sec:xs}, becomes more complicated, since, e.g.\ in this case the sum has to be invariant.
        Of course this also applies to massive coefficient functions from \cref{sec:hq} and while the general construction, e.g.\ for FONLL, is the same, the ingredient become more complicated.
    \item Splitting functions $\vb P$ depend on the PDF scheme and thus (almost) all complications discussed in this work just appear there, \cref{sec:resum}.
    \item TMCs, \cref{sec:TMC}, are universal kinematic corrections which apply to any leading-twist expression and are independent of the PDF scheme.
        Thus, the naive simplicity of \cref{eq:CDIS} is destroyed when considering TMC, as we must for any real-life DIS comparison.
\end{itemize}

Considering PDF schemes other then the default $\msbar$ scheme can sometimes be advantages~\cite{Delorme:2026vln}, even if just for theoretical arguments as we illustrate next.
The difference between \cref{eq:CpQCD} (implicitly in $\msbar$ scheme) and \cref{eq:CDIS} is, by construction, a perturbative object and thus the scheme transformation between these two schemes is so as well.
In particular, we write
\begin{equation}
    \vb f^{\mathrm{DIS}}(\xi,Q^2) = \left(\vb S^{\mathrm{DIS}\leftarrow\msbar}(Q^2) \otimes \vb f^{\msbar}(Q^2)\right)(\xi)
\end{equation}
with a perturbative operator $\vb S^{\mathrm{DIS}\leftarrow\msbar}(Q^2)$.
In this specific case we simply have
\begin{equation}
    \vb S^{\mathrm{DIS}\leftarrow\msbar}(Q^2) = \left(\vb C^{T,\mathrm{DIS}}(Q^2)\right)^{-1} \vb C^{T,\msbar}(Q^2) \label{eq:DIStrafo}
\end{equation}
where the inversion is to be understood in a functional sense.

Now, since \cref{eq:F2DIS} defines an observable\footnote{or one can generalize it such that it actually does~\cite{Candido:2023ujx}} and \cref{eq:CDIS} only contains positive numbers, this implies that $\vb f^{\mathrm{DIS}}(Q^2)$ must also be positive.
Recall that cross section are physical objects and thus positive by definition, PDFs are not physical objects and their positivity is non-trivial in general.
In the DIS scheme, the two objects (quasi) coincide and thus the positivity of DIS PDFs follows.
However, together with the perturbativity property of $\vb S(Q^2)$ also the positivity of $\vb f^{\msbar}$ follows as long as $\vb S(Q^2)$ is actually perturbative~\cite{Candido:2023ujx}.
Checking \cref{eq:DIStrafo} explicitly yields the condition that $\msbar$ PDFs should be positive above $Q^2 \gtrsim \SI{5}{\GeV^2}$, which has practical consequences for PDF fitting.

\section{Summary}
\label{sec:summary}
Fully inclusive DIS cross sections depend on
\begin{itemize}
    \item the experimentally measured kinematic variables, \cref{sec:xs}, $x$, $y$, $Q^2$.
        They are not independent from each other, but linked through the invariant mass of the electron-proton system $s_l$, \cref{eq:defslep}, given by the experiment.
    \item the properties of the electro-weak bosons, \cref{sec:diverse}.
        This includes basically all properties, such as the couplings to both leptons and quarks, specifically the electric charges and the electro-weak charges, or their respective masses.
    \item the discretization of interpolation grids, \cref{sec:interpolation}.
        In practice it is important to ensure the introduced error is negligible with respect to all other sources.
    \item the specific solution method for the RGE of the strong coupling and the PDFs, \cref{sec:resum}.
        Although all solutions must be perturbatively equivalent, in practice they are not numerically identical.
    \item the specific choice of renormalization and factorization scale as well as the applied factorization scale variation scheme, \cref{sec:sv}.
        Although cross section do not depend perturbatively on any of the unphysical scales, they still do so in practice or in other words: they depend on the chosen amount of higher order terms.
    \item the heavy quark masses and the adopted FNS, \cref{sec:hq}.
    \item the target mass and the adopted approximation (if any), \cref{sec:TMC}.
    \item the PDF scheme, \cref{sec:schemes}.
        Although cross section do not depend perturbatively on the PDF scheme, they still do so numerically.
        In practice, we only use the $\msbar$ scheme.
\end{itemize}

This means in turn coefficient functions depend on
\begin{itemize}
    \item the partonic momentum fraction $z$.
    \item the structure function $F_2$, $F_L$, or $F_3$, \cref{sec:xs}.
    \item the perturbative order, \cref{sec:xs}.
    \item the exchanged boson, i.e.\ NC vs.\ CC, \cref{sec:diverse}.
    \item the unique combination of partonic channel, e.g.\ up, down, gluon, etc., and quark-boson coupling, e.g.\ $\qty(g_V^{\gamma q})^2=e_q^2$, $g_V^{\gamma q}g_V^{Z q}$, etc., \cref{sec:nlo}.
    \item the RSL representation, \cref{sec:interpolation}.
    \item the adopted FNS and so potentially on the ratio $Q^2/m^2$, \cref{sec:hq}.
\end{itemize}
Since interpolation (\cref{sec:interpolation}), scale variations (\cref{sec:sv}), and TMC (\cref{sec:TMC}), eventually can be applied a posteriori and on a global level, we can disentangle them from the coefficient functions.

When performing a real-life PDF extraction all these dependencies must be taken into account and \texttt{Yadism}~\cite{Candido:2024rkr,barontini_2026_18758473} provides one specific implementation thereof.

Finally, we can consider the generalizations of the various features:
\begin{itemize}
    \item \cref{sec:diverse} is a reminder that we need to consider all particles or all diagrams in pQCD, see \cref{box:pQCD}.
        The distinction between NC and CC repeats also for the Drell-Yan process at hadron colliders: $h_1 + h_2 \to l + l' + X$.
    \item The RSL representation from \cref{sec:rsl} are a general math property of distributions.
    \item While interpolation grids from \cref{sec:grids} are usually a bonus for fully inclusive DIS, they are a must for practically all other pQCD processes as, typically, they are much more challenging.
    \item The evolution equations from \cref{sec:resum} are general pQCD results and, in particular, the strong coupling $\alpha_s$ and PDFs $\vb f$ are a fundamental part of any pQCD calculation and even beyond.
    \item Scale variation from \cref{sec:sv} are a general pQCD feature, which can and must be applied to any calculation using either the strong coupling $\alpha_s$ or PDFs $\vb f$.
    \item FNSs from \cref{sec:hq} are a general pQCD feature, which must be addressed before computing any diagram.
        The phenomenological impact of choosing a \enquote{good} FNS depends on the considered observable and, in particular, on the relevant scale $\mu^2$.
        Many measurements at the LHC are performed at $\mu = m_Z$ and we can safely assume $n_f=5$ massless quarks.
        However, if a small scale is present, such as e.g.\ in B-meson production~\cite{Cacciari:2012ny,Helenius:2023wkn}, the question becomes non-trivial again.
\end{itemize}

  \chapter{More advanced topics}
\label{chap:outlook}

\chapterquote{Lass die Leute reden und lächle einfach mild,\\
die meisten Leute haben ihre Bildung aus einer normalen Quantenfeldtheorievorlesung}%
{Leser}

As an outlook we briefly discuss some more advanced topics, which are related to DIS.
The list is not exhaustive and the order is arbitrary.

\section[gamma5]{$\gamma_5$}
\label{sec:g5}
When introducing the Dirac matrices $\gamma_\mu$ in QFT, one often also introduces the additional matrix $\gamma_5$ along side, since together they can form a basis of the four-dimensional Dirac space, which is needed for the Dirac equation.
In the naive picture $\gamma_5$ anti-commutes with the normal matrices $\gamma_\mu$, i.e.\
\begin{equation}
    \{\gamma_5,\gamma_\mu\} = \gamma_5 \gamma_\mu + \gamma_\mu \gamma_5 = 0 \label{eq:g5ac}
\end{equation}
it obeys
\begin{equation}
    \tr(\gamma_\mu\gamma_\nu\gamma_\rho\gamma_\sigma\gamma_5) = \varepsilon^{\mu\nu\rho\sigma} \label{eq:g5def}
\end{equation}
and, in addition, when computing traces we often use explicitly the cyclicity of traces, i.e.\
\begin{equation}
    \tr(\gamma_\mu \gamma_\nu \ldots \gamma_\rho) = \tr(\gamma_\nu \ldots \gamma_\rho \gamma_\mu) \label{eq:g5tr}
\end{equation}
However, in dimensional regularization, which is the most common regularization scheme applied in pQCD, those three properties, \cref{eq:g5ac,eq:g5def,eq:g5tr}, can no longer be satisfied at the same time.
This implies a specific prescription needs to be imposed, when dealing with $\gamma_5$ and not only are there several algorithms available, but also they all require special care~\cite{Gnendiger:2017pys}.
Note that the Levi-Civita tensor $\varepsilon^{\mu\nu\rho\sigma}$ is directly related to $\gamma_5$, \cref{eq:g5def}, and thus needs also special attention.
Unfortunately, $\gamma_5$ is explicitly present when considering the NC or CC DIS in the couplings of the weak bosons, $Z$ and $W$.
Moreover, also the introduction of massive quarks explicitly breaks certain symmetries which are present in the massless case~\cite{Hekhorn:2019nlf}.
Since \cref{app:LO} only considers LO, we do not need to worry about these problems.

On the other side, by considering $\gamma_5$ we gain explicitly access to the helicity of a parton, which opens a whole new world: polarized DIS.
In fact we can rerun the arguments of this work completely in parallel to the discussion so far.
\begin{enumerate}
    \item \cref{chap:intro}: polarized DIS gives access to polarized PDFs~\cite{Cruz-Martinez:2025ahf,Nocera:2026est}, which give a different view on hadrons.
    \item \cref{sec:diverse}: various DIS experiments have been run, are run, and will be run. They have various structure functions and both, NC and CC, variants.
    \item \cref{sec:interpolation}: except that all coefficient functions are different, the mathematical structure is of course identical.
    \item \cref{sec:resum}: while $\alpha_s$ is of course universal, polarized PDF obey their own RGE.
    \item \cref{sec:sv}: once we account for the different splitting functions, scale variations remain the same.
    \item \cref{sec:hq}: the ambiguity about the number of light flavors is universal, but the FNS can also here be applied in the same way, e.g.\ FONLL~\cite{Hekhorn:2024tqm}.
    \item \cref{sec:TMC}: the actual TMC kernels need to be re-derived, but the structure remains the same.
\end{enumerate}

\section{Higher orders}
When computing higher order coefficient functions, the number of required diagrams grows very quickly.
This often implies the use of dedicated mathematical and numerical tools.
The currently highest available perturbative order is N$^3$LO, which is indeed the backbone of current approximate N$^3$LO PDF extractions~\cite{NNPDF:2024nan,McGowan:2022nag}.
However, for the massive coefficient functions currently only approximate expressions exist~\cite{Barontini:2026drp}, which implies in a PDF fit we also need to consider Incomplete Higher Order Uncertainties (IHOU) reflecting the uncertainty of this approximation.
Eventually, the experimental precision of the available and future DIS data requires the use of N$^3$LO perturbation theory or, turning the argument around, if we aim for PDFs with percent uncertainties~\cite{NNPDF:2021njg} we need N$^3$LO perturbation theory.
Actually, massless DIS is often the first QCD process becoming available at the next perturbative order~\cite{Basdew-Sharma:2021jhx}.
While all possible partonic channels are already available starting from NNLO, N$^3$LO still brings new coupling combinations into the calculations.

\section{Electroweak corrections}
While increasing the perturbative order in QCD, which is all what we discussed so far, we might also wonder about electroweak corrections.
Allowing the Z boson to be exchanged in parallel or instead of the photon is considered a trivial electroweak correction and, indeed, required to match the experimental data.
Rather, here we are asking about additional electroweak bosons to take part in the underlying diagrams.

However, it turns out that the traditional DIS theory setup, which starts by defining structure functions, \cref{sec:xs}, can not deal with these corrections as can be seen from \cref{fig:DIS}.
If we allow for additional electroweak boson, they may also be exchanged between the leptonic part of the diagram and the hadronic part of the diagram, which implies that we can no longer associate $q$, \cref{eq:defq}, with the momentum of a single vector boson.
Thus we can no longer define neither a leptonic tensor $L_{\mu\nu}$ or a hadronic tensor $W_{\mu\nu}$ and thus no structure functions.

However, it is of course still possible to compute these diagrams and thus to compute the cross section $\sigma$.
The price we have to pay is, we need to give up on our master equation, \cref{eq:fact1}, and use a more complex factorization formula~\cite{Liu:2020rvc,Liu:2021jfp,Qiu:2026fed}, which now involves not only a non-perturbative distribution function for the hadron (the PDF $\vb f$), but also for the lepton.

\section{Photon DIS}
We can add one more particle to our list of targets: the photon.
Quasi-real photons, i.e.\ photons with a small virtuality, contain a non-perturbative hadronic component, e.g.\ quarks in real photons, which can be described by a Photon PDF and which can be determined in Photon DIS~\cite{Chithirasreemadam:2026mqp}.
In this setup a highly virtual photon, \cref{box:DISreqM}, probes a quasi-real photon.
These experiments are typically performed at electron-positron colliders, where each lepton provides one of the photons.
Otherwise we can choose a particular factorization scheme for the Photon PDF, such that the remaining discussion in this work can be directly applied, but we have to account for the modified evolution of Photon PDFs compared to normal hadrons.

~

\end{mainmatter}

\begin{appendices}
  \chapter{Theory exercises}

\section{LO DIS}
\label{app:LO}
\begin{figure}
    \includegraphics[width=.5\textwidth]{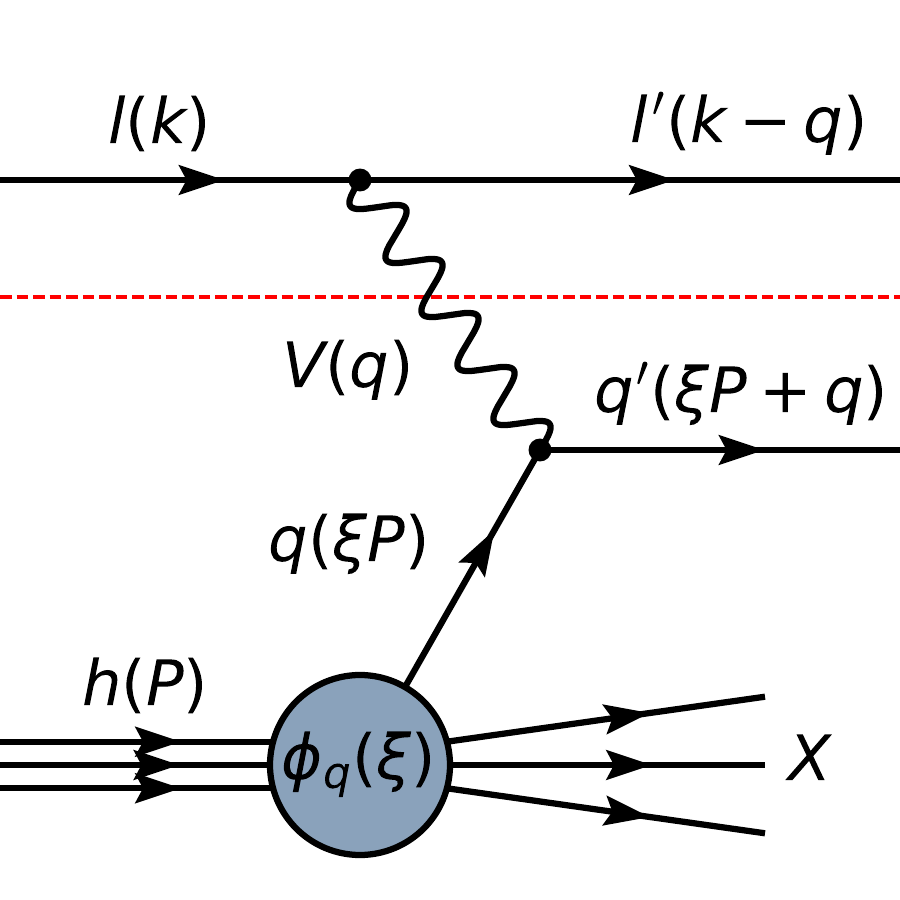}
    \caption{LO DIS Feynman diagram: $l + h \to l' + X$}
    \label{fig:LODIS}
\end{figure}

We are recomputing the LO DIS cross section depicted in \cref{fig:LODIS} in full generality.
As we say in \cref{sec:xs} we can decompose the full cross section into a leptonic tensor $L_{\mu\nu}$ and a hadronic tensor $W_{\mu\nu}$ where the red line indicates the split.
In the collinear framework it is sufficient to compute the partonic (sub-)diagram $l + q \to l' + q'$.
The parton-in-hadron amplitude $\phi_q$ combines upon taking the modulus square of the whole amplitude together with the inclusiveness $X$ into the PDF $f_q(\xi)$.
This leaves us essentially with twice \cref{fig:LO}, once for the leptonic side and once for the quark side.
Since the fermion lines are distinct we can compute them one at a time and if we keep them general enough we only need to do it once.
Note that we first compute the lepton tensor, i.e.\ we keep the vector boson coupling Lorentz-index open and different in the amplitude and its complex conjugate.

In the first step we only deal with the Dirac structures.
We need the following Feynman rules
\begin{itemize}
    \item outgoing fermion $\bar u(k')$
    \item vector boson-fermion interaction $\Gamma_\mu$, which we make concrete below
    \item incoming fermion $u(k)$
\end{itemize}
and we assume massless fermions in this exercise.
Thus we have
\begin{equation}
    \text{\cref{fig:LO}} = \mathcal A_\mu = \bar u(k')\Gamma_\mu u(k)
\end{equation}
where we have suppressed any additional dependency and we need to compute
\begin{equation}
    T_{\mu\nu} = A_\mu A_\nu^* \,.
\end{equation}

We recall some relevant Dirac traces:
\begin{align}
    \tr(\mathbbm 1) &= 4 \\
    \tr(\gamma_\mu) &= 0 \\
    \tr(\gamma_\mu\gamma_\nu) &= 4g_{\mu\nu} \\
    \tr(\gamma_\mu\gamma_\nu\gamma_\rho) &= 0 \\
    \tr(\gamma_\mu\gamma_\nu\gamma_\rho\gamma_\sigma) &= 4(g_{\mu\nu}g_{\rho\sigma} - g_{\mu\rho}g_{\nu\sigma} + g_{\mu\sigma}g_{\nu\rho}) \label{eq:trDirac4}
\end{align}
We also need the simple anti-commutating $\gamma_5$, with properties
\begin{equation}
    \gamma_5 = \frac i {4!}\varepsilon^{\mu\nu\rho\sigma}\gamma_\mu\gamma_\nu\gamma_\rho\gamma_\sigma, \quad \gamma_5\gamma_\mu = - \gamma_\mu\gamma_5, \quad \gamma_5\gamma_5 = \mathbbm 1
\end{equation}
and the corresponding Dirac traces
\begin{equation}
    \tr(\gamma_5) = \tr(\gamma_\mu\gamma_5) = \tr(\gamma_\mu\gamma_\nu\gamma_5) = \tr(\gamma_\mu\gamma_\nu\gamma_\rho\gamma_5) = 0
\end{equation}
and
\begin{equation}
    \tr(\gamma_\mu\gamma_\nu\gamma_\rho\gamma_\sigma\gamma_5) = 4i\varepsilon_{\mu\nu\rho\sigma}\,. \label{eq:trDirac5}
\end{equation}

We start by assuming $\Gamma_\mu = \gamma_\mu$ and averaging over fermion spins $s$ upon which the spinors collapse to just, e.g., $\slashed{k}$.
\begin{exercisebox}[label=box:exLOT1]{Step 1}
    Proof
    \begin{align}
        T^1_{\mu\nu} &= \frac 1 2 \sum_{s} \bar u^{(s)}(k')\gamma_\mu u^{(s)}(k) \left(\bar u^{(s)}(k')\gamma_\nu u^{(s)}(k)\right)^*\\
         &= 2\left(k_\mu k'_\nu + k_\nu k'_\mu - k \cdot k' g_{\mu\nu}\right)
    \end{align}
\end{exercisebox}
\noindent{}$T^1_{\mu\nu}$ becomes the spin-averaged leptonic tensor $L_{\mu\nu}$ for photons when choosing the lepton kinematics.

Next, we give up on averaging and instead consider fermions with definite helicities\footnote{Helicity is more convenient than spin in this context}.
For spinors with a given helicity $\lambda = \pm 1$ we find
\begin{equation}
    u(k,\lambda) \bar u(k,\lambda) = \frac 1 2 \slashed k (1 - \lambda \gamma_5) \,.
\end{equation}
\begin{exercisebox}[label=box:exLOT2]{Step 2}
    Proof
    \begin{align}
        T^2_{\mu\nu} &= \bar u(k',\lambda)\gamma_\mu u(k,\lambda) \left(\bar u(k',\lambda)\gamma_\nu u(k,\lambda)\right)^*\\
         &= T^1_{\mu\nu} + 2 i \lambda \varepsilon_{\mu\nu\rho\sigma} k^\rho k'^\sigma
    \end{align}
\end{exercisebox}
\noindent{}$T^2_{\mu\nu}$ becomes the leptonic tensor $L_{\mu\nu}^\gamma$ for photons when choosing the lepton kinematics.

Next, we want to use a more realistic boson-fermion coupling $\Gamma_\mu$, since eventually we need to consider the interference between a purely vectorial boson, such as the photon, and the $Z$ boson, which has a mixture between a vectorial coupling and an axial vectorial coupling,
\begin{equation}
    \Gamma^{ph}_\mu = g_V^{\gamma f}\gamma_\mu\,, \qquad \Gamma^Z_\mu = \gamma_\mu(g_V^{Z f} - g_A^{Z f} \gamma_5)
\end{equation}
where $g_V^{\gamma f}$ is the coupling strength between the photon and the fermion $f$ and $g_V^{Z f}$ and $g_A^{Z f}$ are the vectorial and axial-vectorial coupling strength between the Z boson and the fermion $f$.
By convention we exclude an explicit factor of the elementary electric charge from the definition of $g_{V,A}^{B f}$.
$T^2_{\mu\nu}$ obtains a trivial factor of $(g_V^{\gamma f})^2$ when using $\Gamma^{ph}_\mu$.
\begin{exercisebox}[label=box:exLOT3]{Step 3}
    Proof
    \begin{align}
        T^3_{\mu\nu} &= \bar u(k',\lambda)\Gamma^Z_\mu u(k,\lambda) \left(\bar u(k',\lambda)\Gamma^{ph}_\nu u(k,\lambda)\right)^*\\
         &= g_V^{\gamma f}(g_V^{Z f} - \lambda g_A^{Z f})T^2_{\mu\nu}
    \end{align}
\end{exercisebox}
\noindent{}$T^3_{\mu\nu}$ becomes the leptonic tensor $L_{\mu\nu}^{\gamma Z}$ for the photon-Z interference when choosing the lepton kinematics and couplings.

Finally, we need the pure Z boson contributions.
\begin{exercisebox}[label=box:exLOT4]{Step 4}
    Proof
    \begin{align}
        T^4_{\mu\nu} &= \bar u(k',\lambda)\Gamma^Z_\mu u(k,\lambda) \left(\bar u(k',\lambda)\Gamma^{Z}_\nu u(k,\lambda)\right)^*\\
         &= (g_V^{Z f} - \lambda g_A^{Z f})^2 T^2_{\mu\nu}
    \end{align}
\end{exercisebox}
\noindent{}$T^4_{\mu\nu}$ becomes the leptonic tensor $L_{\mu\nu}^{Z}$ for the Z bosons when choosing the lepton kinematics and couplings.

The leptonic tensor $L_{\mu\nu}^{W}$ for the W bosons follows from the Z boson case by choosing appropriate couplings $g_{V}^{W f}=g_{A}^{W f}$.

The lepton kinematics are obtained by choosing the incoming momentum $k$ and the the outgoing momentum $k' = k - q$, \cref{fig:LODIS}.
\begin{exercisebox}[label=box:exLOWard]{Step 5: Ward identity}
    Proof
    \begin{equation}
        L^b_{\mu\nu} q^\mu = 0 = L^b_{\mu\nu} q^\nu \quad b \in\{\gamma, \gamma Z, Z, W\}
    \end{equation}
\end{exercisebox}

\begin{exercisebox}[label=box:exLOg]{Step 6}
    Proof
    \begin{equation}
        L^\gamma_{\mu\nu} g^{\mu\nu} = -2 Q^2
    \end{equation}
\end{exercisebox}

We now turn our attention to the hadronic side, which is essentially identical except the fermion is now the quark and we have to choose different incoming and outgoing momenta.
We define $p = \xi P$ and $T^2_{\mu\nu}$ becomes the photon-quark tensor $\hat W^{\gamma q}_{\mu\nu}$ by choosing the incoming momentum $p$ and the outgoing momentum $p'=p+q$, \cref{fig:LODIS}.
\begin{exercisebox}[label=box:exLOPP]{Step 7}
    Proof
    \begin{equation}
        L^\gamma_{\mu\nu} P^{\mu} P^{\nu} = \frac{Q^4}{x^2 y^2}(1-y)
    \end{equation}
\end{exercisebox}
\begin{exercisebox}[label=box:exLOLW]{Step 8}
    Proof
    \begin{equation}
        L^\gamma_{\mu\nu} \hat W^{\mu\nu}_{\gamma q} = 2\qty(g_{V}^{\gamma q})^2\frac{Q^4}{z^2 y^2}\left(2-2y + z y^2\right) \label{eq:LOsigma}
    \end{equation}
    if we average over helicities.
\end{exercisebox}
The phase space, which we have neglected so far, contains the overall momentum conservation delta function, which, eventually, yields the $\delta(1-z)$ of \cref{eq:LOC}.
Effectively, this set $z=1$ in \cref{eq:LOsigma} and thus it is matching to the prefactor of $F_2$ in \cref{eq:dsigmadxdy}.
All remaining factors to obtain the full expression for the (double differential) cross sections are given by considering all factors arising from phase space and other global prefactors, e.g.\ the flux.
In the language of coefficient functions and structure function, which we use in the rest of this work, only the factors $\qty(g_{V}^{\gamma q})^2 = e_q^2$ in \cref{eq:LOsigma} are accounted for in the LO coefficient function for $F_2$ given by \cref{eq:LOC} (and not elsewhere).

\section{RSL}
\label{app:RSL}
\begin{exercisebox}[label=box:RSLD1]{RSL representation}
    Proof
    \begin{equation}
        C(z) = \left(\frac{\ln(1-z)}{1-z}\right)_+ \,\leftrightarrow\, C^R(z) = 0, C^S(z) = \frac{\ln(1-z)}{1-z}, C^L(x) = \frac{\ln(1-x)^2}{2}
    \end{equation}
\end{exercisebox}
\noindent{}Hint: read \cref{sec:rsl}.

\begin{exercisebox}[label=box:rslprod1]{Product of functions 1}
    Give the RSL representation of the product a regular function $r(z)$ and a +-distribution $\qty(p(z))_+$, $C(z) = r(z)\qty[p(z)]_+$
\end{exercisebox}
\noindent{}A popular example found in many text books is
\begin{equation}
    P_{qq}^{(0)}(y) = C_F\left( \frac{1+y^2}{(1-y)_+} + \frac 3 2 \delta(1-y)\right) \,\forall q \label{eq:Pqq}
\end{equation}
where the first term has the requested structure, i.e.\ the +-distribution is only intended for the denominator.

\begin{exercisebox}[label=box:rslprod2]{Product of functions 2}
    Proof
    \begin{equation}
        \left[r(z)p(z)\right]_+ = r(z) \left[p(z)\right]_+ - \delta(1-z) \int\limits_0^1 \! \dd y ~ r(y) \left[p(y)\right]_+
    \end{equation}
    where again $r(z)$ is a regular function and $\qty[p(z)]_+$ is a +-distribution.
\end{exercisebox}
\noindent{}Check that \cref{eq:Pqq} can be also written as
\begin{equation}
    P_{qq}^{(0)}(y) = C_F\left( \frac{1+y^2}{1-y}\right)_+ \,.
\end{equation}
Another (lengthy) example is \cref{eq:NLOCq} vs.\ \cref{eq:NLOCqp}.

\section{Resummation}
\begin{exercisebox}[label={box:LOalphas}]{LO $\alpha_s$ evolution 1}
    Proof \cref{eq:LOalphas} can be written as
    \begin{equation}
        \alpha_s^\text{LO}(\mu_R^2) = \frac 1 {\beta_0 \ln(\mu_R^2/\LQCD^2)}
    \end{equation}
    and given an explicit expression for $\LQCD$.
\end{exercisebox}

Before we go to NLO let us use the more practical notation:
\begin{exercisebox}[label={box:LOalphassol}]{LO $\alpha_s$ evolution 2}
    Show
    \begin{equation}
        \alpha_s^\text{LO}(t_R) = \frac{\alpha_s(0)}{1 + \alpha_s(0) \beta_0 t_R} \label{eq:LOalphast}
    \end{equation}
    (i.e.\ \cref{eq:LOalphas} recast in terms of $t_R = \ln(\mu_R^2/\mu_{R,0}^2)$) is a solution to
    \begin{equation}
        \dv {\alpha_s^\text{LO}(t_R)}{t_R} = -\beta_0 \left(\alpha_s^\text{LO}(t_R)\right)^2
    \end{equation}
    (i.e.\ \cref{eq:beta} at LO recast in terms of $t_R$).
\end{exercisebox}

\begin{exercisebox}[label={box:NLOalphas}]{NLO $\alpha_s$ evolution 1}
    Show
    \begin{equation}
        \alpha_s^\text{NLO}(t_R) = \alpha_s^\text{LO}(t_R) - \frac{\beta_1}{\beta_0} \left(\alpha_s^\text{LO}(t_R)\right)^2 \ln(\alpha_s(0) / \alpha_s^\text{LO}(t_R)) \label{eq:NLOalphast}
    \end{equation}
    is a solution to
    \begin{equation}
        \dv {\alpha_s^\text{NLO}(t_R)}{t_R} = -\beta_0 \left(\alpha_s^\text{NLO}(t_R)\right)^2 -\beta_1 \left(\alpha_s^\text{NLO}(t_R)\right)^3 \label{eq:NLObetat}\,
    \end{equation}
    (i.e.\ \cref{eq:NLObeta} recast in terms of $t_R$) at NLO accuracy, i.e.\ the full solution (which is not known analytically) and $\alpha_s^\text{NLO}(t_R)$ differ by higher order terms.
\end{exercisebox}
\noindent{}Make two steps:
\begin{enumerate}
    \item compute the left-hand-side of \cref{eq:NLObetat} using \cref{eq:NLOalphast}
    \item compute the right-hand-side of \cref{eq:NLObetat} using \cref{eq:NLOalphast}
\end{enumerate}
Show the two sides differ by $\order{\qty(\alpha_s(t_R))^4}$, i.e.\ they agree up to the required accuracy.

\begin{exercisebox}[label={box:NLOalphasLL}]{NLO $\alpha_s$ evolution 2}
    Show that \cref{eq:NLObetat} performs a next-to-leading logarithmic (NLL) resummation.
\end{exercisebox}
\noindent{}The leading logarithmic (LL) resummation is visible in \cref{eq:LOalphasresum}, where for each logarithm there is the adjacent coupling, in other words: for each coupling $\alpha_s(\mu_{R,0})^k$, there is one $\ln^k(\mu_R^2/\mu_{R,0}^2)$ or, equivalently, one $t_R^k$ per $\alpha_s(0)^k$.
At NLL we must in addition resum also terms $\alpha_s(0)^k t_R^{k-1}$, i.e.\ where the logarithm has one power less.
Note that \cref{eq:LOalphasresum} has an overall factor of $\alpha_s(\mu_{R,0}^2)$, which we neglect here since the resummation happens for the solution operator $T$, which we can define in an analogous way to the EKO, \cref{eq:EKO},
\begin{equation}
    \alpha_s(\mu_R^2) = T(\mu_R^2 \leftarrow \mu_{R,0}^2) \alpha_s(\mu_{R,0}^2)\,.
\end{equation}

Make several steps:
\begin{enumerate}
    \item Expand $\alpha_s^\text{LO}(t_R)$ in $t_R$ up to and including $\order{t_R^2}$ using \cref{eq:LOalphasresum}
    \item Insert this expansion into \cref{eq:NLOalphast} and expand $\alpha_s^\text{NLO}(t_R)$ to the same accuracy. Recall $\ln(1 + w \epsilon) = w \epsilon - \frac 1 2 w^2 \epsilon^2 + \order{\epsilon^3}$.
    \item Independently, we make a Taylor expansion of $\alpha_s^\text{NLO}(t_R)$ up to and including $\order{t_R^2}$.
        The first derivative is \cref{eq:NLObetat}.
        Compute the second derivative using \cref{eq:NLObetat}.
    \item Check that the two expression agree at LL, $\alpha_s(0)^k t_R^{k}$ for $k=0,1,2$
    \item Check that the two expression agree at NLL, $\alpha_s(0)^k t_R^{k-1}$ for $k=1,2,3$
    \item Check that the two expression start to disagree at next-to-NLL (NNLL), $\alpha_s(0)^k t_R^{k-2}$ for $k=2,3,4$
\end{enumerate}

\chapter{Coding exercise}
We provide an alternative coding exercise at
\begin{center}
    \url{https://github.com/felixhekhorn/jss-2026/blob/main/coding-exercise.py}
\end{center}
which actually implements the simple case discussed in \cref{chap:intro} and compares it to real HERA data.

Adjust the code in the mentioned places to get a real life DIS comparison.
The necessary Python packages are mentioned at the top of the script and note that you need in addition LHAPDF~\cite{Buckley:2014ana}.

We provide a possible solution at
\begin{center}
    \url{https://github.com/felixhekhorn/jss-2026/blob/main/coding-exercise-solution.py}
\end{center}

\chapter{Solution theory exercises}
\section{LO DIS}
\paragraph{Solution for \cref{box:exLOT1}}
\begin{align}
    T^1_{\mu\nu} &= \frac 1 2 \sum_{s} \bar u^{(s)}(k')\gamma_\mu u^{(s)}(k) \left(\bar u^{(s)}(k')\gamma_\nu u^{(s)}(k)\right)^*\\
        &= \frac 1 2 \tr(\gamma_\mu \slashed{k} \gamma_\nu \slashed{k'})\\
        &= 2\left(k_\mu k'_\nu + k_\nu k'_\mu - k \cdot k' g_{\mu\nu}\right)
\end{align}
where we use \cref{eq:trDirac4} in the second step.

\paragraph{Solution for \cref{box:exLOT2}}
\begin{align}
    T^2_{\mu\nu} &= \bar u(k',\lambda)\gamma_\mu u(k,\lambda) \left(\bar u(k',\lambda)\gamma_\nu u(k,\lambda)\right)^*\\
        &=\frac 1 {2^2} \tr(\gamma_\mu \slashed{k}(1-\lambda \gamma_5) \gamma_\nu \slashed{k'} (1-\lambda \gamma_5))\\
        &=\frac 1 {2^2} \left(\tr(\gamma_\mu \slashed{k}\gamma_\nu \slashed{k'}) - \lambda\tr(\gamma_\mu \slashed{k}\gamma_5 \gamma_\nu \slashed{k'} + \gamma_\mu \slashed{k}\gamma_\nu \slashed{k'} \gamma_5) + \lambda^2\tr(\gamma_\mu \slashed{k}\gamma_5 \gamma_\nu \slashed{k'} \gamma_5)\right)\\
        &=\frac 1 {2^2} \left(\tr(\gamma_\mu \slashed{k}\gamma_\nu \slashed{k'}) - \lambda\tr(\gamma_\mu \slashed{k}\gamma_\nu \slashed{k'}\gamma_5  + \gamma_\mu \slashed{k}\gamma_\nu \slashed{k'} \gamma_5) + \lambda^2\tr(\gamma_\mu \slashed{k}\gamma_\nu \slashed{k'} \gamma_5 \gamma_5)\right) \label{eq:solT2ac} \\
        &=\frac 1 {2} \left(\tr(\gamma_\mu \slashed{k}\gamma_\nu \slashed{k'}) - \lambda\tr(\gamma_\mu \slashed{k}\gamma_\nu \slashed{k'}\gamma_5 )\right) \label{eq:solT2sq}\\
        &= T^1_{\mu\nu} + 2 i \lambda \varepsilon_{\mu\nu\rho\sigma} k^\rho k'^\sigma
\end{align}
where
\begin{itemize}
    \item in \cref{eq:solT2ac} we use the anti-commutativity of $\gamma_5$
    \item in \cref{eq:solT2sq} we use the fact that $\lambda^2$ and $(\gamma_5)^2$ are unity (in their appropriate space)
    \item in the last line we use \cref{eq:trDirac5}
\end{itemize}

\paragraph{Solution for \cref{box:exLOT3}}
\begin{align}
    T^3_{\mu\nu} &= \bar u(k',\lambda)\Gamma^Z_\mu u(k,\lambda) \left(\bar u(k',\lambda)\Gamma^{ph}_\nu u(k,\lambda)\right)^*\\
        &=\frac 1 {2^2} \tr(\gamma_\mu(g_V^{Z f}- g_A^{Z f}\gamma_5) \slashed{k}(1-\lambda \gamma_5) g_V^{\gamma f}\gamma_\nu \slashed{k'} (1-\lambda \gamma_5))\\
        &=\frac 1 {2} g_V^{\gamma f}\tr(\gamma_\mu(g_V^{Z f}- g_A^{Z f}\gamma_5) \slashed{k} \gamma_\nu \slashed{k'} (1-\lambda \gamma_5)) \label{eq:solT3T2}\\
        &= g_V^{\gamma f}\left(g_V^{Z f}T^2_{\mu\nu} - g_A^{Z f}\tr(\gamma_\mu\gamma_5\slashed{k}\gamma_\nu \slashed{k'} (1-\lambda \gamma_5))\right)\\
        &= g_V^{\gamma f}\left(g_V^{Z f}T^2_{\mu\nu} - g_A^{Z f}\tr(\gamma_\mu\slashed{k}\gamma_\nu \slashed{k'} (\lambda - \gamma_5))\right) \label{eq:solT3ac}\\
        &= g_V^{\gamma f}\left(g_V^{Z f}T^2_{\mu\nu} - \lambda g_A^{Z f}\tr(\gamma_\mu\slashed{k}\gamma_\nu \slashed{k'} (1-\lambda\gamma_5))\right) \label{eq:solT3sq}\\
        &= g_V^{\gamma f}(g_V^{Z f} - \lambda g_A^{Z f})T^2_{\mu\nu}
\end{align}
where
\begin{itemize}
    \item in \cref{eq:solT3T2} we use \cref{eq:solT2sq}. Note that the manipulations work for any Dirac matrix $\Gamma_\mu$ there
    \item in \cref{eq:solT3ac} we use the anti-commutativity of $\gamma_5$
    \item in \cref{eq:solT3sq} we use the fact that $\lambda^2$ is unity
\end{itemize}

\paragraph{Solution for \cref{box:exLOT4}}
\begin{align}
    T^4_{\mu\nu} &= \bar u(k',\lambda)\Gamma^Z_\mu u(k,\lambda) \left(\bar u(k',\lambda)\Gamma^{Z}_\nu u(k,\lambda)\right)^*\\
        &=\frac 1 {2^2} \tr(\gamma_\mu(g_V^{Z f}- g_A^{Z f}\gamma_5) \slashed{k}(1-\lambda \gamma_5)\gamma_\nu (g_V^{Z f}- g_A^{Z f}\gamma_5) \slashed{k'} (1-\lambda \gamma_5))\\
        &= g_V^{Z f}/g_V^{\gamma f} T^3_{\mu\nu} - g_A^{Z f}\tr(\gamma_\mu(g_V^{Z f}- g_A^{Z f}\gamma_5) \slashed{k}(1-\lambda \gamma_5)\gamma_\nu \gamma_5 \slashed{k'} (1-\lambda \gamma_5))\\
        &= g_V^{Z f}/g_V^{\gamma f} T^3_{\mu\nu} - g_A^{Z f}\tr(\gamma_\mu(g_V^{Z f}- g_A^{Z f}\gamma_5) \slashed{k}(1-\lambda \gamma_5)\gamma_\nu \slashed{k'} (\lambda - \gamma_5))\label{eq:solT4ac}\\
        &= g_V^{Z f}/g_V^{\gamma f} T^3_{\mu\nu} - \lambda g_A^{Z f}\tr(\gamma_\mu(g_V^{Z f}- \lambda g_A^{Z f}\gamma_5) \slashed{k}(1-\lambda \gamma_5)\gamma_\nu \slashed{k'} (1 - \lambda\gamma_5))\label{eq:solT4sq}\\
        &= (g_V^{Z f} - \lambda g_A^{Z f})^2 T^2_{\mu\nu}
\end{align}
where
\begin{itemize}
    \item in \cref{eq:solT4ac} we use the anti-commutativity of $\gamma_5$
    \item in \cref{eq:solT4sq} we use the fact that $\lambda^2$ is unity
\end{itemize}

\paragraph{Solution for \cref{box:exLOWard}}
If the contraction vanishes, we can neglect all prefactors and just focus on the Lorentz structure.
First, we consider the anti-symmetric component, i.e.\ $T^2_{\mu\nu} - T^1_{\mu\nu}$: since the Levi-Civita tensor is already contracted with both $k$ and $k'$ any further contraction with any linear combination of those, such as $q = k - k'$, must vanish identically
Second, for the remaining symmetric part, $T^1_{\mu\nu}$, we only need to prove one equality since the other is implied.
We need
\begin{equation}
    q = k - k' \Rightarrow q^2 = -2k\cdot k' \Rightarrow 2k\cdot q = 2k \cdot (k-k') = q^2 \land 2k'\cdot q = 2k'\cdot(k-k') = -q^2 \label{eq:solqprod}
\end{equation}
and so
\begin{align}
    T^1_{\mu\nu}q^\mu &= 2\left(k_\mu k'_\nu + k_\nu k'_\mu - (k \cdot k') g_{\mu\nu}\right) q^\mu\\
    &= 2\left((k\cdot q) k'_\nu + k_\nu (k' \cdot q)  - (k \cdot k') q_{\nu}\right)\\
    &= q^2 \left(k'_\nu - k_\nu + q_\nu \right)\\
    &= 0
\end{align}

\paragraph{Solution for \cref{box:exLOg}}
The couplings are trivial $g_{V}^{\gamma l} = 1$, since we removed the global electric charge from their definition.
We do not need to consider the Levi-Civita tensor since we are contracting with a symmetric tensor.
Using \cref{eq:solqprod} we find
\begin{align}
    L^\gamma_{\mu\nu} g^{\mu\nu} &=2\left(k_\mu k'_\nu + k_\nu k'_\mu - (k \cdot k') g_{\mu\nu}\right) g^{\mu\nu}\\
    &=2(k\cdot k')(1 + 1 -4)\\
    &= -2 Q^2
\end{align}

\paragraph{Solution for \cref{box:exLOPP}}
Again, we do not need to consider the Levi-Civita tensor since we are contracting with a symmetric tensor.
We find
\begin{align}
    L^\gamma_{\mu\nu} P^{\mu} P^{\nu} &= 2\left(k_\mu k'_\nu + k_\nu k'_\mu - (k \cdot k') g_{\mu\nu}\right) P^{\mu} P^{\nu}\\
    &= 2 \cdot 2 (k\cdot P) (k' \cdot P)\\
    &= 4 (k\cdot P)^2 - 4 (k \cdot P)(q \cdot P)\\
    &= (2 q \cdot P)^2 \left(\frac 1 {y^2} - \frac 1 y\right)\\
    &= \frac{Q^4}{x^2 y^2}(1-y)
\end{align}
using $k' = k-q$ and \cref{eq:defy,eq:defx}.

\paragraph{Solution for \cref{box:exLOLW}}
For the (spin-averaged) partonic tensor we have
\begin{align}
    \hat W_{\gamma q}^{\mu\nu} &= 2 g_V^{\gamma q}\left(p^\mu p'^\nu + p^\nu p'^\mu - (p \cdot p') g^{\mu\nu}\right)\\
     &= 2g_V^{\gamma q}\left(2 p^\mu p^\nu + p^\mu q^\nu + p^\nu q^\mu - (p\cdot q) g^{\mu\nu} \right)
\end{align}
where we just use $p' = p + q$.
Eventually, we get
\begin{align}
    L^\gamma_{\mu\nu} \hat W^{\mu\nu}_{\gamma q} &= 2g_V^{\gamma q} L^\gamma_{\mu\nu} \left(2 p^\mu p^\nu + p^\mu q^\nu + p^\nu q^\mu - (p\cdot q) g^{\mu\nu} \right)\\
    &= 2g_V^{\gamma q} L^\gamma_{\mu\nu} \left(2\xi^2 P^\mu P^\nu - \xi\frac{Q^2}{2x} g^{\mu\nu}\right) \label{eq:solLWp}\\
    &= 2g_V^{\gamma q} \left(2\frac{Q^4}{z^2 y^2}(1-y) + \frac{Q^2}{2z} 2 Q^2 \right) \label{eq:solLWz}\\
    &= 2\qty(g_{V}^{\gamma q})^2\frac{Q^4}{z^2 y^2}\left(2-2y + z y^2\right)
\end{align}
where
\begin{itemize}
    \item in \cref{eq:solLWp} we use $p=\xi P$, \cref{eq:defx}, and \cref{box:exLOWard}
    \item in \cref{eq:solLWz} we use $\xi = x/z$
\end{itemize}

\section{RSL}
\paragraph{Solution to \cref{box:RSLD1}}
Using \cref{eq:RSLD} we find
\begin{align}
    C_R(z) &= 0\\
    C_S(z) &= \frac{\ln(1-z)}{1-z}\\
    C_L(x) &= -\int\limits_0^x\! \dd z ~ \frac{\ln(1-z)}{1-z} = \int\limits_0^x\! \dd z \ln(1-z) \dv {\ln(1-z)} {z} = \frac 1 2 \ln^2(1-x)
\end{align}

\paragraph{Solution to \cref{box:rslprod1}}
Let $f(\xi)$ be a sufficiently smooth test function. We find
\begin{align}
    &(C\otimes f)(x)\nonumber\\
    &= \int\limits_x^1 \frac{\dd z}{z} f(x/z) r(z) \cdot \left[ p(z) \right]_+\\
    &= \int\limits_0^1 \frac{\dd z}{z} f(x/z) r(z) \cdot \left[ p(z) \right]_+ - \int\limits_0^x \frac{\dd z}{z} f(x/z) r(z) \cdot \left[ p(z) \right]_+\\
    &= \int\limits_0^1\! \dd z \left(\frac{f(x/z)r(z)}{z} - f(x)r(1)\right) \cdot p(z) - \int\limits_0^x\!\dd z\, \frac{ f(x/z) r(z)}{z} \cdot p(z)\\
    &= \int\limits_x^1\! \dd z \left(\frac{f(x/z)r(z)}{z} - f(x)r(1)\right) \cdot p(z) - f(x) r(1) \int\limits_0^x\dd z~ p(z)\\
    &= \int\limits_x^1\! \dd z \left(\frac{f(x/z)(r(z)+r(1)-r(1))}{z} - f(x)r(1)\right) \cdot p(z) - f(x) r(1) \int\limits_0^x\dd z~ p(z)\\
    &= \int\limits_x^1\! \dd z \left(\frac{f(x/z)}{z} - f(x)\right) r(1)\cdot p(z) + \int\limits_x^1\! \dd z \frac{f(x/z)(r(z)-r(1)))}{z} p(z) \nonumber\\
    &\hspace{20pt}  - f(x) r(1) \int\limits_0^x\!\dd z~ p(z)\\
    &= \int\limits_x^1 \frac{\dd z}{ z} f(x/z)  r(1)\cdot \left[p(z)\right]_+ + \int\limits_x^1\! \dd z \frac{f(x/z)(r(z)-r(1)))}{z} p(z) - f(x) r(1) \int\limits_0^x\!\dd z~ p(z)
\end{align}
\begin{equation}
    \Rightarrow C^R(z) = (r(z)-r(1))p(z)\,,~  C^S(z) = r(1)p(z)\,,~ C^L(x) = -r(1)\int\limits_0^x\!\dd z\, p(z)
\end{equation}

\paragraph{Solution to \cref{box:rslprod2}}
\begin{align}
    \int\limits_0^1 \!\dd z~ f(z) \left[r(z)p(z)\right]_+ &= \int\limits_0^1 \dd z \left(f(z) - f(1)\right)r(z)p(z)\\
      &= \int\limits_0^1 \left(f(z)r(z) - f(1)r(1)\right)p(z)~\dd z - f(1)\int\limits_0^1\! \dd z(r(z)-r(1))p(z)\\
      &= \int\limits_0^1\! \dd z~ f(z)\left(r(z) \left[p(z)\right]_+\right) - f(z)\left(\delta(1-z)\int\limits_0^1\! dy~ r(y) \left[g(y)\right]_+\right)
\end{align}

Moreover, we have
\begin{align}
    P_{qq}^{(0)}(y) &= C_F\left( \frac{1+y^2}{1-y}\right)_+ \\
     &=C_F\left( \frac{1+y^2}{(1-y)_+} - \delta(1-y)\int\limits_0^1 \!\dd z~ \frac{1+z^2 -2}{1-z}\right)\\
     &=C_F\left( \frac{1+y^2}{(1-y)_+} + \delta(1-y)\int\limits_0^1 \!\dd z~ 1+z\right)\\
     &=C_F\left( \frac{1+y^2}{(1-y)_+} + \frac 3 2\delta(1-y)\right)
\end{align}

\section{Resummation}
\paragraph{Solution for \cref{box:LOalphas}}
\begin{align}
    \alpha_s(\mu_R^2) &= \frac{\alpha_s(\mu_{R,0}^2)}{1 + \alpha_s(\mu_{R,0}^2)\beta_0\ln(\mu_R^2/\mu_{R,0}^2)}\\
    &= \frac 1 {\beta_0} \frac{1}{1/(\alpha_s(\mu_{R,0}^2)\beta_0) + \ln(\mu_R^2/\mu_{R,0}^2)}\\
    &= \frac 1 {\beta_0} \left\{\ln\left[\mu_R^2/\mu_{R,0}^2\exp(1/(\alpha_s(\mu_{R,0}^2)\beta_0))\right]\right\}^{-1}\\
    &= \frac 1 {\beta_0 \ln(\mu_R^2/\LQCD^2)}
\end{align}
with
\begin{equation}
    \LQCD^2 = \mu_{R,0}^2\exp(-1/(\alpha_s(\mu_{R,0}^2)\beta_0))
\end{equation}

\paragraph{Solution for \cref{box:LOalphassol}}
\begin{align}
    \dv {\alpha_s^\text{LO}(t_R)}{t_R} &= \dv {t_R}  \frac{\alpha_s(0)}{1 + \alpha_s(0) \beta_0 t_R}\\
     &= \frac{\alpha_s(0)}{(1 + \alpha_s(0) \beta_0 t_R)^2} \cdot (-1) \cdot \alpha_s(0) \beta_0\\
     &= -\beta_0 \left(\alpha_s^\text{LO}(t_R)\right)^2
\end{align}

\paragraph{Solution for \cref{box:NLOalphas}}
Step 1: compute LHS.
\begin{align}
    \dv {\alpha_s^\text{NLO}(t_R)}{t_R} &= \dv {t_R} \left(\alpha_s^\text{LO}(t_R) - \frac{\beta_1}{\beta_0} \left(\alpha_s^\text{LO}(t_R)\right)^2 \ln(\alpha_s(0) / \alpha_s^\text{LO}(t_R)) \right)\\
     &= \dv {\alpha_s^\text{LO}(t_R)}{t_R} - \frac{\beta_1}{\beta_0} \left[ 2 \alpha_s^\text{LO}(t_R) \dv {\alpha_s^\text{LO}(t_R)}{t_R} \ln(\alpha_s(0) / \alpha_s^\text{LO}(t_R)) \right. \nonumber\\
     &\hspace{120pt} \left. - \left(\alpha_s^\text{LO}(t_R)\right)^2 \dv{\ln(\alpha_s^\text{LO}(t_R))}{t_R}  \right]\\
     &= -\beta_0 \left(\alpha_s^\text{LO}(t_R)\right)^2 - \beta_1\left[  -2\left(\alpha_s^\text{LO}(t_R)\right)^3\ln(\alpha_s(0) / \alpha_s^\text{LO}(t_R)) + \left(\alpha_s^\text{LO}(t_R)\right)^3 \right]\\
     &= -\beta_0 \left(\alpha_s^\text{LO}(t_R)\right)^2 - \beta_1\left(\alpha_s^\text{LO}(t_R)\right)^3\left[1  -2\ln(\alpha_s(0) / \alpha_s^\text{LO}(t_R))\right]
\end{align}

Step 2: compute RHS.
\begin{align}
    &-\beta_0 \left(\alpha_s^\text{NLO}(t_R)\right)^2 -\beta_1 \left(\alpha_s^\text{NLO}(t_R)\right)^3 \nonumber\\
    &=-\beta_0 \left[\left(\alpha_s^\text{LO}(t_R)\right)^2 - 2 \frac{\beta_1}{\beta_0} \left(\alpha_s^\text{LO}(t_R)\right)^3 \ln(\alpha_s(0) / \alpha_s^\text{LO}(t_R)) \right] \nonumber\\
    &\hspace{20pt} - \beta_1 \left(\alpha_s^\text{LO}(t_R)\right)^3 + \order{(\alpha_s^\text{LO}(t_R))^4}\\
    &= -\beta_0 \left(\alpha_s^\text{LO}(t_R)\right)^2 - \beta_1\left(\alpha_s^\text{LO}(t_R)\right)^3\left[1  -2\ln(\alpha_s(0) / \alpha_s^\text{LO}(t_R))\right]  + \order{(\alpha_s^\text{LO}(t_R))^4}
\end{align}
where we immediately neglect all term $\order{(\alpha_s^\text{LO}(t_R))^4}$.

\paragraph{Solution for \cref{box:NLOalphasLL}}
Step 1: expand $\alpha_s^\text{LO}(t_R)$.
\begin{equation}
    \alpha_s^\text{LO}(t_R) = \alpha_s(0)\left[1 - \alpha_s(0)\beta_0 t_R + (\alpha_s(0)\beta_0)^2 t_R^2 \right] + \order{t_R^3}
\end{equation}

Step 2: expand $\alpha_s^\text{NLO}(t_R)$.
Note that for the hint we use
\begin{equation}
    \ln(\alpha_s(0)/\alpha_s^\text{LO}(t_R)) = \ln(1 + \alpha_s(0)\beta_0 t_R) = \alpha_s(0) \beta_0 t_R - \frac 1 2 (\alpha_s(0)\beta_0)^2 t_R^2  + \order{t_R^3}
\end{equation}
by using \cref{eq:LOalphast}.
Together with step 1 we find
\begin{align}
    \alpha_s^\text{NLO}(t_R) &= \alpha_s(0)\left[1 - \alpha_s(0)\beta_0 t_R + (\alpha_s(0)\beta_0)^2 t_R^2 \right] \nonumber\\
    &\hspace{20pt} - \frac{\beta_1}{\beta_0} \alpha_s^2(0) \left[1 - 2\alpha_s(0)\beta_0 t_R + 3(\alpha_s(0)\beta_0)^2 t_R^2 \right]\nonumber\\
    &\hspace{70pt} \cdot \left[\alpha_s(0) \beta_0 t_R - \frac 1 2 (\alpha_s(0)\beta_0)^2 t_R^2 \right] + \order{t_R^3}\\
    &= \alpha_s(0)\left[1 + \left(-\beta_0\alpha_s(0) - \beta_1\alpha_s^2(0) \right) t_R\right. \nonumber\\
    &\hspace{50pt}\left. +\left(\beta_0^2 \alpha_s^2(0) + \frac 5 2 \beta_1 \beta_0 \alpha_s^3(0) \right)t_R^2 \right]  + \order{t_R^3} \label{eq:solresumNLLexp}
\end{align}
where the term $~\sim \alpha_s^3(0) t_R^2$ receives contributions from two summands.
Note that this is the full list of terms which appear up to $t_R^2$.

Step 3: Taylor-expand $\alpha_s^\text{NLO}(t_R)$.
First, we compute the second derivative:
\begin{align}
    \dv[2]{\alpha_s^\text{NLO}(t_R)}{t_R} &= \dv {t_R} \left( -\beta_0 \left(\alpha_s^\text{NLO}(t_R)\right)^2 -\beta_1 \left(\alpha_s^\text{NLO}(t_R)\right)^3 \right)\\
    &= -2\beta_0 \left(\alpha_s^\text{NLO}(t_R)\right) \dv{\alpha_s^\text{NLO}(t_R)}{t_R} -3\beta_1 \left(\alpha_s^\text{NLO}(t_R)\right)^2 \dv{\alpha_s^\text{NLO}(t_R)}{t_R}\\
    &= \left(-2\beta_0 \alpha_s^\text{NLO}(t_R) - 3\beta_1 \left(\alpha_s^\text{NLO}(t_R)\right)^2\right)\nonumber\\
    &\hspace{50pt} \cdot\left( -\beta_0 \left(\alpha_s^\text{NLO}(t_R)\right)^2 -\beta_1 \left(\alpha_s^\text{NLO}(t_R)\right)^3\right)\\
    &= 2\beta_0^2 \left(\alpha_s^\text{NLO}(t_R)\right)^3 + 5\beta_0\beta_1 \left(\alpha_s^\text{NLO}(t_R)\right)^4 + 3\beta_1^2 \left(\alpha_s^\text{NLO}(t_R)\right)^5
\end{align}
Now, we perform the Taylor-expansion:
\begin{align}
    \alpha_s^\text{NLO}(t_R) &= \alpha_s(0) + \frac 1 {1!}\left.\dv{\alpha_s^\text{NLO}(t_R)}{t_R}\right|_{t_R=0} t_R + \frac 1 {2!}\left.\dv[2]{\alpha_s^\text{NLO}(t_R)}{t_R}\right|_{t_R=0} t_R^2 + \order{t_R^3}\\
    &= \alpha_s(0)\left[1 + \left(-\beta_0\alpha_s(0) - \beta_1\alpha_s^2(0) \right) t_R\right. \nonumber\\
    &\hspace{50pt}\left. +\left(\beta_0^2 \alpha_s^2(0) + \frac 5 2 \beta_1 \beta_0 \alpha_s^3(0) + \frac 3 2 \beta_1^2 \left(\alpha_s^\text{NLO}(t_R)\right)^5 \right)t_R^2 \right]  + \order{t_R^3} \label{eq:solresumNLLTaylor}
\end{align}

Step 4: \cref{eq:solresumNLLexp} and \cref{eq:solresumNLLTaylor} agree at LL, $\alpha_s^k(0) t_R^{k}$ for $k=0,1,2$, and all terms are non-zero.

Step 5: \cref{eq:solresumNLLexp} and \cref{eq:solresumNLLTaylor} agree at NLL, $\alpha_s^k(0) t_R^{k-1}$ for $k=1,2,3$.
Note that the term $\alpha_s^1(0) t_R^{0}$ is trivially 0 in both expressions and thus typically not considered a part of the NLL resummation.

Step 6: \cref{eq:solresumNLLexp} and \cref{eq:solresumNLLTaylor} do not agree at NNLL, $\alpha_s^k(0) t_R^{k-2}$ for $k=2,3,4$.
Note that the terms $\alpha_s^2(0) t_R^{0}$ and $\alpha_s^3(0) t_R^{1}$ are both trivially 0 in both expressions and thus typically not considered a part of the NNLL resummation.
The first non-trivial term $\alpha_s^4(0) t_R^{2}$ appears in the Taylor-expansion \cref{eq:solresumNLLTaylor}, but not in the expansion of \cref{eq:solresumNLLexp}.
This implies that \cref{eq:NLOalphast} is not the true solution to \cref{eq:NLObetat}, as we say above.
However, \cref{eq:NLOalphast} does solve the beta function at the requested accuracy, \cref{box:NLOalphas}, and it does perform the requested NLL resummation.

Note that we merely prove here that the first few terms are available and not the full NLL accuracy, which requires that all terms $\alpha_s^k(0) t_R^{k-1}$ for $k=1,\cdots,\infty$ are present.

\end{appendices}

\begin{backmatter}
\bibliographystyle{utphys}
\bibliography{refs}

\end{backmatter}

\end{document}